\documentclass[preprint,11pt,flushrt]{aastex}
\def\hubunits{\hbox{km \hskip -2pt s$^{-1}$ \hskip -2pt Mpc$^{-1}$}}
\def\kms{{\rm km}\;{\rm s}^{-1}}
\def\msun{{M_\odot}}
\def\Msun{{M_\odot}}
\def\mdot{{\dot{m}}}
\def\Mdot{{\dot{M}}}
\def\epsd{{\epsilon_{\hbox{\tiny 0.1}}}}
\def\Ledd{{L_{\rm Edd}}}
\def\MdotEdd{{\Mdot_{\rm Edd}}}
\def\fon{{f_{\rm on}}}
\def\yr{{\rm yr}}
\def\erg{{\rm erg}}
\def\sec{{\rm sec}}
\def\Lbol{{L_{\rm bol}}}

\def\mdotcrit{{\mdot_{\rm crit}}}
\def\mdotmin{{\mdot_{\rm min}}}
\def\mdotmax{{\mdot_{\rm max}}}

\def\avMdott{{\langle \Mdot(t) \rangle}}
\def\avmdot{{\langle \mdot \rangle}}
\def\avmdotM{{\langle \mdot(M) \rangle}}
\def\avmdott{{\langle \mdot(t) \rangle}}
\def\avmdotMt{{\langle \mdot(M,t) \rangle}}
\def\pstar{{p_{*}}}
\def\nstar{{n_{*}}}
\def\mdotstar{{\dot{m}_{*}}}
\def\Mstar{{M_{*}}}
\def\mdotl{{\dot{m}_{\hbox{\tiny L}}}}
\def\lbrk{L_{\rm brk}}
\def\Lbrk{L_{\rm brk}}

\def\tacc{t_{\rm acc}}
\def\taccz{t_{{\rm acc,}z}}
\def\dunits{{\;{\rm Mpc}^{-3}}}
\def\stq{{short-$t_q$}}
\def\ltq{{long-$t_q$}}

\def\rhobh{\rho_{\rm bh}}
\def\Mmed{M_{\rm med}}
\def\Mbh{M_{\rm bh}}
\def\Lhost{L_{\rm host}}
\def\Lhostmed{L_{\rm host,med}}
\def\Mmin{M_{\rm min}}
\def\Mmax{M_{\rm max}}

\def\be{\begin{equation}}
\def\ee{\end{equation}}

\begin{document}

\title{Accretion Driven Evolution of Quasars and Black Holes: Theoretical
Models}


\author{Adam Steed and David H. Weinberg}
\affil{Department of Astronomy, The Ohio State University, 
Columbus, OH 43210}

\email{\sf asteed,dhw@astronomy.ohio-state.edu}

\shorttitle{Evolution of Quasars and Black Holes}
\shortauthors{Steed \& Weinberg}

\begin{abstract}
We present a flexible framework for constructing physical models of quasar
evolution that can incorporate a wide variety 
of observational constraints,
such as multi-wavelength luminosity functions, estimated masses and
accretion rates of active black holes, space densities of quasar host
galaxies, clustering measurements, and the mass function of black holes in the
local universe.  The central actor in this formulation is the accretion rate
distribution $p(\mdot|M,z)$, the probability that a black hole of mass $M$ at
redshift $z$ accretes at a rate $\mdot$ in Eddington units.  Given a model of
accretion physics that specifies the radiative efficiency and SED shape as a
function of $\mdot$, the quasar luminosity function (QLF) is determined by
a convolution of $p(\mdot|M,z)$ with the black hole mass function $n(M,z)$.
In the absence of mergers, $p(\mdot|M,z)$ also determines the full evolution
of $n(M,z)$, given a ``boundary value'' of $n(M)$
at some redshift.  If $p(\mdot|z)$ is
independent of mass, then the asymptotic slopes of the QLF match the asymptotic
slopes of $n(M)$, and $n(M)$ evolves in a self-similar fashion, retaining its
shape while shifting to higher masses.  Matching the observed decline of the
QLF ``break'' luminosity at $z<2$ requires either a shift in $p(\mdot|z)$ that 
increases the relative probability of low accretion rates 
or an evolving mass dependence of $p(\mdot|M,z)$ that preferentially shuts off
accretion onto high mass black holes at low $z$.  
These two scenarios make different predictions for the 
masses and accretion rates of active black holes.  If the first 
mechanism dominates, then
the QLF changes character between $z=2$ and $z=0$, shifting from a sequence of
black hole mass towards a sequence of $L/\Ledd$.  We use our framework to 
compare the predictions of five models that illustrate different assumptions
about the quasar population: two dominated by unobscured thin-disk accretion
with short and long quasar lifetimes, respectively, one with a 4:1 ratio of
obscured to unobscured systems, one with substantial black hole merger activity
at low redshift, and one with substantial low redshift growth in radiatively
inefficient flows.  We discuss the observational advances that would be most
valuable for distinguishing such models and for pinning down the physics that
drives black hole and quasar evolution.
\end{abstract}

Submitted to The Astrophysical Journal, \today.

\section{Introduction}

The study of the quasar and AGN populations has been transformed in
recent years by ambitious new optical 
(e.g., \citealt{boyle00,schneider02,wolf03}),
X-ray (e.g., \citealt{brandt01,giacconi02,anderson03,ueda03}), and radio 
(e.g., \citealt{white00}) surveys,
by the recognition that low efficiency accretion modes may become
important when the accretion rate itself is low 
(e.g., \citealt{narayan98} and references therein),
by detailed studies of low luminosity AGN in the local universe
(e.g., \citealt{ho01}), and, perhaps most of all, by the accumulating
evidence that supermassive black holes are ubiquitous in the
bulges of present-day galaxies 
(e.g., \citealt{richstone98,merritt00,gebhardt00}).
The dynamical studies of nearby galaxies strengthen the 
long-standing hypothesis 
that quasars are powered by black hole accretion
\citep[e.g.,][]{lyndenbell69,rees78}, and the ``demography'' of the
local black hole population provides a powerful constraint on models
of quasar evolution and its connection to galaxy evolution.
These developments have inspired increasingly sophisticated theoretical
models that place quasar evolution in the context of hierarchical
clustering models for the formation of dark matter halos and
galaxies \citep[e.g.,][]{kauffmann00,cavaliere02,haimanloeb,wyithe03}.

This paper presents a physically motivated calculational framework
that is intermediate in complexity between such {\it ab initio} models
of the quasar population and older descriptions in terms of
``luminosity evolution'' or ``density evolution.''
The central actor in our formulation is the accretion probability
distribution $p(\mdot|M,z)$, the probability that a black hole
of mass $M$ at redshift $z$ is accreting mass at a rate $\mdot$
in Eddington units (discussed in \S\ref{sec:framework} below).
The key supporting players are the black hole mass function $n(M,z)$
and a physical model of accretion that predicts the radiative
efficiency for a given $\mdot$.\footnote{Henceforth,
we will usually drop the explicit dependence on $z$ and refer only
to $p(\mdot|M)$ or $n(M)$, but these should always be understood
to refer to the distribution at some particular redshift.}
An example of an accretion model would be thin-disk accretion
with efficiency $\epsilon \equiv L/\Mdot c^2 \sim 0.1$ when
$\mdot \sim 1$, changing to low efficiency advection-dominated (ADAF)
accretion when $\mdot$ is below
some critical value.
At a given redshift, $p(\mdot|M)$ and $n(M)$ together determine
the quasar luminosity function, and $p(\mdot|M)$ also determines
the accretion driven growth of the black hole population, and hence
the evolution of $n(M)$.  Thus, given physical assumptions about radiative
efficiencies and a ``boundary condition'' specifying $n(M)$ at one
redshift, the history of $p(\mdot|M)$ determines the complete evolution
of the black hole population and the quasar luminosity function.
An essential caveat is that mergers of black
holes following mergers of their host galaxies could alter $n(M)$
independently of the accretion characterized by $p(\mdot|M)$.

The simplest scenario connecting black hole and quasar evolution is that
black holes ``shine as they grow'': a luminous quasar is powered by
a black hole radiating at near-Eddington luminosity with efficiency
$\epsilon \sim 0.1$, and no significant growth occurs in a non-luminous phase.
In this case, the bolometric luminosity function at a given redshift is just
$\Phi(L) = \fon n(L/l) l^{-1}$ , where $l$ is the (universal)
ratio of Eddington luminosity to black hole mass and $\fon$
is the fraction of black holes that are accreting.
In a time interval $\Delta t$, black holes on average increase their mass by
a factor $\exp(\fon \Delta t/t_g)$, 
where $t_g = M/(\Ledd/\epsilon c^2) = 4.5\times 10^7(\epsilon/0.1)\;\yr$ is the
$e$-folding time for growth at the Eddington luminosity
(\citealt{salpeter64}; see discussion in \S\ref{sec:framework}).   
The mass density in black holes at the present day 
is simply related to the emissivity of the quasar population
integrated over luminosity and redshift, 
$\rho_{\rm BH} = \int_0^{t_0} U(t) dt / \epsilon c^2$ 
(\citealt{soltan82}, updated by, e.g.,
\citealt{chokshi92,richstone98,yu02,fabian03}).
In our language, this is a model in which all active black holes have
the same radiative efficiency and 
$p(\mdot|M)=\fon\delta_D(\mdot-1)$, where $\delta_D$ is
the Dirac-delta function.  The evolution of quasars and black
holes is determined by a boundary condition on $n(M)$ and the
redshift history of the active fraction $\fon(z)$.  
The ``quasar era'' $z\sim 2-4$ when the
emissivity of the population peaks is also the era in which
today's black holes grew to their current mass.

This simple scenario may not be too far from the truth, but the 
possible complications raise a number of questions.
Did today's black holes gain a significant fraction of their mass
through low-$\mdot$, low efficiency, ADAF-type accretion, thus
growing at low luminosity?  Have black hole mergers substantially
altered $n(M)$, leaving the integrated 
density $\rho_{\rm BH}$ fixed but changing the relative numbers
of high and low mass black holes?
Are some quasars accreting mass at super-Eddington rates,
radiating at high luminosity but low efficiency in 
``smothered,'' optically thick ADAF modes 
\citep{katz77,begelman78,abramowicz88}?
Do some quasars radiate substantially above the 
Eddington limit \citep{begelman02}?
Is a significant fraction of quasar activity obscured by gas and dust,
as hypothesized in synthesis models of the X-ray background 
(e.g., \citealt{setti89,comastri95,fiore99,fabian99,gsh01}),
thus redistributing 
bolometric luminosity from the optical-UV-soft X-ray to the far-IR?
Are low luminosity AGNs powered mainly by low mass black holes
radiating at Eddington luminosity, by more massive black holes
with thin-disk efficiencies but sub-Eddington accretion rates,
or by still more massive black holes with sub-Eddington accretion
rates {\it and} low efficiency?  The methods developed here provide
useful tools for addressing these questions, allowing us to construct
concrete, quantitative models that answer them in different ways,
then examine how observational data might distinguish among such models.

Our framework complements, but by no means replaces, the {\it ab initio}
approach that connects the evolution of quasars and black holes to that
of the underlying dark halo and galaxy populations.  This approach has
yielded many valuable insights, including the recognition 
that the rise of the quasar population probably traces the formation of 
the first dark halos large enough to host massive black holes, that the
rapid decline of the population at low redshift probably reflects the
combined impact of declining galaxy interaction rates and decreasing
gas supplies in quasar hosts, and that the clustering of quasars
with themselves or with galaxies can provide a valuable diagnostic 
of typical quasar lifetimes
\citep[e.g.,][]{efstathiou88,haehnelt93,hnr98,salucci99,kauffmann00,
cavaliere00,haiman01,martini01,mhn01,wyithe03}.
In combination with semi-analytic models of galaxy formation,
these quasar evolution models can also predict the properties and environments
of quasar hosts and the relation between the properties of present day
galaxies and the masses of their central black holes.
However, the models necessarily rely on specific assumptions about the
mechanisms that trigger quasar activity and the accretion rates
that these mechanisms produce.  To put things in our terms, the
{\it ab initio} 
models adopt a particular set of hypotheses about quasar activity
in order to predict $p(\mdot|M,z)$ from first principles.

Our framework is designed to model observational data in a flexible
way with relatively few assumptions, while retaining the basic physical
picture of black hole accretion that underlies nearly all modern 
interpretations of the quasar population.  One hope is that 
measurements of the luminosity function and the local black hole
mass function will eventually allow us to integrate backwards in time
and determine $p(\mdot|M,z)$ empirically, drawing on a variety of
observations to test the assumptions that enter such a reconstruction.
We may find that the data are not powerful enough to tie down
$p(\mdot|M,z)$ without some {\it a priori} constraints on its expected form,
but that finding in itself would be a valuable, if disappointing, lesson.
More generally, we hope to illuminate the connections between black
hole evolution and quasar activity and learn what observations can
and cannot tell us about these connections.

The \cite{hnr98} paper has had the strongest impact
on our thinking about these issues, but the most direct antecedent
that we know of to the approach taken here is the lucid paper
of \cite{small92}.  They adopted a similar description of black
hole evolution and its connection to quasar activity, and they
applied this description to the observational data available at the
time.  Advances in the observational data and the theoretical models
of accretion make this an opportune time to revive and extend this
approach.  \cite{yu02} have recently used a similar method in
assessing constraints on black hole accretion and mergers,
though their assumptions and goals are more restrictive than ours ---
in particular, they assume that quasars radiate at Eddington
luminosity and thus that $p(\mdot)$ consists of $\delta$-functions
at $\mdot=1$ and $\mdot=0$.

The difference in the form of $p(\mdot)$ is one of the most significant
differences between the models presented in this paper and most
models of the quasar population in the literature.  These typically
assume that $p(\mdot)$ for $\mdot>0$
is sharply peaked at some value close to Eddington,
such as a $\delta$-function (e.g., \citealt{small92})
or a spike followed by an exponential decline 
(e.g., \citealt{hnr98,kauffmann00}).
A sharply peaked $p(\mdot)$ could arise physically if the
central black hole typically plays a large role in controlling
its own fuel supply, through feedback or influence on stellar dynamics.
However, we think it is more likely that fueling is driven by
galactic scale events --- galaxy mergers, interactions, and bar
formation, for example --- that are minimally influenced by the
central black hole, and therefore do not ``know'' that they should
feed it at any particular rate.  In particular, it seems reasonable
that for every major event that leads to Eddington-like fueling
of a central black hole there are many minor events that
fuel it at a sub-Eddington rate.  The particular functional
forms that we adopt here to represent this scenario, a power-law or
broken power-law $p(\mdot)$ between some $\mdotmin$ and some
$\mdotmax$, are arbitrary, and chosen largely for
mathematical convenience, but they reflect this general thinking
about the process of quasar fueling.
In the long run, one goal of our effort is to test observationally whether 
$p(\mdot)$ is in fact a broad function or a peaked function, which
would in turn have implications for the mechanisms of quasar fueling.
At a qualitative level, the wide $L/\Ledd$ distribution of AGN activity
in the local universe (e.g, \citealt{ho01}) seems to support
the idea of a broad distribution of accretion rates. 

In this paper we will keep our contact with observations
relatively loose, focusing instead on presenting our framework
in a clear way and illustrating how observations might distinguish
among different scenarios for the quasar population.  
We will carry out a detailed analysis of multi-wavelength measurements of the
quasar luminosity function and constraints on active black hole
masses and the local black hole mass function a subsequent paper.  
We adopt a cosmological model with $\Omega_m=1$ and 
$h \equiv H_0/100\;\hubunits=0.5$ because all of the observational
papers include results for this model, and not always for the
low density, $\Lambda$-dominated, $h\sim 0.7$ model favored by
recent cosmological observations.  Cosmology affects our models
indirectly through its influence on observationally inferred
luminosity functions and directly through the time-redshift
relation.  We would not expect a change of cosmological model
to have any qualitative impact on our results, and we expect that
the quantitative impact could be compensated by modest changes
to mass scales and accretion rates, especially since the age of the universe
is similar for $\Omega_m=1$, $h=0.5$, and for $\Omega_m=0.3$,
$\Omega_\Lambda=0.7$, $h=0.7$.

We present the definitions and key equations of our framework
in the following section.  We then present mathematical results
for luminosity functions in \S\ref{sec:lumfunction} and for luminosity
and black hole mass evolution in \S\ref{sec:evolution},
focusing on analytically solvable cases that illustrate general points.
In \S\ref{sec:illus} we construct five illustrative models
of the quasar population, each designed to match the observed
evolution of the optical luminosity function but differing from
one another in the typical quasar lifetime or in the importance
of mergers, ADAF growth, or obscuration. We show how measurements
of the black hole mass function, luminosity functions at other wavelengths, 
the masses and accretion rates of active black holes, and the space
density of host galaxies might distinguish among these scenarios.  
We summarize our results in \S\ref{sec:summary}.  This is a long
paper, and a reader who wants to get quickly to the main points can read 
\S\ref{sec:framework}, look through the figures and captions,
and read \S\ref{sec:summary}.  The definitions of the
models illustrated in Figures~\ref{fig:bbandevol}--\ref{fig:spacedens}
are summarized in Table~\ref{tbl:parameters} and its accompanying caption.

\section{Framework}
\label{sec:framework}

\subsection{Definitions and assumptions}
\label{sec:defs}

We define $l$ to be the ratio of a black
hole's Eddington luminosity to its mass,
\be
l \equiv {\Ledd\over M} = 
{4\pi G m_p c \over \sigma_T} = 1.26 \times 10^{38} \erg\;\sec^{-1}\msun^{-1}.
\label{eqn:ldef}
\ee
We often scale the radiative efficiencies to the value $0.1$ 
that is typically adopted
for thin-disk accretion,
\be
\epsilon \equiv {L \over \Mdot c^2} = 0.1\epsd,
\label{eqn:epsdef}
\ee
where $\Mdot$ is the mass accretion rate and $L$ is the
bolometric luminosity.  We define the Eddington
accretion rate to be the mass accretion rate for which a black hole
{\it with radiative efficiency $\epsd=1$} has Eddington luminosity,
\be
\MdotEdd \equiv {\Ledd \over 0.1 c^2} \approx 22
  \left({M \over 10^9 \msun}\right) \msun\;\yr^{-1},
\label{eqn:MdotEdddef}
\ee
and the dimensionless accretion rate
\be
\mdot \equiv {\Mdot\over \MdotEdd}\ .
\label{eqn:mdotdef}
\ee
This definition of $\MdotEdd$ in terms of a fixed, thin-disk efficiency
is common in the literature, but not universal.
A black hole accreting at $\MdotEdd$ grows in mass exponentially,
with an $e$-folding timescale
\be
t_s \equiv {M \over \MdotEdd} = 4.5\times 10^7 \yr.
\label{eqn:tsdef}
\ee
For $\epsd=1$, $t_s$ is equal equal to the \cite{salpeter64} timescale 
for growth at the Eddington luminosity; note, however, that we define
$t_s$ to be $4.5\times 10^7\,\yr$ independent of efficiency and of $L/\Ledd$.
With these definitions, a black hole of mass $M$ accreting at
a dimensionless rate $\mdot$ with radiative efficiency $\epsilon=0.1\epsd$
has bolometric luminosity
\be
L = 0.1\epsd \Mdot c^2 = \epsd \mdot l M\ .
\label{eqn:luminosity}
\ee

At each redshift, we define the black hole mass function $n(M)$
such that $n(M)dM$ is the comoving space density of black holes
in the mass range $M \rightarrow M+dM$; the units of $n(M)$ are
thus number per comoving volume per unit mass.
In our plots, we usually show $Mn(M)$, the comoving space density
(with units Mpc$^{-3}$) in an interval $d\ln M$, rather than $n(M)$ itself.
We define the accretion probability distribution $p(\mdot|M)$
such that $p(\mdot|M)d\mdot$ is the probability that a black
hole of mass $M$ has an accretion rate in the range
$\mdot \rightarrow \mdot+d\mdot$.  
In our subsequent calculations, we will frequently consider 
the restricted and analytically convenient class of models
in which $p(\mdot|M)=p(\mdot)$, i.e., with accretion
probability distribution independent of mass.
Since more massive black holes reside in more massive galaxies
that have larger internal gas supplies and can cannibalize
larger companions, this assumption could be a reasonable approximation
to the real universe (see further discussion in \S\ref{sec:massdepdeclining}
below).  However, it is at best a convenient 
approximation, and we will try to be mindful of its limitations.

Many theoretical models of quasar evolution specify $p(\mdot)$ implicitly
through a typical ``light curve'' $L(t)$ that follows each triggering
event.  The probability $p(\mdot)d\mdot$ can be identified with the
fraction of time that $L/\Ledd$ is in the range 
$\epsd\mdot \rightarrow \epsd(\mdot+d\mdot)$.  However, the
relation between light curves and $p(\mdot)$ distributions is many-to-one,
even for constant efficiency.  For example, if every quasar lights
up at the Eddington luminosity and declines exponentially thereafter,
with $\epsd=1$ throughout, then there is constant accretion probability
per logarithmic interval of $\mdot$, and thus $p(\mdot) \propto \mdot^{-1}$
for $\mdot \leq 1$.  But the same $p(\mdot)$ could correspond to an 
``on-off'' activity model where the luminosity of each quasar
is constant for a fixed time while 
the number of quasars per luminosity interval is
proportional to $(L/\Ledd)^{-1}$.  

The model of accretion physics specifies the probability 
that a black hole of mass $M$ and accretion
rate $\mdot$ has bolometric radiative efficiency $\epsd$.
Since most of the properties of accretion flows scale in a fairly
simple way with mass, it is reasonable to expect that this probability
distribution is independent of $M$.  To simplify our calculations,
we will also assume that all black holes of the same $\mdot$ have
the same $\epsd$, i.e., we will assume that the probability 
distribution of $\epsd$ has zero width at given $\mdot$.
Outside of a small range of $\mdot$ where black
holes might cycle between thin-disk accretion and an ADAF-type flow,
this assumption seems plausible, though one could imagine that 
a range of black hole spins could induce a range of $\epsd$ values
even at fixed $\mdot$.  We will generally assume that 
$\epsd = 1$ in a range $\mdotcrit \leq \mdot \leq 1$,
where $\mdotcrit$ is a critical value of the accretion rate below
which thin-disk accretion gives way to some lower efficiency mode.
For stellar mass black holes, which exhibit transitions among accretion
modes, $\mdotcrit$ may be as high as 0.09 \citep{esin97}, 
but for supermassive black holes the value is more uncertain
and probably lower (R.\ Narayan, private communication).
We will adopt $\mdotcrit=0.01$.
Below this threshold, we will assume $\epsd = (\mdot/\mdotcrit)$,
a scaling motivated by ADAF models \citep{narayan98}.

If the accretion rate is determined by large scale gas flows 
beyond the influence of the black hole itself, then there is
no particular reason that $\mdot$ should not exceed one.
In such situations, we will assume that the accretion proceeds in 
a lower efficiency, ``smothered'' mode 
\citep{katz77,begelman78,abramowicz88}, so that
the black hole radiates at the Eddington luminosity ---
in other words, $\epsd = \mdot^{-1}$ when the accretion rate
is super-Eddington.  
It is not clear that these smothered
accretion modes are stable enough to exist in nature,
and it may be that black holes in this situation
drive the excess gas out of the nucleus altogether rather
than accepting mass at a super-Eddington rate, thus regulating the
accretion rate so that $\mdot$ never exceeds one.
The distinction between these two pictures is usually not important
for our purposes, provided that the black holes radiate at the Eddington
luminosity in either case, but it does make a difference if 
super-Eddington accretion makes a significant contribution to
black hole mass growth. It has also been suggested \citep{begelman02} that 
black holes can radiate substantially above the Eddington luminosity (an order
of magnitude or more) --- this {\it would} make a difference to our
predictions, as we discuss briefly in \S\ref{sec:summary}.

To summarize, we generally assume that the bolometric radiative efficiency
is 
\be
\epsd(\mdot)= \left\{ \begin{array}{cc}
                      (\mdot/\mdotcrit) & \mbox{for $\mdot < \mdotcrit$} \\
                      1 & \mbox{for $\mdotcrit \leq \mdot \leq $1} \\
                      \mdot^{-1} & \mbox{for $\mdot > 1$},
                      \end{array}
              \right.
\label{eqn:effofmdot}
\ee
with $\mdotcrit=0.01$.
We will also usually assume that $p(\mdot)$ is non-zero only over
some range $\mdotmin$ to $\mdotmax$, and in some simplified cases we will
choose this range so that only the thin-disk regime with $\epsd(\mdot)=1$
contributes.  To calculate luminosities in particular wavebands, we will
incorporate luminosity fractions $F_\nu$, which in some cases we allow
to depend on $\mdot$ or to vary from one black hole to another. 
We discuss our assumptions about $F_\nu$ as they arise.

\subsection{The luminosity function and the black hole evolution equation}
\label{sec:qlfevoleqn}

The QLF is obtained from the black hole mass function and the
accretion probability distribution by a straightforward counting argument.
Black holes in the mass range $M\rightarrow M+dM$ with 
accretion rate $\mdot$ correspond to quasars with luminosity
in the range $L \rightarrow L+dL$ with
$L=\epsd\mdot l M$ and $dL = (\epsd\mdot l) dM$ (eq.~\ref{eqn:luminosity}).
The comoving space density of black holes in this mass range
is $n(M)dM$, and the corresponding density of quasars with 
luminosity $L\rightarrow L+dL$, is obtained by multiplying
by the accretion probability $p(\mdot|M)$ and integrating over $\mdot$:
\be
\Phi(L) dL = \int_\mdotmin^\mdotmax d\mdot \;
  p(\mdot | M) n(M) {dL\over \epsd\mdot l}, 
  \qquad M={L \over \epsd \mdot l},
\label{eqn:lumfun}
\ee
where the integral covers the full range over which $p(\mdot|M)$
is non-zero.  
One can obtain a mathematically equivalent expression by 
identifying the luminosity range $dL$ 
with the range of accretion rates $d\mdot=dL/(\epsd l M)$ at fixed $M$,
then integrating over masses.
The above argument is slightly complicated by the possibility
that $\epsd$ depends on $\mdot$, but provided that $\epsd \mdot$ is a
monotonic function of $\mdot$, the probability
density transformation $p(\mdot)d\mdot = p(\epsd\mdot)d(\epsd\mdot)$
still leads to equation~(\ref{eqn:lumfun}).
For our standard assumption about super-Eddington accretion,
$\epsd\mdot=1$ for $\mdot>1$, the luminosity function of
super-Eddington accretors is 
\be
\Phi_{\rm SE}(L) dL = n(L/l){dL\over l} \int_1^\mdotmax d\mdot \;
  p(\mdot | M=L/l),
\ee
so equation~(\ref{eqn:lumfun}) remains valid when $\mdotmax > 1$.
Allowing a range of efficiencies at fixed $\mdot$ would broaden
the luminosity function, since one would then convolve $n(M)$
with $p(\epsd\mdot)$ instead of $p(\mdot)$ to obtain $\Phi(L)$.

Equation~(\ref{eqn:lumfun}) gives the bolometric luminosity function
in terms of $n(M)$ and $p(\mdot|M)$ at a particular redshift.
If accretion is the only process contributing to black hole 
growth, then the evolution of $n(M)$ is determined by a simple
continuity equation,
\be
\frac{\partial n(M,t)}{\partial t}=-\frac{\partial (n \avMdott)}{\partial M}
=-\frac{1}{t_{s}}\frac{\partial (nM\avmdotMt)}{\partial M},
\label{eqn:acc}
\ee
which follows from considering the rate at which black holes are
leaving and entering a mass range $M\rightarrow M+dM$ \citep{small92}.
(The last equality follows from
equations~[\ref{eqn:mdotdef}] and~[\ref{eqn:tsdef}], which imply that 
$\Mdot = \mdot M/t_s$.)
The important simplification that follows from the continuity argument
is that the evolution of $n(M)$ depends on $p(\mdot|M)$ only
through the mean accretion rate,
$\avmdotMt =\int d\mdot\; \mdot\; p(\mdot|M,t)$.
In any given time interval, some black holes
will grow faster than average and some will grow slower, but the
mass function is an average over the full population, and its evolution
depends only on the average rate at which black holes move from
one mass range to another.

If $\avmdott$ is independent of $M$, then it can be moved out of the
derivative on the r.h.s. of equation~(\ref{eqn:acc}), and in this
case the solution to the evolution equation is remarkably simple:
\be
n(M,t)=F \left( \frac{M}{M_{*}}\right)\frac{M_{*,i}}{M_{*}}, \qquad
M_*(t) = M_{*,i} \exp 
         \left( \int_{t_{i}}^{t} \avmdott \frac{dt}{t_{s}} \right),
\label{eqn:accgrowth}
\ee
where $F(x)$ is an arbitrary function and $M_*$ is any fiducial scale
in the mass function.
This solution can be verified by direct substitution into 
equation~(\ref{eqn:acc}), but its physical basis is easy to see.
Since $\avmdott$ is independent of mass, and only $\avmdot$ matters
rather than the full distribution $p(\mdot)$, the evolution is
equivalent to that of a population in which all black holes 
accrete mass at the rate $\Mdot = \avmdott\MdotEdd = \avmdott M/t_s$.
In this case, every black hole grows by the exponential factor
on the r.h.s. of equation~(\ref{eqn:accgrowth}), so the scale of the
mass function simply shifts; the normalization drops in proportion
to $1/M_*$ because the width of the differential mass range $dM$
occupied by a given set of black holes increases as the black
hole masses themselves increase.
The simplicity of the solution~(\ref{eqn:accgrowth}) makes models
in which $p(\mdot)$ is independent of $M$ more tractable than
others, so this restricted class of models is useful for gaining
intuition and illustrating different types of behavior.

The more general equation for the evolution of $n(M,t)$ is
\begin{eqnarray}
\frac{\partial n(M,t)}{\partial t}&=&
  -\frac{1}{t_{s}}\frac{\partial (nM\avmdotMt)}{\partial M}
    +C(M,t)-D(M,t) \nonumber \\
  &+&\int_{0}^M dM' n(M-M',t)n(M',t)\Gamma(M-M',M',t) \nonumber \\
  &-&n(M,t)\int_{0}^{\infty} dM'n(M',t)\Gamma(M,M',t) ~,
\label{eqn:evol}
\end{eqnarray}
where the last two terms represent formation of black holes of mass $M$
by mergers of black holes of mass $M'$ and $M-M'$ and loss of black
holes of mass $M$ by mergers with other black holes,
and the creation ($C$) and destruction ($D$) terms allow for 
processes that are neither smooth accretion nor mergers.
Equation~(\ref{eqn:evol}) is essentially the same equation
that \cite{murali02} use to model the evolution of the galaxy
mass function by accretion and mergers (see their equation A1).
The genuine destruction of supermassive black holes seems an unlikely
prospect, but they could be removed from the population of
galactic nuclei by ejection in three-body encounters 
(e.g., \citealt{valtonen94}),
which would effectively count as destruction for our purposes.
We will not consider this possibility in our models here,
but it could be a significant effect if there is a long delay
between the merger of galaxies and the merger of their central
black holes (see, e.g., \citealt{madau03}).
We will also assume that there is no black hole creation in
the mass and redshift range that we consider, i.e., all
of the black holes are already present at the highest redshift
in our calculations, and their masses change only by accretion and mergers.
A calculation that included the formation of ``seed'' black
holes would need to incorporate the creation term $C(M,t)$,
but here we treat the mass distribution of these seeds
as the initial condition for evolution of $n(M)$.
In the context of our models, a ``seed'' black hole is one that forms
at low efficiency by a process that is not well described by the same 
$p(\mdot)$ governing most accretion driven growth --- e.g., by the
collapse of a supermassive star, a relativistic gas disk, or 
a relativistic star cluster.

If the accretion, creation, and destruction terms are ignored,
equation~(\ref{eqn:evol}) becomes the ``coagulation equation,''
which has been widely studied in the context of planetesimal
growth (see \citealt{lee00} and references therein) 
and applied on occasion to star formation and galaxy clustering
(e.g., \citealt{silk78,silk79,murray96,sheth98}).  
Unfortunately, the coagulation equation has
analytic solutions only for rather specialized forms of the
collision rates $\Gamma(M,M',t)$ that do not seem particularly
applicable to black hole evolution, and even these solutions
no longer apply once accretion is also important.
The usefulness of equation~(\ref{eqn:evol}) is therefore largely
conceptual, and as a basis for numerical calculations given some
physically motivated forms of the collision rates.
In this paper, we will restrict our consideration of mergers
to idealized recipes that are, we hope, sufficient to illustrate
their generic effects.

\section{Luminosity Functions}
\label{sec:lumfunction}
\subsection{Power-Law p($\mdot$) for $\mdot$ in the Thin-Disk Range}
\label{sec:powerlawpofmdot}

 The functions $p(\mdot | M)$ and $n(M)$ could in principle have complicated 
forms, but it proves useful to consider some simple forms and determine  
qualitative behaviors that would hold true in more general cases.  We 
begin by 
implementing our framework in a specific case where we consider only the 
range of
accretion rates that corresponds to thin-disk accretion, i.e. 
$\mdotcrit<\mdot<1$ and thus $\epsd=1$.
We also assume that the probability that a black hole
accretes at a rate $\mdot$ is independent of its mass, i.e. 
$p(\mdot|M)=p(\mdot)$.  We first consider 
a truncated power-law,
\be
\begin{array}{cc}
p(\mdot)=\pstar \mdot^{a}, &  \mbox{ $\mdotmin<\mdot<\mdotmax$.}
\end{array}
\label{eqn:powpofmdot}
\ee
[There is also a $\delta-$function at $\mdot=0$, representing inactive
black holes, so that $p(\mdot)$ integrates to one.]
In this section, 
we adopt $\mdotmin=\mdotcrit=0.01$ (see \S\ref{sec:defs}) and $\mdotmax=1$.  
We adopt a broken power-law form for the black hole mass function,
\be
n(M)= \left\{ \begin{array}{cc}
        \nstar \left(\frac{M}{\Mstar}\right)^{\alpha} & \mbox{ $M < \Mstar$, } 
	\\
        \nstar \left(\frac{M}{\Mstar}\right)^{\beta}  & \mbox{ $M > \Mstar$. }
     \end{array}
      \right.
\label{eqn:dblpownofm}
\ee
Henceforth, we will frequently refer to the 
normalization of $n(M)$ in terms of $n_* M_*$, the number density 
of objects within a logarithmic interval around $M_*$. 

The convolution integral~(\ref{eqn:lumfun}) for the luminosity function
breaks into three regimes because the range of accretion rates is finite.
Luminosities 
$L < \epsd \mdotmin l \Mstar = 0.01 l \Mstar $ cannot be generated by black 
holes with masses $M>\Mstar$ because they would require an accretion rate 
below $\mdotmin$.  Thus, only the low mass end of the black 
contributes to this luminosity regime,
\be
\Phi(L)=\int_{\mdotmin}^{\mdotmax} \pstar \mdot^a \nstar 
  \left({L \over \epsd \mdot l \Mstar} \right)^\alpha {1 \over \epsd l \mdot} 
    d\mdot,  \qquad 
 \mbox{ $L < 0.01 l \Mstar$.}
\label{eqn:simplelowlum}
\ee
Similarly, luminosities $L > \epsd \mdotmax l \Mstar = l \Mstar $ 
cannot be generated by black holes with masses $M<\Mstar$, so the
integral for this regime is the same as 
equation~(\ref{eqn:simplelowlum}) but with $\alpha$ replaced by $\beta$.
For the intermediate regime, black holes with masses above and below
$M_*$ contribute.  It is convenient to define the accretion rate $\mdotl$
at which an $\Mstar$ black hole has luminosity $L$, 
\be
\mdotl \equiv {L \over \epsd l \Mstar}~,
\label{eqn:mdotl}
\ee 
so that the $\Phi(L)$ integral breaks into pieces contributed
by the high and low ends of the black hole mass function:
\begin{eqnarray}
\Phi(L) &=& \int_{\mdotmin}^\mdotl \pstar \mdot^a \nstar 
  \left({L \over \epsd \mdot l \Mstar} \right)^\beta 
  {1 \over \epsd l \mdot} d\mdot \nonumber \\ 
&+& \int_{\mdotl}^{\mdotmax} \pstar \mdot^a \nstar 
  \left({L \over \epsd l \mdot \Mstar} \right)^\alpha 
  {1 \over \epsd l \mdot} d\mdot,  \qquad
\mbox{ $0.01 l \Mstar < L < l \Mstar$.} 
\label{eqn:simplemedlum}
\end{eqnarray}
The solution for the QLF is
\be
\Phi(L) = \left\{ \begin{array}{cc}
\frac{\nstar \pstar}{l}{1-0.01^{a-\alpha} \over (a-\alpha)} 
  \left({L \over l \Mstar}\right)^{\alpha} & 
\mbox{ $L < 0.01 l \Mstar$,} \\ [3mm]

\frac{\nstar \pstar}{l} \left[ \frac{\beta-\alpha}{(a-\beta)(a-\alpha)}
\left({L \over l \Mstar}\right)^a - {0.01^{a-\beta} \over a-\beta}
  \left({L \over l \Mstar}\right)^{\beta} 
+ {1 \over a-\alpha} \left( {L \over l \Mstar}\right)^{\alpha}\right] & 
  \mbox{ $0.01 l \Mstar < L < l \Mstar$,} \\ [3mm]

\frac{\nstar \pstar}{l}
{1 - 0.01^{a-\beta} \over(a-\beta)} \left({L \over l \Mstar}\right)^{\beta} &
\mbox{ $L > l \Mstar$,}
 \end{array}
 \right.
\label{eqn:sol:powpofmdot}
\ee
where the values $\mdotmin=0.01$, $\epsd=1$, and $\mdotmax=1$ have 
been explicitly included.  
Note that we have scaled luminosities to the Eddington luminosity
$lM_*$ of an $M_*$ black hole.

The most comprehensive observational analysis of the optical quasar luminosity 
function at $z \la 3$ is that of \cite{boyle00}, based on
the 2dF Quasar Redshift Survey and the Large Bright Quasar Survey.  
They find that the rest-frame $B$-band
luminosity function can be adequately fit by a double power-law
function of the form
\be
\Phi(L_B) \propto \left[\left(\frac{L_B}{\lbrk}\right)^{\alpha_L} + 
                        \left(\frac{L_B}{\lbrk}\right)^{\beta_L}\right]^{-1},
\label{eqn:boyle}
\ee
with the break luminosity $\lbrk$ evolving with redshift.
(We denote the break luminosity $\lbrk$ rather than $L_*$
to keep clear the distinction between this observational quantity
and the characteristic parameters of our models, $M_*$ and $\mdotstar$.)

\begin{figure}
\epsscale{0.9}
\plotone{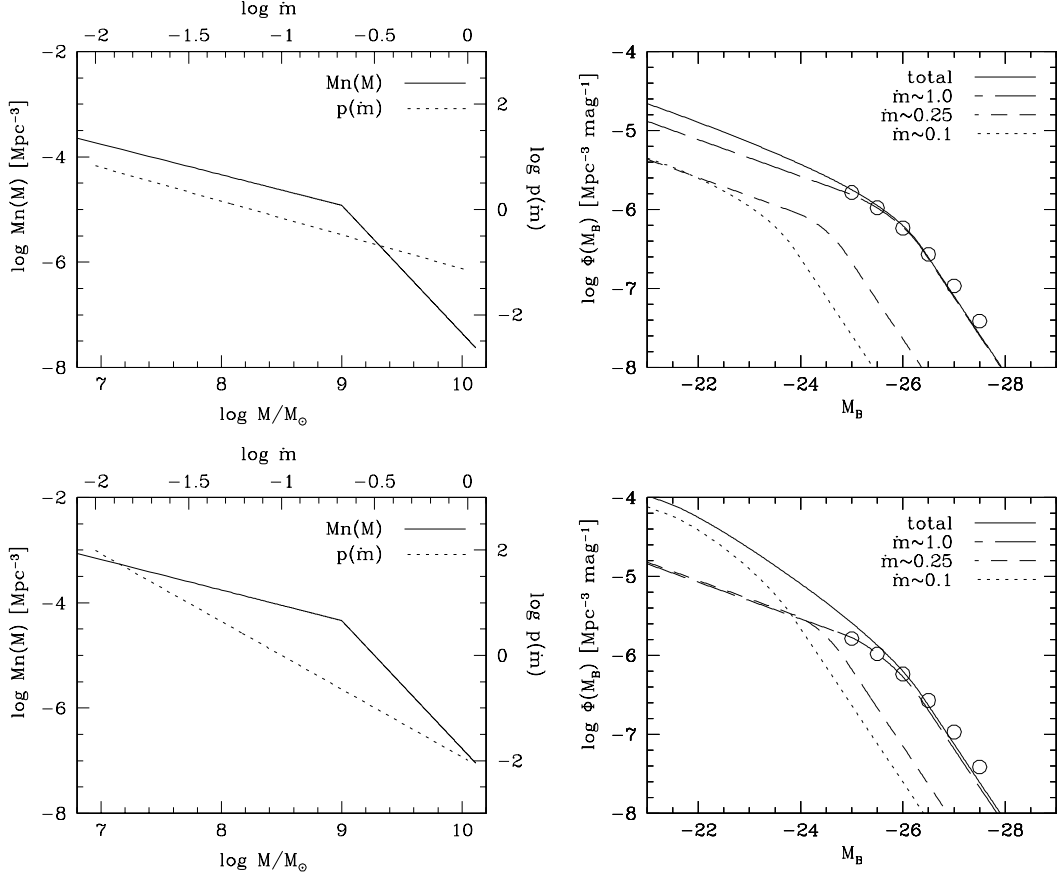}
\caption{
QLF for models with a double power-law black hole mass function and a single
power-law $p(\mdot)$, with thin-disk accretion only.  Left panels show the
input $p(\mdot)$ (dotted line, top and right axis labels, dimensionless) and
mass function (solid line, bottom and left axis labels).
In all Figures, we plot $Mn(M)$,
the number per comoving Mpc$^3$ per $\ln M$ interval, rather than the
mass function $n(M)$ itself.  Right panels show the corresponding
$B$-band luminosity function, plotted in number per comoving Mpc$^{3}$
per magnitude against $B$-band absolute magnitude.  Solid lines show the
total luminosity function, while long-dashed, short-dashed, and dotted
lines show the contributions from accretion rates in the ranges
$0.01 < \mdot < 0.1$, $0.1 < \mdot < 0.25$, and $0.25 < \mdot < 1.0$,
respectively.  Open circles show the double power-law fit to the
\cite{boyle00} QLF measurements at $z=2$, over the range of absolute magnitudes
probed by the data.  Top and bottom rows show results for 
$p(\mdot)$ power-law slopes
$a=0$ and $a=-1.2$, respectively, for the same black hole mass function.
This and all subsequent figures assume an $\Omega_m=1$ cosmology with
$h=0.5$.
}
\label{fig:simplelum}
\end{figure}

Figure~\ref{fig:simplelum} shows QLFs computed by 
equation~(\ref{eqn:sol:powpofmdot}) for two different choices of the $p(\mdot)$
index, $a=-1$ (top) and $a=-2$ (bottom).  
Left hand panels show the input mass functions
and $p(\mdot)$, and right hand panels show the corresponding QLF.  
Instead of the mass function
$n(M)$ itself, we plot $Mn(M)$, the space density of black holes
per $\ln M$ interval, since this quantity is easier to interpret.
We adopt $n(M)$ slopes $\alpha=-1.5$ and $\beta=-3.4$ 
to match the low and high end slopes of the \citet{boyle00} QLF.
Note that the slopes of the QLFs in 
Figure~\ref{fig:simplelum} are actually $\alpha+1$ and $\beta+1$ 
because they are plotted in terms of magnitudes rather than luminosities.
The QLFs have been converted from bolometric emission into absolute $B$-band 
magnitudes by using $L_{\rm bol}/\nu_{B}L_{\rm B}=10.4$ \citep{elvis94}
to calculate the $B$-band luminosity and then using 
$M_{B}=-2.5(\log\,L_{B}-32.67)+5.48$ to convert the luminosity into an 
absolute magnitude, where 5.48 is the absolute 
magnitude of the sun in the $B$ band and 10$^{32.67}$ erg s$^{-1}$ is the 
solar luminosity in the $B$ band.  However, it is important to keep in mind 
that these are scaled bolometric luminosity 
functions and correspond to true optical luminosity functions only if this 
$L_{\rm bol}/\nu_{B}L_{\rm B}$ ratio is constant from quasar to quasar
(see \S\ref{sec:wavelength}).
The points in the right hand panels correspond to Boyle et al.'s
fit (eq.~\ref{eqn:boyle}) at $z\sim 2$, and they cover only the 
range of magnitudes actually observed at this redshift.

The amplitude of the QLF is directly proportional to $\pstar$ and to the mass 
function normalization $\nstar$.  The value of $\pstar$ determines 
the probability that a given black hole will be ``on'' at some non-zero 
luminosity.  For the $a=-2$ model (lower panel), we have somewhat
arbitrarily chosen $\pstar$ so that this probability,
$\int_\mdotmin^\mdotmax p(\mdot)d\mdot$, is equal to one.
Though all of the black holes are active in this case, most of 
the activity is at accretion rates near $\mdotmin$.
We then choose a mass function normalization 
$n_*M_* = 4.6\times 10^{-5} {\rm Mpc}^{-3}$ so
that the model QLF matches the \citet{boyle00} data point at the
break luminosity.  For the $a=-1$ model, we have kept the same $n(M)$
and again chosen $\pstar$ to match the observed QLF at the break.
Note that equation~(\ref{eqn:sol:powpofmdot}) reveals a complete degeneracy 
between 
$\nstar$ and $\pstar$ in the QLF.  Thus, at a fixed time, it is impossible to
determine from the QLF alone if there is a high number density of black 
holes of 
which a small fraction are accreting or a low number density of black 
holes of which a large fraction are accreting.  We will see later that 
the evolution of the QLF breaks this degeneracy to some degree.

The QLF in the upper right panel of Figure~\ref{fig:simplelum} demonstrates 
one of the important general features of this solution: the slopes of the 
low and high luminosity ends of the QLF are determined by the slopes of the 
low and high mass ends of the black hole mass function 
(see eq.~\ref{eqn:sol:powpofmdot}).
This behavior follows whenever $p(\mdot)$ is independent of $M$ and the range 
of accretion rates is finite. 
This result can be understood by considering that a small range $d\mdot$ at 
$\mdot=1$ gives 
$\Phi(L) = n(M) (\epsd l)^{-1} p(\mdot=1) d\mdot$ with $M=L/(\epsd l)$,
which is a 
simple mapping of the black hole mass function.  The same range
$d\mdot$ at a lower $\mdot$ gives a contribution to the QLF mapped to 
luminosities fainter by a factor of $\mdot$ and up or down by a factor 
proportional to $p(\mdot)/p(\mdot=1)$. 
The total QLF is just the sum of these transformed black hole mass functions.
Since we have low and high luminosity regimes in which only one slope of $n(M)$
contributes, the sum in these regimes will be a sum of power-laws with the same
slope, resulting in a QLF with the same slope.
The mid-range luminosity is more complicated because both parts of the black 
hole mass function are contributing.  The relative contributions from the 
low and high end of the black hole mass function depend on the relative 
probability of low
and high accretion rates.  This can be seen in the middle part of 
equation~(\ref{eqn:sol:powpofmdot}), which shows that the slope of the 
probability
distribution as well as the slopes of $n(M)$ affect the
shape of the QLF in this luminosity regime.  
The agreement of asymptotic slopes between $n(M)$ and $\Phi(L)$
motivates our choice of a double power-law $n(M)$, though we will
see in \S\ref{sec:nofmz} that this choice has some problems when
compared to local observations.

The $a=-1$ model in Figure~\ref{fig:simplelum}
corresponds to equal probabilities of accretion in each
logarithmic interval of $\mdot$ between $\mdotmin$ and $\mdotmax$,
but the emissivity of the population is dominated by accretion rates
close to $\mdotmax$ because luminosities are proportional to $\mdot$.
The model QLF is in reasonably good agreement with the \citet{boyle00} data.
Dotted, short-dashed, and long-dashed lines in the QLF panels 
show the contribution
from accretion rates in the ranges $0.01<\mdot<0.1$, 
$0.1<\mdot<0.25$, and $0.25<\mdot<1.0$ respectively.  
For $a=-1$, the QLF is dominated by black holes with near-Eddington
accretion rates at all luminosities.  The curves for lower 
accretion rates are shifted horizontally to lower luminosities,
with slight vertical shifts because the three bins are not equal
logarithmic intervals.

The $a=-2$ model has equal contributions to the emissivity from each
logarithmic interval of $\mdot$.  The slow transition between the
low and high luminosity regimes yields a worse fit to the 
\cite{boyle00} QLF.  High accretion rates still dominate the 
high end of the QLF, but low accretion rates dominate the low end.
In general, low $\mdot$ black holes 
can more easily make a significant contribution below the break in the 
luminosity 
function because a shift ``left'' can be more easily compensated by a 
shift ``up,''
especially when the slope of the low end of the mass function is shallow.     
However, above the break in luminosity, it is difficult for low $\mdot$ 
accretors to make a contribution.  For example, a $p(\mdot)$ slope
$a \sim -3.3$ would be required to make the
contribution of low and high accretion rates comparable
at high luminosities in Figure~\ref{fig:simplelum}.

\subsection{Double Power-Law p($\mdot$) with ADAF and Super-Eddington Accretion}
\label{sec:dblpowpofmdot}

Our assumptions for this section are similar to those of
\S\ref{sec:powerlawpofmdot}, except that we consider a wider range of 
allowed accretion rates and adopt a double power-law form of $p(\mdot)$.  
Specifically, we consider accretion rates in the range $\mdotmin=10^{-4} 
< \mdot < \mdotmax= 10$, which allows for ADAF, thin-disk, and super-Eddington 
accretion modes, for which we adopt the efficiencies given in 
equation~(\ref{eqn:effofmdot}). The functional 
form of $n(M)$ we consider here is the double power-law expressed in 
equation~(\ref{eqn:dblpownofm}), and the form of $p(\mdot)$ is analogous, 
\be
p(\mdot)=\left\{ \begin{array}{cc}
  \pstar \left({\mdot \over \mdotstar}\right)^{a} & \mbox{ $\mdot < \mdotstar$,} \\
  \pstar \left({\mdot \over \mdotstar}\right)^{b} & \mbox{ $\mdot > \mdotstar$,}
          \end{array}
        \right.
\label{eqn:dblpowpofmdot}
\ee
where $\mdotstar$ is the characteristic accretion rate at the break in 
$p(\mdot)$.  The solution for the QLF in this case is given in
Appendix~\ref{sec:appx1}.

\begin{figure}
\epsscale{1.0}
\plotone{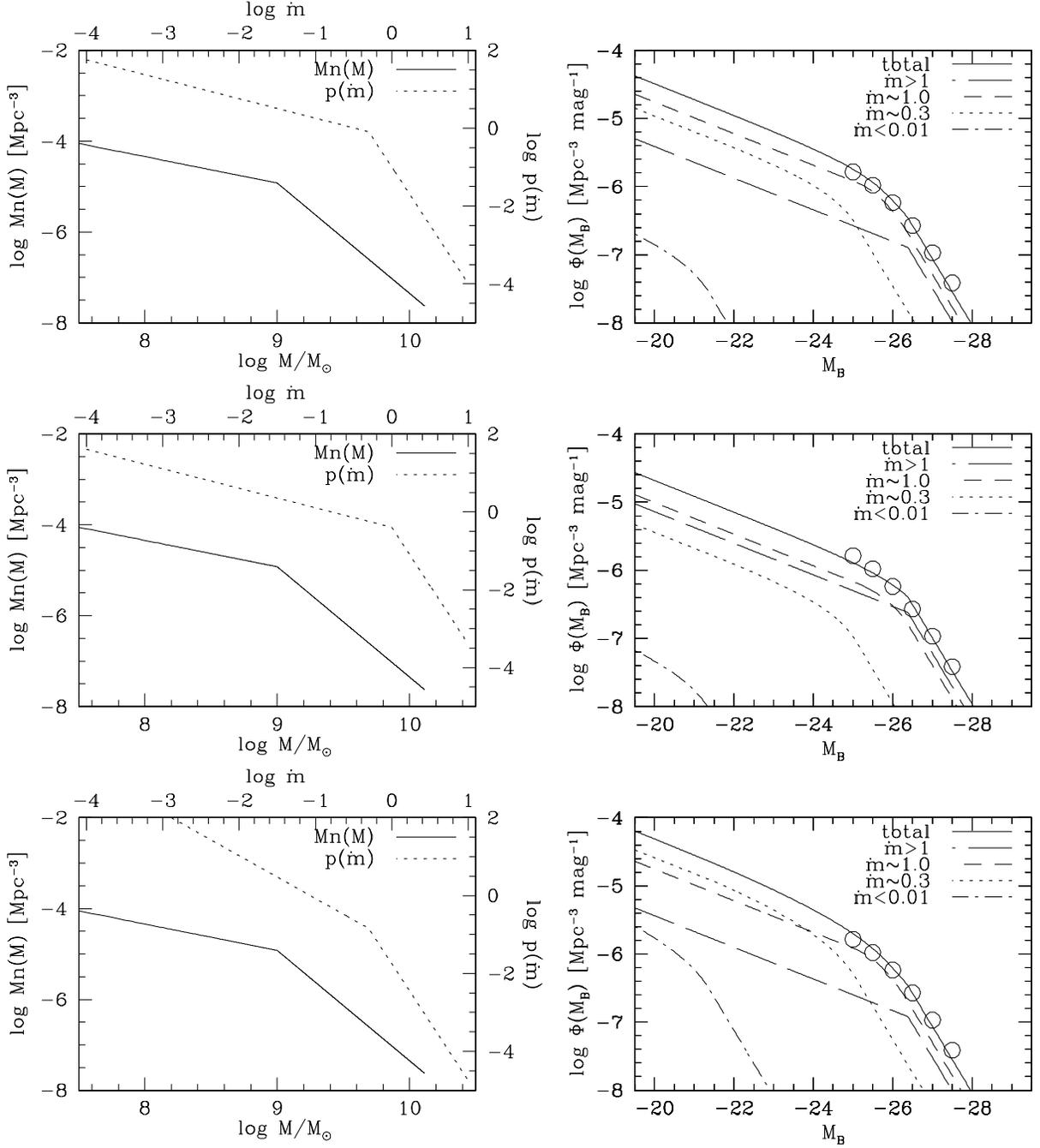}
\caption{
Like Fig.\ 1, but for double power-law $p(\mdot)$ covering a wider
range of accretion rates, including values in the ADAF regime and
super-Eddington regime.  Efficiencies as a function of $\mdot$
are calculated according to eq.~(\ref{eqn:effofmdot}).
In the right panels, solid lines show the total QLF, and other lines
show the contributions from the ranges $\mdot>1$ (super-Eddington,
long-dashed), $0.3 < \mdot < 1$ (thin-disk, high accretion rate, short-dashed),
$0.01 < \mdot < 0.3$ (thin-disk, low accretion rate, dotted),
and $10^{-4} < \mdot < 0.01$ (ADAF, dot-dashed).
Relative to the case in the top row, the middle and bottom rows
show cases with a higher characteristic accretion rate
($\mdotstar=1$ vs.\ $\mdotstar=0.5$)
and a steeper slope at low accretion rates ($a=-1.1$ vs. $a=-0.5$),
respectively.
}
\label{fig:dblpowlum}
\end{figure}

Figure~\ref{fig:dblpowlum} is analogous to 
Figure~\ref{fig:simplelum}, but for the double power-law $p(\mdot)$.
In the upper panels, we adopt $a=-0.5$, $b=-3$, and $\mdotstar=0.5$.
At high luminosity, the QLF is dominated by the higher accretion rates in the 
thin-disk mode, but there is also a significant contribution from the 
super-Eddington mode.  
At low luminosity, the contribution from higher accretion 
rates in the thin-disk mode dominates over lower accretion rates 
in the thin-disk 
mode, with the super-Eddington mode becoming much less significant.
The contribution of the ADAF mode to the QLF is barely noticeable, and
this is typically the case in our models because low $\mdot$ and 
low $\epsd$ combine to push black hole luminosities down to 
$\sim 10^{-2} - 10^{-6}$ of Eddington, which is usually below
observed luminosities.  ADAF contributions can be more significant
at low redshifts and X-ray wavelengths, as discussed in \S\ref{sec:wavelength}
below.

The middle panel shows the effect of increasing $\mdotstar$ to 1.0,
with $\pstar$ decreased to keep $\Phi(\Lbrk)$ fixed.  Super-Eddington
accretion is now common enough that it dominates the high end of the 
QLF, with a comparable but smaller contribution from thin-disk accretion
with $0.3 < \mdot < 1.0$.  Lower luminosities are dominated by thin-disk 
accretion, with high accretion onto lower mass black holes more
important than low accretion onto high mass black holes.

The bottom panels show a case with $\mdotstar=0.5$ and a 
low end slope of $a=-1.1$, which weights $p(\mdot)$ 
more strongly to low accretion rates. The high luminosity 
end of the QLF in the bottom panel has contributions similar to those 
in the upper panel, but the low luminosity end is now dominated by lower 
accretion rates in the thin-disk mode.  
The turnover of the QLF in the lower panel is 
less sharp than in the upper panel because of the change of the relative 
fraction of 
high and low accretion rates contributing to the QLF over the transitional 
range of luminosity. 
The contribution from the ADAF mode is higher than in the top panel,
but it remains small at all luminosities.

As noted earlier, with the luminosity function alone there is a complete
degeneracy between the normalizations $\nstar$ and $\pstar$,
subject only to the limitation that the fraction of black holes 
accreting at a given time not exceed 100\%.  There is also a partial
degeneracy between the characteristic values $M_*$ and $\mdotstar$:
increasing $\Mstar$ shifts the break in the QLF to higher luminosity,  
but reducing $\mdotstar$ makes lower accretion rates dominate the QLF, 
shifting the break back down.
With our assumption that the efficiency $\epsd(\mdot)$
decreases for $\mdot>1$, this 
tradeoff is limited to a factor of a few, since
changes in $\mdotstar$ also affect the shape of the turnover in the QLF.
We will see later that this tradeoff is even more restricted
when the evolution of the QLF is considered.  

\subsection{Mass Distribution of Active Black Holes}
\label{sec:massdist}

The masses of active black holes can be estimated using
reverberation mapping (e.g., \citealt{wandel99,onken02}), 
the widths of lines such as H$\beta$ and CIV 
\citep[e.g.,][]{laor98,vestergaard03,corbett03}, or indirectly from the
properties of host galaxies \citep[e.g.,][]{dunlop03},
the variability power spectrum \citep{czerny01}, or the 
spectral energy distribution \citep{kuraszkiewicz00}.
The distribution of active black hole masses at a given luminosity
depends on both the underlying black hole mass function $n(M)$
and the distribution of accretion rates $p(\mdot)$.  While
the necessary measurements are challenging, especially at high
redshift, they can play a critical role in discriminating among
models that make very similar predictions for the luminosity function.

If there is a maximum value of the product $\epsd \mdot$ (e.g., unity
if luminosities cannot exceed Eddington, as in eq.~\ref{eqn:effofmdot}),
then black holes with $M<(L/l)[(\epsd\mdot)_{\rm max}]^{-1}$
cannot contribute to the QLF at luminosity $L$ because they cannot
shine brightly enough.  Conversely, black holes with 
$M>(L/l)[(\epsd\mdot)_{\rm min}]^{-1}$ are always brighter than $L$,
when they are active at all.
The contribution to the QLF from black 
holes with masses in the allowed range is the product of the number density 
of black holes with mass $M$ and the probability of having an accretion rate in
 the range $\mdot \rightarrow \mdot + \Delta \mdot$ that yields 
luminosity $L \rightarrow L +\Delta L$.  
Thus, for a quasar of luminosity $L$, the relative
probability that its black hole has mass $M_1$ or $M_2$, if both
masses are in the allowed range, is
\be
{p(M_1|L) \over p(M_2|L)} ~=~
{ n(M_1) p\left(\mdot={L \over \epsd l M_1}\right) {\Delta L /M_1 \over \epsd l}
  \over 
n(M_2) p\left(\mdot={L \over \epsd l M_2}\right) {\Delta L /M_2 \over \epsd l} }
  ~=~ \left({M_1 \over M_2}\right)^{\alpha-(a+1)} ~,
\label{eqn:relmassdist}
\ee
where $(\Delta L / \epsd l M )=\Delta \mdot $.
The rightmost equality applies for constant $\epsd$ and power-law
forms of $p(\mdot)$ and $n(M)$, with slopes of $a$ and $\alpha$ 
respectively.  
For this case we see that if $\alpha < a+1$ then
the rising low end of the mass function results in low mass
black holes with high accretion rates dominating the active population
at luminosity $L$.
Conversely, if $\alpha > a+1$, then low accretion rates are common
enough that high mass black holes with low $\mdot$ dominate.
A single power-law can only approximate $n(M)$ or $p(\mdot)$ over
a finite range, but this example gives insight
into the more general case and allows one to judge whether quasars
of luminosity $L$ are likely to be dominated by masses near a break
in $n(M)$, or by accretion rates near a break in $p(\mdot)$ or $\epsd(\mdot)$.
For example, the high luminosity regime 
Figure~\ref{fig:dblpowlum} is dominated by Eddington luminosity
black holes (near a break in $\epsd$) because of the steep slope
of $n(M)$ at high masses.

For a statistical quantity that is easier to measure, it is usually
desirable to integrate equation~(\ref{eqn:relmassdist}) to obtain
the distribution of black hole masses
for a specified range in luminosities.  We will show in \S\ref{sec:illus}
that such statistics can discriminate between models that yield similar 
QLFs over the range of observed luminosities but have significantly 
different parameter values.

\subsection{Mass Dependence of p($\mdot$)}
\label{sec:massdeppofmdot}

A general mass-dependent distribution of accretion rates can be written
in the form $p(\mdot|M) = p_0(\mdot)D(M|\mdot)$,
where $p_0(\mdot)=p(\mdot|M_0)$ at an arbitrarily chosen mass scale $M_0$
and the function $D$ encodes the mass dependence,
with $D(M_0|\mdot)\equiv 1$.
To understand the potential influence of mass dependence, we will
consider the restricted case in which the function
$D(M)$ is independent of $\mdot$, and
$p(\mdot|M)$ is thus a separable function.
In this class of models, the relative probabilities of high and
low accretion rates are independent of mass, but the overall
duty cycle can have an arbitrary mass dependence.
Recall that the general expressions for the QLF (eq.~\ref{eqn:lumfun})
and the distribution $p(M|L)$ of black hole masses at a given luminosity
(eq.~\ref{eqn:relmassdist}) involve $p(\mdot|M)$ and $n(M)$ only
through the product $p(\mdot|M)n(M)$. 
Therefore, for any QLF and $p(M|L)$ generated by a black hole mass
function $n(M)$ and a mass-independent $p(\mdot)$, there is a family
of models with mass function $n'(M)=n(M)/D(M)$ 
and mass-dependent accretion rate distribution $p'(\mdot|M)=p(\mdot)D(M)$
that predicts the same QLF and $p(M|L)$,
for any choice of the function $D(M)$. 
The mass dependence of $p(\mdot)$ therefore introduces a
rather serious degeneracy into models of the luminosity function,
which cannot be broken by measurements of the distribution of {\it active} 
black hole masses.  

The key difference within this class of degenerate models is the
relation between the luminosity function and the underlying black 
hole mass function $n(M)$.  In particular, we have shown in 
\S\ref{sec:powerlawpofmdot} and \S\ref{sec:dblpowpofmdot}
that for a mass-independent $p(\mdot)$ the low and high luminosity
slopes of the QLF match the low and high mass slopes of $n(M)$.
This is no longer the case if $p(\mdot)$ depends on $M$.
For example, with $D(M) \propto M^x$ and a double power-law $n(M)$,
the QLF has asymptotic slopes $\alpha+x$ and $\beta+x$, rather than
$\alpha$ and $\beta$.  Thus, a measurement of the mass function of
all black holes, not just the active ones, is crucial to 
diagnosing the mass dependence of accretion rates.

Our discussion above focuses on the QLF at a particular redshift, 
and the degeneracy applies if $p(\mdot|M)$ and $n(M)$ can be
chosen at will.  If one considers a range of redshifts over
which black holes grow by a substantial factor, then the predictions
of models with different mass dependence of $p(\mdot|M)$ are 
likely to diverge, since the mass-dependence of growth rates
will change the shape of $n(M,z)$ from one model to the next
(see \S\ref{sec:massdepdeclining}).
Thus, this degeneracy should be less serious in a complete
evolutionary model of the population.
Furthermore, a measurement of $n(M)$ at $z=0$ may be sufficient
to diagnose the mass dependence of $p(\mdot|M)$ at higher redshift.

\subsection{Wavelength Dependent Efficiencies}
\label{sec:wavelength}

Equation~(\ref{eqn:lumfun}) gives the bolometric luminosity function 
$\Phi(L)$ in terms of the accretion rate distribution, the black hole
mass function, and the efficiency $\epsd$, which may itself be a
function of $\mdot$.  If all quasars have the same spectral energy
distribution (SED), then the translation to the luminosity function
in a band at frequency $\nu$ is straightforward:
\be
\Phi(L_\nu)=\int_{\mdotmin}^{\mdotmax} p(\mdot) 
            n\left(M={L_\nu \over \epsd \mdot l F_{\nu}}\right) 
	    { 1 \over \epsd \mdot l F_{\nu}} d\mdot ~,
\label{eqn:mwlumfunc}
\ee
where $F_\nu \equiv L_\nu/\Lbol$ is the fraction of the quasar's
bolometric luminosity that emerges in the $\nu$-band.
(Note that we are using subscript-$\nu$ to represent a finite band,
not a monochromatic flux density.)
We have so far presented results for the rest-frame $B$-band luminosity
function, assuming that all accreting black holes have the broad-band
SED estimated by \cite{elvis94}.  
For a universal SED, the luminosity functions in all bands are just 
shifted versions of the bolometric luminosity function, so they
all have the same shape.

The story is more interesting if some accreting black holes 
have radically different SED shapes.  Here we will consider
two representative examples, an ``obscured'' accretion mode in 
which optical, UV, and soft X-ray radiation are absorbed by gas 
and dust near the nucleus
and re-radiated in the far-IR, and an ADAF mode that has a
high ratio of X-ray flux to optical flux (in addition to a 
reduced bolometric efficiency).
Obscured accretion is thought to play an important role in 
producing the X-ray background, and typical synthesis models in the
literature have a $\sim 4:1$ ratio of obscured to unobscured sources 
(e.g., \citealt{comastri95,fabian99}).
More recent results from {\it Chandra} show that the obscured fraction
is probably lower than this, especially at high luminosities
(e.g., \citealt{barger02,ueda03}).
ADAFs are expected on theoretical grounds to have depressed UV/optical
emission relative to X-ray and far-IR \citep{narayan98}, 
and many low luminosity AGN in the nearby universe, including Sgr A$^*$ in
the Galactic Center, appear to have these broader SED shapes 
(e.g., \citealt{ho99}).
There are, of course, other possibilities for SED variations,
including a steady change of SED shape with black hole mass
caused by the lower characteristic temperatures around higher
mass black holes, a change of SED shape in the super-Eddington regime,
and perhaps a transition within the $\sim 0.1-1\Ledd$ regime
as the accretion disk grows in importance relative to the hard
X-ray corona.

The first task is to calculate values of $F_\nu$ for the model SEDs.
We will consider luminosity functions in the rest-frame $B$-band,
soft (0.5-2 keV) and hard (2-10 keV) X-ray bands, and a 
``far-IR'' band that we take to cover the range $10-1000\micron$.
Since most X-ray studies work with observed-frame fluxes
(though see Cowie et al.\ [\citeyear{cowie03}], 
Steffen et al.\ [\citeyear{steffen03}], and Ueda et al.\ [\citeyear{ueda03}]
for recent efforts to measure evolution
of the rest-frame 2-8 keV luminosity function), we also consider
the soft and hard X-ray bandpasses redshifted to $z=0.5$, 1, and 2.
We assume that our far-IR band is wide enough that all 
high-energy radiation absorbed in obscured systems is re-radiated
somewhere within it.  Unfortunately, realistic experiments are
likely to probe a narrower band, for which the predictions may
be quite sensitive to assumptions about dust temperatures and
departures from a blackbody spectrum.  We will not consider
radio luminosities here.  To the extent that the radio-quiet/radio-loud
dichotomy is a simple effect of orientation or black hole spin,
it could be treated as a
stochastic variation analogous to our treatment of obscuration.
However, radio luminosity may also be connected to accretion rate
or to black hole mass \citep{dunlop03}. 

\begin{deluxetable}{ccccccccccc}
\tablewidth{0pt}
\tablecaption{ Fractional Bolometric Output 
\label{tbl:fnu}}
\startdata \hline
         & $F_{2-10}$ & $F_{3-15}$ & $F_{4-20}$ & $F_{6-30}$ & $F_{0.5-2}$ 
	 & $F_{0.75-3}$ & $F_{1-4}$ & $F_{1.5-6}$ & $F_{B-{\rm band}}$ 
	 & $F_{\rm FIR}$ \\ \hline
     Thin-Disk & 0.026 & 0.028 & 0.030 & 0.031 & 0.020 & 0.020 & 0.020 & 0.022 & 0.025 & 0.17  \\
     Obscured & 0.008 & 0.015 & 0.020 & 0.027 & 1$\times 10^{-9}$ & 1.3$\times 10^{-5}$ & 2.9$ \times 10^{-4}$ & 2.6$\times 10^{-3}$ & 0.00 & 0.79  \\
     ADAF & 0.084 & 0.089 & 0.093 & 0.099& 0.057 & 0.061 & 0.064 & 0.068 & 0.012 & 0.03
\enddata
\tablecomments{Values of the inverse bolometric correction, 
$F_{\nu-{\rm band}}$, where 
$L_{\nu-{\rm band}}=F_{\nu-{\rm band}} L_{\rm Bol}$.  Values include 
intrinsic X-ray ranges at $z=2, 1,$ and 0.5 that correspond to the observed 
soft and hard bands at $z=0$, as well as the rest-frame $B$-band and FIR band.
}
\end{deluxetable}

Table~\ref{tbl:fnu} summarizes our values of $F_{\nu}$.
We assume that unobscured quasars with accretion rates 
$\mdot > \mdotcrit= 0.01$
have the mean radio quiet SED of \cite{elvis94}, and we obtain $F_\nu$
by integrating over the appropriate wavelength bands.
Simple power-law extrapolations were used in regions without observations, 
and the high 
energy SED was extended to 30 keV by using a power-law with 
photon index $\Gamma_X=1.9$.
For obscured quasars, we assume the same ``underlying'' SED and
compute absorption in the X-rays by taking a mean X-ray photon 
index of $\Gamma_X=1.89$ and an obscuring
column density of $N_{H}=3\times10^{23}$ cm$^{-2}$,
which represents the median value used in the X-ray
background sythesis models of \citet{comastri01}.  
We further assume that the high gas column is accompanied by enough
dust to completely extinguish the $B$-band flux; this assumption
may be inaccurate, as some observed systems appear to have significant
optical/UV flux despite strong absorption in the X-ray.
Finally, we assume that all of the energy absorbed from the optical
to the soft X-rays, about 52\% of the bolometric energy in the
\cite{elvis94} SED, is reradiated in the FIR.  This assumption seems
physically plausible, though it is also possible that some or most
of the absorbed energy is channeled into driving an outflow and
never radiated at all.

As in our previous calculations, we assume that accretion rates
$\mdot < \mdotcrit=0.01$ lead to reduced efficiency, ADAF flows,
but now we assign these flows a different SED.
The appropriate SED for ADAF systems is quite uncertain,
and we have elected simply to take the nucleus of M81 as representative
of all black holes accreting in this mode. 
We calculate $F_\nu$ values from the observations tabulated in 
\citet[tables 2 and 9 and \S 2.1]{ho99}, using the estimated
$\Lbol$ and directly observed $L_{\nu}$ when available and otherwise
using the observed monochromatic $\nu L_{\nu}$ values and an 
appropriate $\alpha_{\nu}$ to integrate over the waveband of interest. 
Since the observed 
FIR flux value of M81 is an upper limit, we used the model of M81 presented in 
\citet[][fig.~1]{quataert99} to estimate $F_{\rm FIR}$ for the ADAF mode. 

With the results of Table~\ref{tbl:fnu} in hand, we can again calculate
multi-wavelength luminosity functions using equation~(\ref{eqn:mwlumfunc}),
but now the results must be computed separately for the three
modes (thin-disk, obscured, ADAF) and added together.
We assume that super-Eddington accretion has a thin-disk SED;
this seems unlikely, but we do not have much idea of what to 
assume in its stead, and it makes little difference to our results.

\begin{figure}
\epsscale{1.0}
\plotone{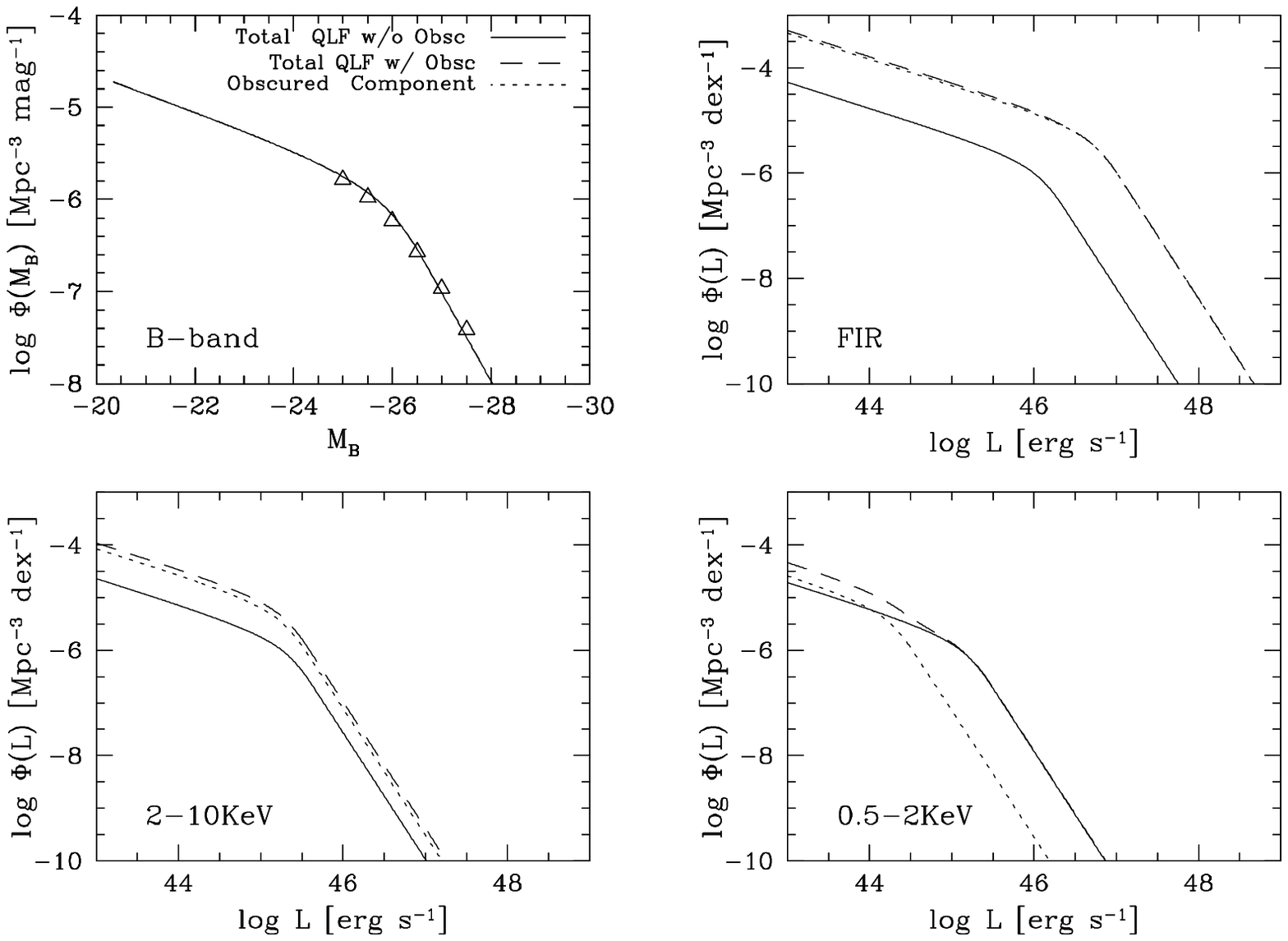}
\caption{
Potential influence of obscuration on multi-wavelength QLFs at $z\sim 2$.
Solid lines show $B$-band, FIR, 0.5-2 keV, and 2-10 keV luminosity
functions of a model in which all quasars have the \cite{elvis94} SED,
with $p(\mdot)$ and $n(M)$ chosen to match the \cite{boyle00}
$B$-band results (triangles in upper left).
Dashed lines show QLFs for a model in which 20\% of quasars have
the \cite{elvis94} SED and 80\% have an obscured SED corresponding
to $N_H = 3\times 10^{23} {\rm cm}^{-2}$ (see $F_\nu$ values
in Table~\ref{tbl:fnu}).  Dotted lines show the contribution of
obscured systems in this model (note that we assume complete
optical obscuration, hence no contribution in $B$).
X-ray luminosities are {\it observed-frame} at $z=2$.
}
\label{fig:obsc.4panel}
\end{figure}

Figure~\ref{fig:obsc.4panel} illustrates the potential influence of
a large obscured quasar population on the multi-wavelength QLF at
$z=2$.  Solid lines show results for a model with no obscured
quasars, with the same black hole mass function and $p(\mdot)$
used in upper panels of Figure~\ref{fig:dblpowlum}. 
Dashed lines show a model in which these unobscured quasars are only
20\% of the total --- i.e., we multiply $p(\mdot)$ by four but 
assign 80\% of the systems with $\mdot > 0.01$ the obscured
SED of Table~\ref{tbl:fnu}.  Since the obscured SED has no $B$-band
flux, the $B$-band luminosity function is unchanged (top left panel).
However, obscuration in the observed-frame 2-10 keV band (rest-frame
6-30 keV) is minimal, so the hard X-ray luminosity function is
nearly a factor of five higher in amplitude, with the contribution
of obscured quasars (dotted line) dominating at all luminosities.  
For the 0.5-2 keV band (rest-frame 1.5-6 keV) the situation is
more complicated, since obscuration suppresses flux in this band
by nearly a factor of ten.  At high luminosities, unobscured quasars
dominate the QLF because the greater numbers of obscured quasars
are not enough to compensate for their reduced fluxes.  However,
the obscured population boosts the faint end of the soft X-ray
QLF by about a factor of two, with obscured and unobscured 
systems making roughly equal contributions.  

The most dramatic effect of the obscured population is to boost the
FIR luminosity function by a large factor, since with our assumptions
the FIR band contains 79\% of the bolometric flux of obscured systems
but only 17\% of the bolometric flux of unobscured systems.
The combination of more systems and more flux per system boosts
$\Phi(L_{\rm FIR})$ by almost two orders of magnitude at
high luminosities, with the FIR QLF totally dominated by obscured
systems at every redshift.  Note that our treatment of $F_{\rm FIR}$
implicitly assumes that obscured and unobscured systems are two
distinct populations, one with high gas columns and one without.
It is also possible, as assumed by \cite{sazonov03}, that the
difference between obscured and unobscured systems is simply one
of orientation, and that even systems that are unobscured along
our line of sight have most of their optical, UV, and soft X-ray
emission absorbed by a dusty torus and re-radiated isotropically 
in the FIR.  In this case, $F_{\rm FIR}$ would be essentially 
the same for both populations, so at a given FIR luminosity
obscured and unobscured systems would be represented in their
global ratio (i.e., 4:1 in our model).
The joint FIR-optical or FIR-soft X-ray luminosity functions
would distinguish these two scenarios.

Figure~\ref{fig:obsc.4panel} shows that a large population of
obscured quasars can substantially alter the relation between
$B$-band, X-ray, and FIR luminosity functions, as one would expect.
The difference in FIR would persist at all redshifts.
At low redshifts, on the other hand,
the 0.5-2 keV band is almost completely suppressed
by a column density $N_H=3\times 10^{23}{\rm cm}^{-2}$, and the 2-10 keV
band is significantly suppressed, so the effect of an obscured 
population on the luminosity function in these bands is reduced.

\begin{figure}
\epsscale{0.5}
\plotone{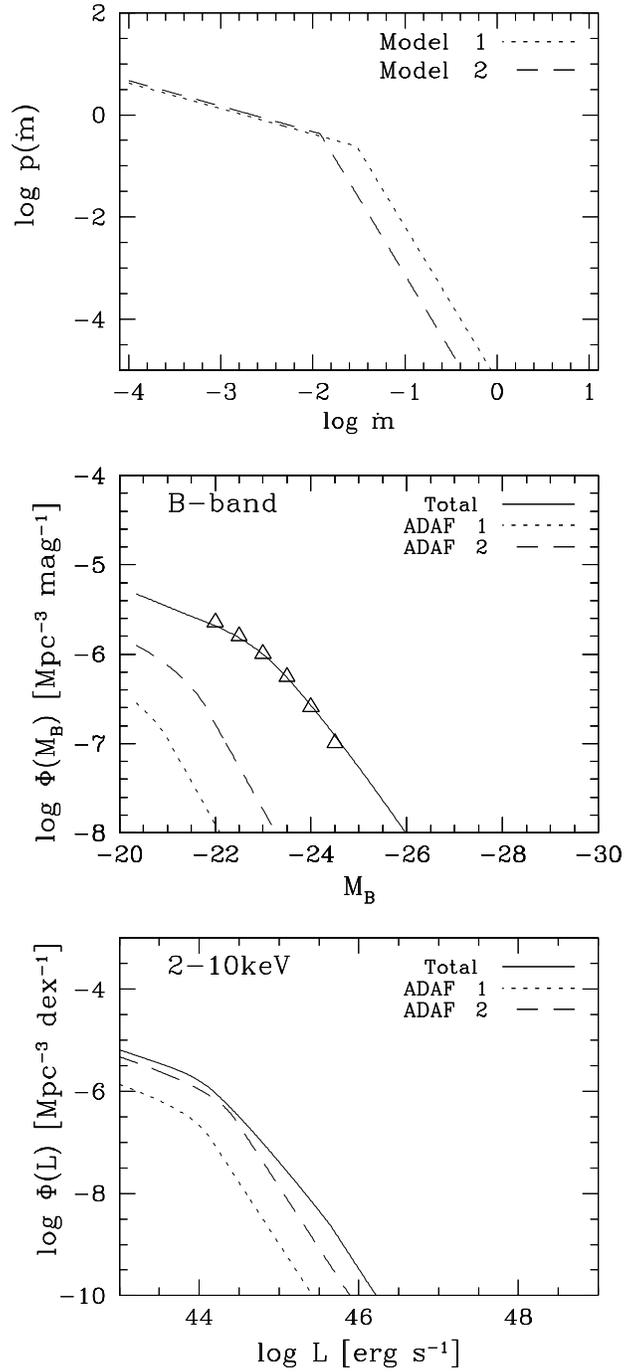}
\caption{
Potential influence of ADAFs on multi-wavelength QLFs at $z\sim 0.5$.
The top panel shows $p(\mdot)$ distributions for two models with
different $\mdotstar$ (the values of $M_*$ are also different).
The total QLFs for the two models are nearly identical in both
$B$-band (middle panel) and 2-10 keV (bottom panel); solid lines
show the total QLFs of Model 1.  Dotted and dashed lines show the
contributions of ADAF accretion ($\mdot < 0.01$) for the two
models, which are small in $B$-band but substantial at 2-10 keV
for Model 2.
}
\label{fig:adaf0.5}
\end{figure}

Although we include the ADAF mode in our calculations
for Figure~\ref{fig:obsc.4panel}, we find that ADAF systems make
no significant contribution to the QLF at any luminosity likely
to be observed at $z\sim 2$, for any plausible choice of our 
model parameters.  However, the situation could be different at low redshift,
in part because observations reach to lower luminosities,
but even more because (as we discuss in \S\ref{sec:declining} below)
matching the observed QLF evolution requires a shift of $p(\mdot)$
towards lower characteristic accretion rates at low redshifts,
thus giving systems with $\mdot < \mdotcrit$ more chance to compete.
Figure~\ref{fig:adaf0.5} shows two models with choices of $p(\mdot)$
and $n(M)$ that give reasonable matches to the \cite{boyle00}
$B$-band QLF at $z \sim 0.5$.  The upper panel shows the two
$p(\mdot)$ distributions, which are double power-laws with 
$\mdotstar=0.03$ and 0.012, respectively.
The black hole mass functions (not shown) have corresponding
$M_*$ values of $1.21 \times 10^9 \Msun$ and $2.5 \times 10^9 \Msun$.

Solid lines in the middle and bottom panels show the predicted
luminosity functions for the first model in rest-frame $B$-band
and observed-frame $2-10$ keV, respectively.  Total luminosity
functions for the second model are nearly identical.
However, the relative contribution of the ADAF mode, shown
by the dotted and dashed lines in these panels, is quite 
different between the two models and between the two bands.
In $B$-band, ADAF contributions to the luminosity function are
strongly suppressed, even for $\mdotstar=0.012$,
because of the low optical flux of the ADAF SED.
However, the relatively high hard X-ray flux allows ADAF 
accretion to dominate the low end of the luminosity function
for the $\mdotstar=0.012$ model and to make a significant
contribution even at high luminosities.  In the $\mdotstar=0.03$
model, on the other hand, ADAF accretion is a subdominant
contribution to the QLF at all luminosities.  
Results for the FIR and $0.5-2$ keV luminosity functions are
qualitatively similar to those for $B$-band and $2-10$ keV,
respectively, as one would expect from the values of $F_\nu$
in Table~\ref{tbl:fnu}.  Measurements
of the optical and X-ray luminosity functions alone would
not distinguish the two models shown in Figure~\ref{fig:adaf0.5},
but the low-$\mdotstar$ model
predicts that X-ray selected quasars (largely ADAF systems)
should have systematically
different SED shapes from optically selected quasars 
(mostly thin-disk systems), 
while the high-$\mdotstar$ model predicts that thin-disk
SEDs dominate both populations.

\section{Evolution}
\label{sec:evolution}

We now turn to evolutionary calculations, applying the basic principles
of \S\ref{sec:qlfevoleqn}.  We assume that the accretion physics ---
the dependence of $\epsd$ and SED shape on accretion rate --- is
independent of redshift, although the distribution of accretion
rates itself evolves.  This assumption seems reasonable, 
since the ``microphysics'' has no direct knowledge of the age of 
the universe, but one
could imagine that systematic changes in the host galaxy population
might affect the influence of gas or dust obscuration on SED shapes,
and perhaps even that galaxy mergers could alter the fraction 
of spinning black holes with higher bolometric efficiencies.
With our assumption and a specified model of the accretion 
physics, $p(\mdot|M,z)$ determines the evolution of $\Phi(L)$ in all 
wavebands,  since it both determines the evolution of $n(M)$ and specifies 
the probability that black holes of a given mass shine at a given luminosity.  
However, mergers can alter the evolution of the QLF 
by changing $n(M)$ independently of $p(\mdot|M,z)$.

We focus in this section on the optical luminosity function, which is 
observed to rise by a large factor between $z\sim 5$ and $z\sim 3$
\citep{who,schmidt95,fan01} and decline by a large factor
between $z\sim 2$ and $z\sim 0$ \citep{schmidt68,boyle00}.
At $z \la 2.5$, \cite{boyle00} find a break in the luminosity
function (eq.~\ref{eqn:boyle})
that evolves towards lower luminosities at lower redshifts,
a form of ``luminosity evolution'' 
that cannot be described by a simple vertical shift in amplitude
(``density evolution'').
We will devote considerable attention to 
the implications of this result, though we should note that
\citet{wolf03} reach a more ambiguous conclusion about the
need for luminosity evolution, using a data set (COMBO-17) that reaches
still lower luminosities.  At $z \ga 2.5$, current observations
probe mainly the high luminosity end of the QLF, where the
data are adequately described by a single power-law.

We begin our discussion below with a few remarks on the
definitions of quasar lifetimes and duty cycles.
We then present evolutionary calculations for a number of
specific models designed to illustrate general points, working
within the 
class of models discussed in \S\ref{sec:dblpowpofmdot}: double power-law
$p(\mdot)$, double power-law $n(M)$, and $\epsd(\mdot)$ as defined in 
equation~(\ref{eqn:effofmdot}).  
In \S\ref{sec:accevol}, we consider cases in which accretion alone drives
the evolution of $n(M)$, looking first at the declining phase of 
QLF evolution ($z \leq 2$) and then at the growing phase.
In \S\ref{sec:mergers}, we use simple models to illustrate the
potential impact of mergers on the evolution of $n(M)$ and the QLF. 

\subsection{Lifetimes and duty cycles}
\label{sec:lifetimes}

One of the key elements in models of quasar evolution is the 
typical quasar lifetime, or, nearly equivalent,
the  duty cycle of quasar activity (see 
\citeauthor{martini03} [\citeyear{martini03}] for a review
of observational estimates).   
In a simple ``on-off'' model of the quasar population, where 
a black hole is either shining at a fixed fraction of its
Eddington luminosity or not accreting at all, it is 
clear what these concepts refer to: 
at a given redshift, the duty cycle is the fraction of black
holes that are active at any one time,
and the typical lifetime is the integral of the duty cycle over the age of 
the universe.  However, if $p(\mdot)$ is broad, and in particular
if there is a tail of increasing probability towards low $\mdot$,
then defining a black hole to be ``on'' if it is accreting at
{\it any} non-zero rate may not be particularly useful, since
changes to $p(\mdot)$ that have negligible observational effect
(such as changing the lower cutoff $\mdotmin$) may have a large
impact on the implied duty cycle or lifetime.  Such a 
definition is also difficult to relate to black hole growth or
emissivity.

For our purposes, it is more useful to consider the accretion weighted 
lifetime
\begin{equation}
\tacc \equiv \int_{t_i}^{t_f} \mdot(t)\, dt~.
\label{eqn:tacc}
\end{equation}
The ratio of $\tacc$ to the Salpeter lifetime 
($t_s=4.5\times 10^7$ yr, cf. eq.~\ref{eqn:tsdef})
gives the number of $e$-folds of mass growth, i.e.,
\begin{equation}
M_f = M_i \exp\left[{\tacc \over t_s}\right].
\label{eqn:mgrowth2}
\end{equation}
Typically, the initial time $t_i$ would refer to some time after the formation
of ``seed'' black holes but before the main epoch of mass accretion,
and $t_f$ would refer to $z=0$.  However, since we model the declining
and growing phases of quasar evolution separately below, we will 
generally choose $t_i$ and $t_f$ to correspond either to the redshift
interval $z=2-0$ or to the redshift interval $z=5-2$.
A useful way to characterize the mean accretion rate at a given
redshift is 
\begin{equation}
\taccz \equiv \avmdot H^{-1}(z)~.
\label{eqn:taccz}
\end{equation}
The ratio $\taccz/t_s$ is the number of $e$-folds of mass growth
that would occur if the mean accretion rate were to stay constant
for the Hubble time $H^{-1}(z)$.

For constant efficiency $\epsd$, weighting the lifetime by $\mdot$
is equivalent to weighting by $L/(\epsd\Ledd)$, so 
the accretion weighted lifetime is simply related to the luminosity weighted
lifetime.  Alternatively, one can weight activity by the ratio
of a black hole's luminosity $L(t)$ to its {\it final} Eddington
luminosity $lM_f$.  In this case, the weighted lifetime
(for constant efficiency and no black hole mergers) is 
$t_s \epsd (1-M_i/M_f),$ which approaches the Salpeter lifetime
in the limit that the black hole mass grows by a large factor. 

\subsection{Pure Accretion Driven Evolution}
\label{sec:accevol}
\subsubsection{Declining Phase with Mass Independent $p(\mdot)$}
\label{sec:declining}

For pure accretion driven evolution, only the first term of 
equation~(\ref{eqn:evol})
enters into the evolution of $n(M,t)$.  We start by considering the
case with $p(\mdot)$ independent of mass, so that the ``self-similar''
solution given in
equation~(\ref{eqn:accgrowth}) applies.  For a double power-law $p(\mdot)$, 
the evolution of the QLF depends on the time evolution of the normalization
$p_{*}(t)$ and the characteristic accretion rate $\mdot_{*}(t)$.  
We initially consider the declining phase of QLF evolution and take both 
functions to be power-laws of time: 
\be
p(\mdot|t)=\left\{ \begin{array}{ll}
         \pstar(t) \left( { \mdot \over \mdotstar (t)}\right)^a & 
	   \mdot<\mdotstar(t) \\
         \pstar(t) \left( { \mdot \over \mdotstar (t)}\right)^b & 
	   \mdot>\mdotstar(t)
         \end{array}
           \right., \qquad 
	   \pstar (t)=p_{*,i} \left({t \over t_i}\right)^{\gamma_p}, \quad
	   \mdotstar(t)=\mdot_{*,i} \left({t \over t_i}\right)^{\gamma_m} ~,
\label{eqn:timedeppofmdot}
\ee
where $t_i$ represents the time from which the initial QLF is evolved, and 
$p_{*,i}$ and $\mdot_{*,i}$ correspond to the values of $\pstar$ and 
$\mdotstar$ at time $t=t_i$.  

For $\gamma_m=0$, ``pure $p_*$ evolution,'' the relative probability of given 
accretion rates remains fixed over time, but the overall probability of 
accreting per unit time 
declines.  This is analogous to ``pure density evolution''
models of the QLF, though the 
masses of the black holes continue to evolve.  
For $\gamma_p=0$, ``pure $\mdot_*$ evolution,'' 
the relative probabilities of given accretion rates change over
time as the break in $p(\mdot)$ evolves.  This is analogous to 
``pure luminosity evolution,'' though again, 
the black hole mass function is evolving along with the 
evolution of $p(\mdot)$.  Physically, $p_*$ evolution could be connected 
to a decline
in the frequency of galaxy interactions as the universe gets older.  Evolution
of $\mdot_*$ could arise from declining gas fractions of galaxies, or a 
decline in their ability to funnel gas to the center as they become larger 
and more stable.

The evolution of the black hole mass function can be determined by using  
equation~(\ref{eqn:accgrowth}) to express $n(M,t)$ as
\be
n(M,t)= \left\{ \begin{array}{ll}
               \nstar \left({M \over \Mstar (t)} \right)^\alpha 
	         {M_{*,i} \over \Mstar(t)} & M<\Mstar(t) ~, \\
               \nstar \left({M \over \Mstar (t)} \right)^\beta 
	         {M_{*,i} \over \Mstar(t)} & M>\Mstar(t) ~,
 \end{array}
 \right. \qquad 
M_*(t) = M_{*,i} \exp 
  \left( \int_{t_{i}}^{t} \avmdott \frac{dt}{t_{s}} \right) ~,
\label{eqn:timedepnofm}
\ee
where $M_{*,i}$ is the mass corresponding to the break in the black hole mass 
function at a time $t=t_i$.  Since we assume that $p(\mdot)$ is independent 
of mass, the slopes of the QLF at low and high luminosities
match the slopes of $n(M)$, and thus we choose the \cite{boyle00}
slopes $\alpha=-1.5$ and $\beta=-3.4$ for $n(M)$. 
We use $M_{*}=10^{9}M_{\odot}$ as in \S\ref{sec:dblpowpofmdot}.  
For $p(\mdot)$, the values $a=-0.5$, $b=-3.0$, 
$\mdot_{*,i}=0.9$ then give a reasonable fit to the data at $z=2$.  
For our choice of $p(\mdot|t)$, the integral for $\Mstar(t)$ can
be done analytically.

\begin{figure}
\epsscale{0.9}
\plotone{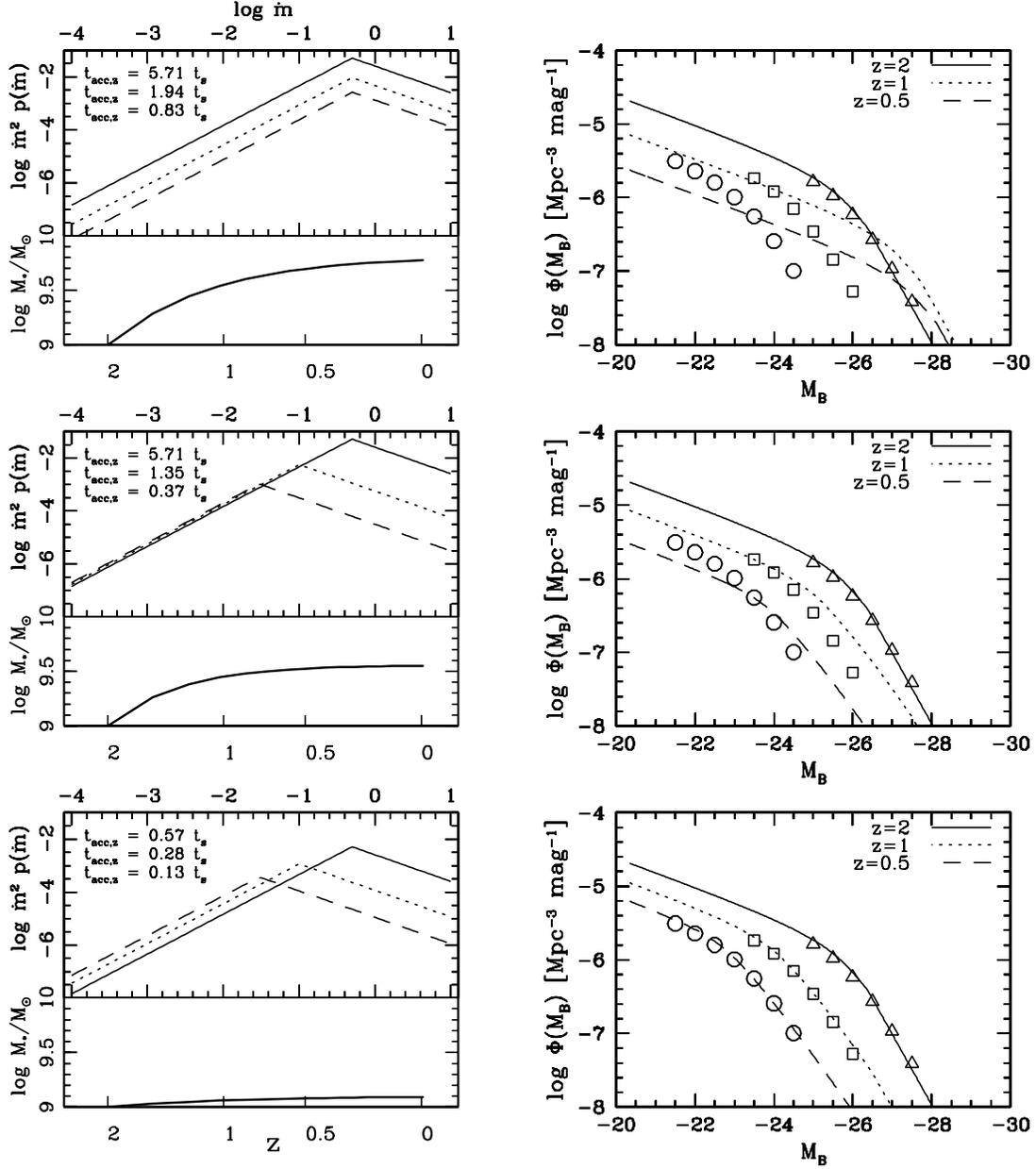}
\caption{
Three models of the declining phase of QLF evolution, from $z=2$ to
$z=0.5$.  For each model, left hand panels show $\mdot^2 p(\mdot)$ at 
$z=2, 1,$ and $0.5$ in the upper window (solid, dotted, and dashed lines, 
respectively), and $\Mstar(z)$ in the lower window.
Each panel also lists the accretion lifetime $\taccz = \avmdot H^{-1}(z)$
in units of the Salpeter 
timescale $t_s=4.5 \times 10^7$ yr, for $z=2, 1,$ and $0.5$ (top to bottom).
Right hand panels show the corresponding QLFs at the three redshifts.
Points show the fits of \cite{boyle00} at $z=2$ (triangles), $z=1$ (squares),
and $z=0.5$ (circles) over roughly the absolute magnitude range
probed by their data.  The top panels show a model in which $\mdotstar$
stays fixed and only the normalization $\pstar$ declines with time.
The middle panels show a model with the same initial $p(\mdot)$ but
declining evolution of $\mdotstar$.  The bottom panels show a similar
model that starts with a higher black hole space density and lower 
normalization of $p(\mdot)$, and consequently less black hole growth.
Matching the observed evolution of $\Lbrk$ towards lower luminosity
requires a decrease in the characteristic accretion rate $\mdotstar$,
not merely a decrease in duty cycle.
}
\label{fig:dblpowevol}
\end{figure}

Figure~\ref{fig:dblpowevol} uses equations~(\ref{eqn:timedeppofmdot}) and
(\ref{eqn:timedepnofm}) to determine the black hole 
mass function and probability function at redshifts $z=2$, 1, and 0.5.
The left hand panels show $\Mstar(t)$ in the lower windows and $p(\mdot|t)$ 
in the upper windows.
The right hand panels show the evolution of the QLF generated by the functions 
in the corresponding left hand panels, using the methods and assumptions
described in \S\ref{sec:dblpowpofmdot}.  To reduce the dimensionality of the
parameter space, we assume a value of $\gamma_m$ and then find a value 
for $\gamma_p$ that makes the evolved, $z\sim0.5$ QLF match the 
\cite{boyle00} data at the break luminosity $\Lbrk$.
As a characterization of the mean accretion level, we list in the left
hand panels the values of the local accretion weighted lifetimes $\taccz$
(eq.~\ref{eqn:taccz}), in units of the Salpeter time $t_s$, at 
$z=2$, 1, and 0.5.  These ratios $\taccz/t_s$ give the number of
$e$-folds of black hole growth that would occur in a Hubble time
$H^{-1}(z)$ if $p(\mdot)$ stayed constant.
In contrast to earlier figures,
we plot $\mdot^2 p(\mdot)$ rather than $p(\mdot)$ itself,
because this product gives the contribution to black hole
growth (and emissivity of the quasar population) per logarithmic
interval of $\mdot$.  For our adopted forms of $p(\mdot)$, the
largest contribution to growth and emissivity always comes from accretion
rates near $\mdotstar$.

The upper panels of Figure~\ref{fig:dblpowevol} show an example
with $\gamma_m=0$, so that the normalization of $p(\mdot|t)$ evolves
but the shape does not.  Although we require the predicted QLF to
pass through the observed $\Phi(\Lbrk)$ at each redshift,
the shapes of the model luminosity functions at $z=1$ and $z=0.5$
disagree strongly with the data.
Generally, evolution of $\pstar$ alone cannot reproduce the type
of QLF evolution found by \cite{boyle00} at low redshift,
because the growth of $M_*$ combined with a fixed {\it  relative}
distribution of accretion rates shifts the predicted QLF horizontally 
to a higher break luminosity, while the observed break luminosity declines
with time.  Reducing $p_*$ reduces
the fraction of accreting black holes, but that only produces
a vertical drop of the QLF, which cannot fully 
compensate for the shift to a higher break luminosity.  

Since ``pure density evolution'' models fail to describe the \cite{boyle00}
data, it is no surprise that our analogous ``pure $p_*$ evolution'' models
also fail.  However, Figure~\ref{fig:dblpowevol} shows that black hole
growth exacerbates the failures of a model in which only the frequency
of fueling activity declines with time, since such models generically
predict an increase in the break luminosity with time.  We conclude
that, if $p(\mdot)$ is independent of mass, then it must evolve in
a way that increases the relative probablilities of low accretion rates
in order to make the QLF evolve to lower break luminosities.
We will consider an alternative explanation --- that $p(\mdot|M)$
has a mass dependence that evolves with time --- 
in \S\ref{sec:massdepdeclining}.  However, a decline in typical values
of $\mdot$ does seem a plausible consequence of decreasing gas
fractions and increasing stability of host galaxies, and such
changes in galaxy properties are likely to play an important role
in producing the observed form of QLF evolution.
Furthermore, even if gas fueling rates remain fixed in physical units,
they decline in Eddington units as black hole masses increase,
thus driving $\mdot$ values down if black holes grow by a
significant factor.
In their semi-analytic model of the quasar and host galaxy population,
\cite{kauffmann00} find that they must account for the decreasing
gas supplies and longer dynamical timescales of host galaxies at
low redshift, in addition to the decreasing frequency of mergers,
to explain the observed evolution of the QLF.  In terms of our models,
the first two effects are analogous to decreases in $\mdotstar$,
while the last is analogous to a decrease in $\pstar$.

The models in the middle and bottom rows of Figure~\ref{fig:dblpowevol}
incorporate declining $\mdotstar$ and match the observed QLF evolution
better.  The case in the middle row starts with the same $p(\mdot)$ and
$n(M)$ at $z=2$, but it has
$\gamma_m=-2.7$, which moves the break in $p(\mdot|t)$ to lower
accretion rates as time progresses.
Though the match to the data is not perfect, it is much better than
before, with the break in $\Phi(L)$ shifting to lower luminosities
as the QLF becomes increasingly dominated 
by the contribution from lower accretion rates.
The model in the bottom row starts at $z=2$ with a black hole space
density a factor of ten higher and an accretion duty cycle a factor
of ten lower ($\nstar$ and $\pstar$ increased and decreased by ten,
respectively).  The QLFs at $z=2$ are identical because of the exact
degeneracy between $\nstar$ and $\pstar$ discussed in 
\S\ref{sec:powerlawpofmdot}.  However, the reduction in $\pstar$
lowers the mean accretion rate $\avmdot$, which in turn leads to 
less black hole growth: this model has $\taccz<t_s$ at all $z<2$,
and the characteristic mass $M_*$ hardly grows at all.
The shift to lower accretion rates coupled 
with the smaller amount of black hole growth over time 
yields QLF evolution in good agreement with the data.

Figure~\ref{fig:dblpowevol} shows that the degeneracy between 
$\nstar$ and $\pstar$ is broken once evolution is taken into account.
If the black hole density is low, then each black hole must accrete
more in order to match the observed QLF, and this accretion leads to
more rapid evolution of $n(M)$.  In the case shown in 
Figure~\ref{fig:dblpowevol}, the model with less black hole growth
matches the data better.  However, it is possible to start with the
initial conditions of the model in the middle panels and match the 
data nearly as well by dropping $\mdotstar$ more rapidly,
with $\gamma_m=-3.7$ instead of $-2.7$.
Thus, changes to $\nstar$ and $\pstar$ can be partly compensated
by changes to other parameters.
Nonetheless, evolution narrows the range of the $\nstar\pstar$
degeneracy because models with too much black hole growth
(too low $\nstar$) cannot yield a declining break luminosity for
any plausible evolution of $p(\mdot)$.
Furthermore, as discussed in \S\ref{sec:massdist} and further
in \S\ref{sec:illus} below, models with different $\nstar$ predict
different distributions of active black hole masses and accretion
rates, different space densities of host galaxies,
and of course different underlying $n(M)$, even when
they match the same, evolving QLF.

Similar remarks apply to the partial degeneracy between $M_*$
and $\mdotstar$ discussed in \S\ref{sec:dblpowpofmdot}.
Combinations of $M_*$ and $\mdotstar$ that yield similar 
QLFs at $z\sim 2$ will have more growth of $n(M)$,
and thus different evolution, if $M_*$ is lower and $\avmdot$
consequently higher.

\subsubsection{Declining Phase with Mass-Dependent $p(\dot{m})$}
\label{sec:massdepdeclining}

As shown in \S\ref{sec:massdeppofmdot}, mass dependence of $p(\mdot)$
can break the link between the shape of the black hole mass function
and the shape of the luminosity function, adding considerable freedom
to models of the QLF.  In an evolutionary calculation, one must also
account for the influence of mass-dependent growth on the shape
of the black hole mass function.  
If the more numerous, low mass black holes 
have a higher probability of being active,
then they grow faster than the
high mass black holes, and the mass function steepens.
Conversely, if high mass black holes are more active, then they
grow faster and the mass function becomes shallower with time.
It is intuitively useful to think of this behavior in graphical terms.
On a log-log plot, a mass-independent $p(\mdot)$ causes the mass
function to shift horizontally in a coherent fashion, maintaining
its shape.
Faster growth of low mass black holes allows the low end of $n(M)$
to translate faster and ``catch up'' with the high end.  
Conversely, faster growth
of high mass black holes
allows the high end of $n(M)$ to stretch away from the low end.
As always (eq.~\ref{eqn:acc}),
it is only the mean accretion rate $\avmdotM$ that matters
for determining the evolution of $n(M)$, and one can calculate the
evolution exactly by
assuming that all black holes of a given mass accrete at this rate.

To describe the evolution of $n(M)$ mathematically,
we define $g(M,t) = M_i/M$, where $M_i$ and $M$ represent black hole
masses at times $t_i$ and $t$, respectively.
Matching number densities in the equivalent mass intervals at the two
times then implies that $n(M)dM=n_i(M_{i})dM_{i}$, 
and thus
\be
n(M)=n_i(M_i){dM_i \over dM}=n_i(M_i)\left[g(M,t) + 
     M {\partial g \over \partial M}\right]~.
\label{eqn:massdepgrow}
\ee
If $g(M,t)$ is independent of mass, then we have simple remapping
of $M_i \rightarrow M$ and renormalization by $g(M,t)$, recovering
the result~(\ref{eqn:accgrowth}).  For mass-dependent $\avmdot$,
the value of $n(M)$ at mass $M$ may be higher or lower than the
mass-independent case depending on the sign of $\partial g/\partial M$.
Equation~(\ref{eqn:massdepgrow}) allows the 
numerical calculation of $n(M)$ for any specified $p(\dot{m}|M,z)$, 
since this determines $g(M,t)$.  However, the exponential relation
between the growth factor and $\int \avmdot dt$ generally means
that any simple analytic form of $n(M)$ is lost once black holes
grow significantly.

We have previously adopted mass-independent $p(\mdot)$ as a mathematical
convenience, but consideration of black hole growth suggests that
this choice is not completely arbitrary.  Suppose that the black holes
in some mass range have a high $\avmdot$ relative to their peers
because they tend to reside in galaxy hosts that can feed
them more efficiently.  These black holes $e$-fold in mass more
rapidly than others, and their fueling rates {\it in Eddington units}
therefore drop more quickly (or grow more slowly), bringing them
back into line.  This regulating mechanism suppresses mass-dependence
of $\avmdot$, causing $n(M)$ to change
 shape until it approaches the ``fixed point''
solution of mass-independent $\avmdot$, after which it evolves in a
self-similar fashion.
Constancy of $\avmdot$ does not necessarily imply constancy of $p(\mdot)$,
but this regulation argument suggests that $p(\mdot)$ might be 
approximately mass-independent in an average sense
at redshifts near the peak of quasar activity.

The regulating mechanism may lose force at low redshift, when 
black holes no longer grow by substantial factors and gas merely
trickles onto full grown systems.
Thus, we might expect stronger mass dependence of $p(\mdot)$ in 
the declining phase of quasar evolution.
We have shown in \S\ref{sec:declining} that
matching the observed shift of $\Lbrk$ to lower luminosity requires a decline
in characteristic $\mdot$ values if $p(\mdot)$ is independent of mass.
However, mass-dependent $p(\mdot)$ offers another possibility:
activity could decline preferentially in more massive black holes
between $z=2$ and $z=0$,
thus driving the break luminosity down as the typical mass of {\it active}
systems declines.

To create a model along these lines, it is helpful first to consider
the case where there is no evolution of $n(M)$ at all, so that one
can infer the required $p(\mdot|M,z)$ from a
simple graphical argument relating the vertical shift of $\Phi(L)$ to the 
horizontal shift of $\Lbrk$.
We consider a double power-law QLF with slopes $\alpha$ and $\beta$ 
below and above $\Lbrk$ respectively, and we assume pure luminosity 
evolution with $\Phi(\Lbrk)={\rm constant}$ and $\Lbrk(t_2)<\Lbrk(t_1)$
for $t_2>t_1$.  For maximum contrast with the model in 
\S\ref{sec:declining}, we assume that the mass dependence of $p(\mdot)$
enters only in the normalization
$p_*(M)$, not in the slopes or in the characteristic accretion rate
$\mdotstar$.
In other words, the relative distribution of accretion rates remains the same 
at all times for all active black holes,
but the probability of a black hole being active at all 
depends on its mass, in a redshift-dependent manner.
Relating the amplitudes of $\Phi(L)$ at times
$t_1$ and $t_2$ to the horizontal shift of $\Lbrk$ from time $t_1$ to $t_2$ 
then implies
\be
{\Phi(L,t_2) \over \Phi(L,t_1)} = \left\{ \begin{array}{ll}
                                 k^{\alpha} & L < \Lbrk(t_2) \\
                                 k^{\beta} & L > \Lbrk(t_1)
				 \end{array}
                                 \right., \qquad k={\Lbrk(t_2) \over \Lbrk(t_1)}
				 ~,
\label{eqn:massdepratio}
\ee
with a more complicated dependence in the range 
$\Lbrk(t_2)<L<\Lbrk(t_1)$.  Since 
$n(M)$ is constant, any shift in the QLF must be produced by a shift in 
$p_*(M)$, and thus,
\be
{p_*(M,t_2) \over p_*(M,t_1)}
      = \left\{ \begin{array}{ll}
                      k^{\alpha} & M < M_{\rm min} \\
                      k^{\beta} & M > M_{\rm max} \\
                      k^{\alpha} + (k^{\beta} - k^{\alpha})U(M) & 
		         M_{\rm min} < M < M_{\rm max} ~,
				 \end{array}
                                 \right. 
\label{eqn:massdeppofmdot}
\ee
where $M_{\rm min}$ is the minimum mass of a black hole that can generate 
$L=\Lbrk(t_2)$ and $M_{\rm max}$ is the maximum mass of a black hole that can 
generate $L=\Lbrk(t_1)$. 
Here $U(M)$ is some function that goes from
zero to one as mass goes from $M_{\rm min}$ to $M_{\rm max}$, 
and it can be tuned to reproduce the observed $\Phi(L)$ in the break region.
With equation~(\ref{eqn:massdeppofmdot}) as a starting guess, we can
find by numerical iteration a solution with mass-dependent $p_*(M)$
that reproduces the observed evolution and self-consistently 
incorporates the implied growth in $n(M)$.

\begin{figure}
\epsscale{1.0}
\plotone{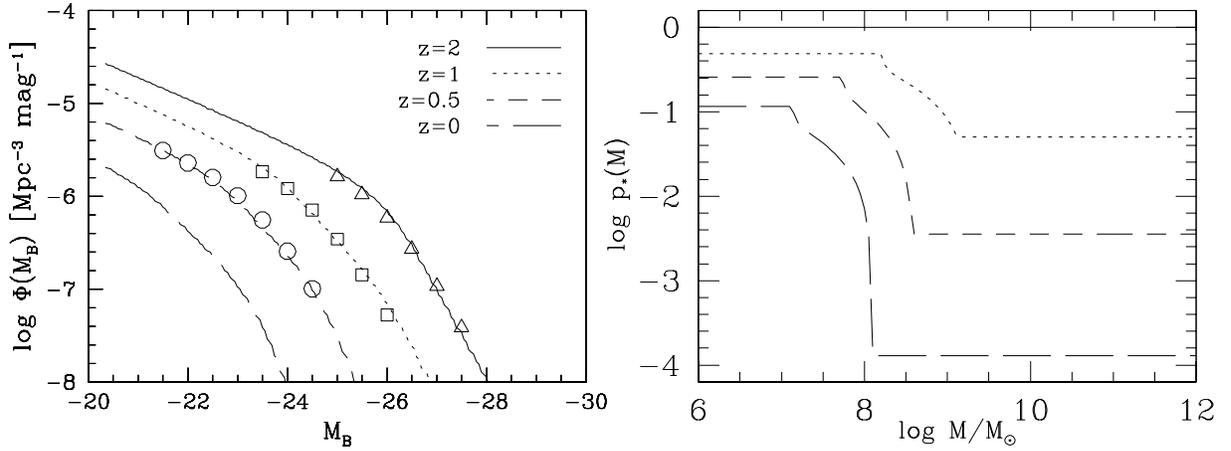}
\caption{
An alternative model for the declining evolution of the QLF, in which
the characteristic accretion rate $\mdotstar$ remains constant but the
mass dependence of $p(\mdot|M)$ evolves to suppress activity 
preferentially in high mass black holes at lower redshifts.
As in Figure~\ref{fig:dblpowevol}, triangles, squares, and circles in
the left hand panel represent the \cite{boyle00} observational fits
at $z=2$, 1, and 0.5, respectively, while lines show the model
predictions (including an extrapolation to $z=0$).
The right hand panel shows the evolving mass dependence of the
normalization $\pstar(M)$, which is proportional to the duty cycle
of black holes of mass $M$.  The model assumes our usual double
power law forms of $n(M)$ and $p(\mdot)$, with $M_*=10^9 M_\odot$
and a mass-independent $p(\mdot)$ at $z=2$, and $\mdotstar=0.5$ at
all redshifts.  
}
\label{fig:massdepevol}
\end{figure}

Figure \ref{fig:massdepevol} shows a model in which we assume that $p(\mdot)$ 
is independent of mass at $z=2$ and that subsequent evolution of the 
QLF (left panel) is driven by the mass 
dependence of $p(\dot{m}|M)$ (right panel).  We choose a 
normalization of the black hole mass function, 
$n_*M_*=1.2 \times 10^{-4}\dunits$, 
that corresponds to a fairly
short quasar lifetime, $\tacc = 7 \times 10^6$ yr (from $z=2$ to $z=0$), so 
that the mass-dependent growth does not severely distort the double 
power-law form of $n(M)$ that our initial guess at $p(\dot{m}|M)$ assumes.  
This model fits the \citet{boyle00} data as well as
the model with mass-independent $p(\mdot)$ 
shown in the bottom panel of Figure~\ref{fig:dblpowevol}.  
With regard to $\Phi(L)$ alone, the two models are effectively
degenerate, but $\Lbrk$ in the mass-dependent model evolves to 
lower luminosities because high mass black holes are 
less likely to be active at low redshift.  Thus, relative to 
the mass-independent model, this model predicts that luminous AGN
at low redshift consist primarily of low mass black holes with
high $L/\Ledd$, and it predicts a narrower range of $\mdot$ values.  We compare
the two models' predictions for the mass distribution of active
systems in \S\ref{sec:illus} below (Fig.~\ref{fig:shortvsmassdep}).

\subsubsection{Growing Phase}
\label{sec:growingphase}

We now turn to the redshift interval $z \sim 5$ to $z\sim 2$, during which
$\Phi(L)$ first grows rapidly, then reaches a plateau between 
$z\sim 3$ and $z\sim 2$ (see, e.g., \citealt{who,schmidt95,pei95}).
In the range $z=3.6-5.0$, \cite{fan01} provide the best measurements
of the bright end of the luminosity function, while 
\cite{wolf03} give constraints at fainter luminosities.
To cover the gap between $z=3.6$ and the \cite{boyle00}
measurements at $z=2$, we use the measurements of \cite{who},
with a median redshift $z \approx 3.25$.  
We use the $\Omega_m=1$ cosmological model, since all of these papers 
give results for this case, and we adopt $h=0.5$ so that the
age of the universe is realistic.
We convert the \cite{who}
space densities and absolute magnitudes from $h=0.75$
to $h=0.5$, and we convert their AB($\lambda$1216\AA) magnitudes
to $B$-band magnitudes using $M_{B}=$AB($\lambda$1216\AA)$-0.605$,
assuming $f_\nu \propto \nu^{-0.5}$ for the conversion
of AB($\lambda$4400\AA) to AB($\lambda$1216\AA).
Our goal here is not to model the data in detail but merely to
illustrate what kinds of $p(\mdot)$ and accompanying $n(M)$ evolution
can fit the general trends.  

For simplicity, we restrict ourselves to the class of models in which
$p(\mdot)$ is a double power-law independent of mass and the characteristic
accretion rate $\mdotstar$ is constant from $z=5$ to $z=2$.
At $z=2$, we assume a double power-law $n(M)$ with $M_*=10^9 M_\odot$
and slopes $\alpha=-1.5$ and $\beta=-3.4$.
The evolution of the QLF is then determined 
by the evolution of the amplitude of $p(\mdot)$, 
which we assume has a piecewise power-law 
form, $p_*(t)\propto t^{\gamma_{p}}$ in the interval between
each of the redshifts where we match the QLF,
though we allow $\gamma_{p}$ to be different from one interval to another.
Our general procedure is to take a model that fits the QLF data at $z=2$, 
then evolve it to higher redshift, iteratively solving for
values of $\gamma_{p}$ in each redshift interval so as to match
the observed amplitude of the QLF at $M_B \sim -26.5$, using the
\cite{who} data at $z=3.25$ and the 
\citet{fan01} data at $z=3.75, 4.15,$ and $4.7$.
Iteration is necessary because we must calculate the time integrated
mean accretion rate for the given $\gamma_p$ and reduce black hole
masses by the corresponding factor, before calculating the QLF
with the evolved $p(\mdot)$.
Because mass-independent $p(\mdot)$ leads to self-similar evolution
of $n(M)$ and identical asymptotic slopes for $n(M)$ and $\Phi(L)$
(see \S\ref{sec:powerlawpofmdot}), this restricted
class of models cannot explain
the apparent change in the bright-end slope of the QLF 
(see \citealt{fan01}) between $z\sim 4$ and $z\sim 2$.

\begin{figure}
\epsscale{1.0}
\plotone{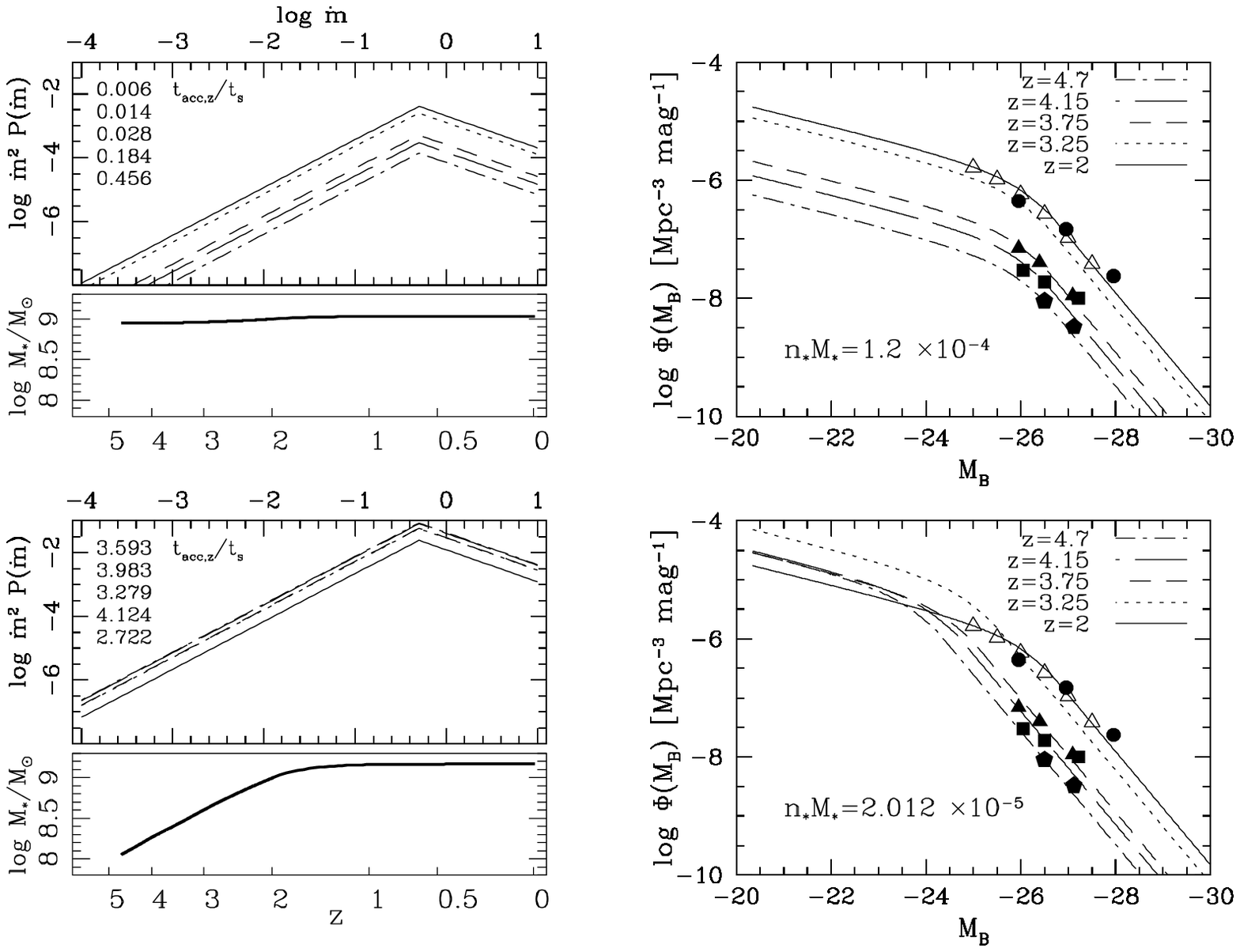}
\caption{Like Figure \ref{fig:dblpowevol}, but for the growing phase of 
QLF evolution from $z\sim5$ to $z\sim2$.  
Points in the right hand panels are based on \cite{fan01} at 
$z\sim 4.7$, 4.15, and 3.75 (pentagons, squares, filled triangles),
on \cite{who} at $z\sim 3.25$ (circles), and on \cite{boyle00} at $z\sim 2$
(open triangles).
Model parameters are chosen to reproduce the amplitude of the
observed QLF at $M_B \sim -26.5$.  The top row shows a model with
a high black hole space density at $z=2$, $n_* M_* = 1.2\times 10^{-4}\dunits$,
in which case there is little growth of black hole masses between 
$z=5$ and $z=2$ and the QLF evolution is driven by increasing $p(\mdot)$.
The bottom row shows a model with a lower black hole space density and,
therefore, more accretion per black hole.  In this model, the growth of
the QLF is driven mainly by the increasing masses of black holes.
}
\label{fig:growth}
\end{figure}

Figure~\ref{fig:growth} shows two qualitatively different solutions
that reproduce the observed evolution of the QLF amplitude.
The first of these solutions, shown in the top panels, has a relatively 
high normalization of the black hole mass function at $z=2$, with
space density $n_* M_* = 1.2 \times 10^{-4}\dunits$. 
The high space density leads to a low normalization of $p(\mdot)$
at each redshift, and consequently to little growth of black hole masses;
the lifetimes $\taccz$ are shorter than $t_s$ at all redshifts,
and the value of $M_*$ increases only slightly over the entire range
$z=4.7$ to $z=0$.  (Note that we show the low redshift evolution of
$M_*$ inferred by continuing the model past $z=2$ using the 
Boyle et al. [\citeyear{boyle00}] data as in the previous section.)
The growth in the amplitude of the QLF is therefore driven 
by the steadily rising amplitude of $p(\mdot)$, with no significant
contribution from growth in $n(M)$ itself.

While this high space density solution fits the QLF data 
reasonably well by construction, it seems rather implausible on
physical grounds.  The minimal evolution of $M_*$ means that
all of the observed quasar activity represents a negligible
contribution to the growth of black holes.  Instead, all of
the supermassive black holes must have been assembled in some
unseen manner before $z=5$, and the observed evolution of the
QLF reflects a gradual ``turning on'' of these black holes
at lower redshift.  In such a model, the ``seed'' black holes
formed at $z>5$ are essentially the same black holes that are
present today, though mergers may have converted many low mass
black holes into a smaller number of high mass systems.

The bottom panels show a solution near the opposite extreme,
with a lower space density $n_* M_* = 2.012\times 10^{-5}\dunits$ at $z=2$.
Matching the QLF now requires a higher normalization of $p(\mdot)$
at each redshift, and the accretion weighted lifetimes are in the range
$\taccz \sim 0.4-2.7 t_s$.
There is roughly a factor of ten growth in $M_*$ between 
$z=5$ and $z=2$, and this growth plays a central role in the
evolution of the QLF.  In contrast to the high space density case,
the amplitude of $p(\mdot)$ is approximately
constant from $z=5$ to $z=3$, with a slight drop to $z=2$.
In this solution, therefore,
the observed quasar activity traces the growth of the black
hole population from much smaller seeds present at $z=5$,
with a roughly constant distribution of accretion rates in
Eddington units during the growing phase of quasar evolution.
The break in the QLF moves to higher luminosity as $M_*$ grows,
so measurements probing to lower luminosities at high redshift
could distinguish this solution from the short lifetime solution
shown in the top panels.

If the $n(M)$ normalization is reduced just slightly further, below 
$n_*M_* \approx 2.01 \times 10^{-5}\dunits$, then the solution
for the evolving $n(M)$ requires unphysical, negative densities
at $z=5$.  Thus, within our assumptions, there is a minimum
allowed space density of black holes, which corresponds to the
limit in which the observed QLF traces all of the accretion onto
the black hole population.  We can see why this is so
by recalling Soltan's (\citeyear{soltan82}) argument that the
integrated bolometric emissivity of the quasar population determines,
for an assumed efficiency, the mass density 
$\rhobh = \int_0^\infty Mn(M) dM$ of the black hole population.
We have adopted a form for $n(M)$, and the evolutionary calculation
allows us to, in effect, correct the observed emissivity for
the contribution from lower luminosity systems.  Since we have
also specified the efficiency $\epsilon(\mdot)$ and the 
bolometric correction $\Lbol/L_B$, matching the observational
data points determines $\rhobh$ at $z=2$ and 
thereby fixes the normalization of $n(M)$.
Our results thus show that a suitable extension of the \cite{soltan82}
argument, aided by some auxiliary assumptions and a measurement
of the QLF at $z\sim 2$, can predict the
black hole mass function $n(M)$ itself, not just the integral $\rhobh$.
We will explore this idea in future work, with attention to the 
sensitivity of the predictions to the auxiliary assumptions and to 
uncertainties in the observational data.

\subsection{Mergers}
\label{sec:mergers}

The general equation~(\ref{eqn:evol}) for the
evolution of the black hole mass function has a very limited set of
analytic solutions, even if one considers only the merger terms
and ignores accretion driven growth.
Realistic calculations incorporating merger driven growth will 
probably need to be done numerically, with some {\it a priori}
model (based on galaxy merger trees, for example) for what merger 
rates should be.  Here we will investigate some simple, analytically
solvable cases that can provide insight into the generic effects of mergers 
on the black hole mass function.  

First, we consider a binary merger model in which there is a probability $f$
per unit time that a given black hole merges with another black hole of
equal mass to form a new black hole of twice the original mass.
In the absence of accretion, a counting argument yields the
equation governing $n(M,t)$,
\be
{\partial n(M,t) \over \partial t}= 
                          -fn(M,t) + {1\over 4}fn\left({M \over 2},t\right) .
\label{eqn:simplemerge}
\ee
The first term on the r.h.s. is the sink representing loss of black
holes of mass $M$ to mergers, and the second is the source representing
creation by mergers of systems with mass $M/2$, with a factor of $1/4$
to account for the replacement of two black holes by one and the factor
of two growth in the $dM$ interval.
The only solution to (\ref{eqn:simplemerge})
in which $n(M)$ maintains its shape over time is a
pure power-law of the form
\be
n(M,t)= \nstar (t) \left({M \over \Mstar}\right)^{\alpha} ,
\label{eqn:powerlawnofm}
\ee 
where we have included all the time dependence in $\nstar$ because 
the effects of
$\nstar$ and $\Mstar$ are degenerate for a pure power-law.  
With this form, the solution to~(\ref{eqn:simplemerge}) is
\be
\nstar (t) = \nstar (t_i) \exp \left[ \int_{t_i}^t 
  \left(2^{-(\alpha+2)}-1\right) f(t) dt\right]. 
\label{eqn:simplesolution}
\ee
If $f(t)$ is constant, then $\nstar (t)$ evolves exponentially in time, while 
$f(t) \propto t^{-1}$ yields $\nstar (t)$ evolving as a power-law with 
slope of $\left(2^{-(\alpha+2)}-1\right) t_i f(t_i)$.  

The important general feature of the solution~(\ref{eqn:simplesolution}) is
that the {\it sign} 
of the evolution depends on the slope $\alpha$ of the mass function.  
For the critical value $\alpha = -2$, mergers do not change $n(M)$
at all, because the source and sink terms in equation~(\ref{eqn:simplemerge})
balance.  If $\alpha < -2$ then $n(M)$ increases with time, and if 
$\alpha > -2$ then $n(M)$ decreases with time.  For a steep $n(M)$,
the black holes added to a given range of mass from 
mergers of lower mass objects exceed the number lost to higher masses.  
For a shallow $n(M)$, on the other hand, mergers consume more black
holes in a given mass range than they create.
While the solution~(\ref{eqn:simplesolution}) is specific to our 
restricted model, different behavior for steep and shallow slopes of $n(M)$ 
follows from mass conservation,
so we expect it to hold quite generally.

The pure power-law $n(M)$ adopted above cannot hold for all masses because
the implied total 
mass density of black holes would be infinite.  
However, the behavior of the single 
power-law solution gives insight into 
the more general case of a mass function that changes slope from low
to high masses.  
For example, a double power-law has a steep high mass end 
where mergers drive $n(M)$ up with time and a shallow low
mass end where mergers drive $n(M)$ down.
Over the range in mass near the break, the shape changes 
as the number of high mass black holes grows and low mass black holes decreases,
making the break smoother and shifting it to lower masses.
Again, we expect these effects of mergers to be fairly generic.

For a pure power-law $n(M)$, we can also include the accretion term of 
equation~(\ref{eqn:evol}) in the calculation, obtaining the solution
\be
{\dot{n} \over n}={\dot{n}_* \over \nstar}= 
     -{\langle \mdot (t) \rangle \over t_s}(1+\alpha) + 
     \left(2^{-(\alpha+2)}-1\right) f(t) ~.
\label{eqn:accreteandmerge}
\ee
The sign of the accretion term also depends on $\alpha$, 
but here the critical slope is $-1$ instead of $-2$, 
reflecting the fact that accretion adds mass to the black 
hole population while mergers do not.
Equation~(\ref{eqn:accreteandmerge}) can be integrated analytically
if accretion and merger rates have same time dependence,
$\langle \mdot(t) \rangle = \langle \mdot(t_i) \rangle h(t)$ and 
$ f(t) =f(t_i) h(t)$, with $h(t)$ an arbitrary function having $h(t_i)=1$.
The solution is
\be 
\nstar(t) =\nstar(t_i)\, {\rm exp} 
  \left( \left[-\frac{\langle \mdot(t_i) \rangle}{t_s}(1+\alpha) + 
  \left(2^{-(\alpha+2)}-1\right) f(t_i)\right] \int_{t_i}^t h(t) dt \right).
\ee
Up to factors that are typically of order unity, the relative importance
of accretion and mergers depends on the value of $ \langle \mdot(t_i) \rangle$ 
relative to $t_s f(t)$, the average number of mergers per Salpeter time. 
However, the
pre-factors can greatly diminish one term or the other close to the critical
slopes $\alpha=-1$ or $\alpha=-2$, and for $ -1 > \alpha > -2 $ accretion
and mergers affect $n(M)$ in opposite directions.

For our illustrative model calculations in \S\ref{sec:illus}, we do not want
to assume a pure power-law $n(M)$, and we will therefore adopt another
very simple prescription for mergers, similar to that of \citet{richstone98}.
We assume that accretion has negligible impact on $n(M)$ after some time $t_1$,
which is true if $\langle \mdot(t_1) \rangle t_1 \ll t_s$ 
and $\langle \mdot(t) \rangle$ 
falls as $t^{-1}$ or faster.  At some later time, $t_2$, every black 
hole with initial mass $M_1$ is assumed to 
have merged with $(f_m-1)$ other black holes of 
mass $M_1$ to make one black hole of mass $M_2=f_m M_1$.  
The black hole mass functions at $t_1$ and $t_2$ are related by
the transformation 
\be
n_2(M_2)dM_2 = {1 \over f_m}\, n_1 \left(M_1={M_2 \over f_m}\right) 
{dM_1 \over f_m},
\ee
or simply 
\be
n_2(M) = f_m^{-2} n_1(M/f_m).
\label{eqn:verysimplemerger}
\ee
In this model as in our previous model, mergers drive
$n(M)$ up when the logarithmic slope is $\alpha<-2$ and
down when $\alpha > -2$.
If $n(M)$ is a double power-law at time $t_1$, then at time $t_2$ it
is still a double power-law, with $M_*$ larger by a factor $f_m$
and the normalization lower by a factor $f_m^2$.

In both of our merger models, we assume that mergers conserve the total
mass of the black hole population, and simply redistribute it from 
low mass systems to high mass systems.  However, it is also possible
for mergers to {\it decrease} the total mass of the population, at
least those black holes that reside at the centers of galaxies and
have the potential to become quasars.  This can happen if multiple
mergers produce triple or quadruple systems that lead to ejection
of one or more of the black holes from the galaxy
(e.g., \citealt{valtonen94}).  Even a merging binary system can
potentially be ejected by a gravitational radiation ``rocket'' effect,
though this seems more likely to be important in shallow potential
wells hosting low mass black holes (see \citealt{redmount89};
\citealt{madau03}).
Finally, a binary could remain at the galaxy
center but radiate a significant fraction of the mass 
of its progenitors in gravity waves during the merger event
(\citealt{yu02} and references therein).
We will not consider any of these possibilities in detail here,
but we note that in all these cases the critical slope at which 
$n(M)$ grows rather than declines would be steeper than $-2$, since
a given bin would lose mass to mergers at the same rate as before
but would gain mass at a lower rate.

\section{Illustrative Scenarios}
\label{sec:illus}

We now utilize the framework developed above to construct models
that illustrate different plausible scenarios for the evolution
of the quasar population.
The simplest scenario is that the luminous quasar population is 
dominated by black holes accreting with thin-disk efficiency $\epsd\approx 1$, 
all radiating with a ``standard'' SED (e.g., \citealt{elvis94}),
and that the growth of black holes is driven by the observed
accretion.  The key parameter of this scenario is the typical quasar 
lifetime, which is linked in turn to the space density of black holes.
However, there are many potential variations on this theme,
including the possibility
that a large fraction of quasar activity is obscured, 
that black hole growth is substantially affected by low redshift mergers, or
that substantial black hole growth occurs though low efficiency ADAF accretion.
Our models here illustrate each of these physically distinct
possibilities, including long and short quasar lifetimes for the simplest 
scenario.  For simplicity, we will 
consider only models with $p(\dot{m}|z)$ independent of mass.
We examine only the regime from $z=2$ 
to $z=0$ and choose parameters so that each model approximately reproduces 
the observed optical QLF.  Our goal is to determine what other observables, 
such as the QLF in other bands, the masses of active black holes at different 
luminosities, the black hole mass function itself, and the
space density of host systems are most
likely to discriminate among these scenarios.  
We include some comparisons to recent estimates of these observables,
but our emphasis is mainly on the differences among the models themselves,
since we have not made any adjustments to model parameters to try to
match these other data.

\subsection{Model Parameters}

The parameters of the five models are summarized in Table~\ref{tbl:parameters}.
Our baseline parameters are similar to those used in the evolutionary
calculations of \S\ref{sec:declining}.  We adopt a double power-law $n(M)$
with slopes $\alpha=-1.5$ and $\beta=-3.4$, 
which are required to match the asymptotic slopes of the \cite{boyle00}
QLF for cases where $p(\dot{m})$ is independent of mass.  
We adopt $M_* \approx 10^9M_\odot$, making the Eddington luminosity
$lM_*$ close to the observed break luminosity at $z=2$.
Except for the ADAF model (discussed below), we adopt
a double power-law $p(\dot{m})$
with parameters $a=-0.5$ and $b=-3.0$.  
We start with a characteristic accretion rate $\mdotstar=0.5$
and evolve it to lower redshift as $\mdotstar \propto t^{\gamma_m}$.
The normalization $p_*$ evolves as $t^{\gamma_{p}}$,
with different $\gamma_p$ values from $z=2$ to $z=1$, $z=1$ to 0.5,
and $z=0.5$ to 0.
The values of $\gamma_m$ for each model are chosen
to give a reasonable match to the observed evolution of $\Lbrk$
given the model's predicted growth of $n(M)$, and
the values of $\gamma_p$ are then chosen by matching the
observed amplitude of the QLF.  
Table~\ref{tbl:evol} summarizes the quantitative evolution of the five
models, giving the values of $M_*$, the comoving black hole
mass density $\rhobh$, and the mean accretion rate $\avmdot$
at redshifts 2, 1, 0.5, and 0, and the accretion weighted lifetime
$\tacc = \int_2^z \langle \mdot(t)\rangle dt$ at $z=1,$ 0.5, and 0.

\begin{deluxetable}{lrrrrr} 
\tablewidth{0pt}
\tablecaption{Input Model Parameters
               \label{tbl:parameters}
	     }
\startdata \hline
                & Short-$t_q$ & Long-$t_q$ & Obscured & Merger & ADAF \\ \hline
$M_{*}(z=2)$     &     1      &     1      &    1     &   1    & 1.25 \\  
$n_* M_* (z=2)$ &$1.2\times10^{-4}$&
 $2.012\times10^{-5}$&$1.2\times10^{-4}$&$1.2\times10^{-4}$&$3.0\times10^{-5}$ \\
$\mdotstar(z=2)$  & 0.5  &  0.5  & 0.5   &   0.5  & 0.5 \\
$p_* (z=2)$       & 0.016   &  0.097   & 0.08    &   0.016   & 0.043 \\ 
$\gamma_m (z=2-0)$ & $-2.7$    &  $-3.9$   & $-3.2$    &   $-3.9$    & $-3.1$ \\
$\gamma_p (z=2-1)$  & 2.84   &  4.51   & 3.07   &   5.33   & 1.60  \\
$\gamma_p (z=1-0.5)$& 2.72    &  5.74   & 3.67    &  5.71     & 2.79  \\
$\gamma_p (z=0.5-0)$& 3.75    &  8.11   & 5.32    &  8.83    & 3.50  \\
\enddata 
\tablecomments{Defining parameters of the five illustrative models discussed
in \S\ref{sec:illus}.  All models assume a double power-law $n(M)$ with
$\alpha=-1.5$, $\beta=-3.4$, and a mass-independent, double power-law
$p(\mdot)$ with $a=-0.5$, $b=-3$, except that the ADAF model has
$a=-1.5$ and a factor of 16 boost to $p(\mdot)$ in the range
$10^{-4} \leq \mdot \leq 10^{-2}$.  Models start at $z=2$ with the
tabulated values of $M_*$ (in $10^9 M_\odot$), 
$n_* M_*$ (in comoving Mpc$^{-3}$ for $\Omega_m=1$, $h=0.5$), $\mdotstar$,
and $\pstar$.  Over the indicated redshift intervals, $\mdotstar$
evolves as $t^{\gamma_m}$ and $\pstar$ as $t^{\gamma_p}$.  The obscured
model has a 4:1 ratio of obscured to unobscured systems.
The merger model incorporates a factor $f_m=1.8$ growth by equal mass
mergers in each redshift interval $z=2-1$, $1-0.5$, and $0.5-0$.
Note that in all five models the amplitude
of $p(\mdot)$ drops with time at all $\mdot$ even though the $\gamma_p$
values are positive, since $\mdotstar$ decreases rapidly with time.
}
\end{deluxetable}

\begin{deluxetable}{lrrrrr} 
\tablewidth{0pt}
\tablecaption{Evolutionary Values
\label{tbl:evol}}
\startdata \hline
 & Short-$t_q$ & Long-$t_q$ & Obscured & Merger & ADAF \\ \hline
$M_*/10^9 M_\odot $ & & & & & \\
$z=2$ & 1 & 1 & 1 & 1 & 1.25 \\
$z=1$ & 1.14 & 1.95 & 1.73 & 2.06 & 4.23 \\
$z=0.5$ & 1.19 & 2.36 & 1.98 & 3.92 & 5.65 \\
$z=0$ & 1.24 & 3.28 & 2.25 & 7.92 & 7.40 \\
$\rhobh/10^5 M_\odot {\rm Mpc}^{-3}$ & & & & & \\
$z=2$   & 3.26 & 0.55 & 3.26 & 3.26 &  1.02 \\
$z=1$   & 3.72 & 1.07 & 5.63 & 3.74 &  3.44 \\
$z=0.5$ & 3.88 & 1.29 & 6.45 & 3.94 &  4.61 \\
$z=0$   & 4.02 & 1.79 & 7.34 & 4.42 &  6.03 \\
$\avmdot$& & & & & \\
$z=2$ &0.0054 & 0.032 & 0.027& 0.0054 & 0.071 \\
$z=1$ & 0.0014 & 0.0054 & 0.0043 & 0.0015 & 0.0084 \\
$z=0.5$&$4.4\times 10^{-4}$ & 0.0022 & 0.0014 & $6.1\times 10^{-4}$ & 0.0033 \\
$z=0$&$1.6\times 10^{-4}$ & 0.0027 & $7.1\times 10^{-4}$ & 0.0011 & 0.0015  \\
$\tacc/10^7 {\rm yr}$& & & & & \\
$z=2-1$& 0.59 & 3.01 & 2.46 &  0.62 & 5.45  \\
$z=2-0.5$& 0.79 & 3.86 & 3.08 & 0.86 & 6.79  \\
$z=2-0$& 0.95 & 5.35 & 3.66 &  1.38 & 8.00 \\
\enddata 
\end{deluxetable}

\begin{figure}
\epsscale{1.0}
\plotone{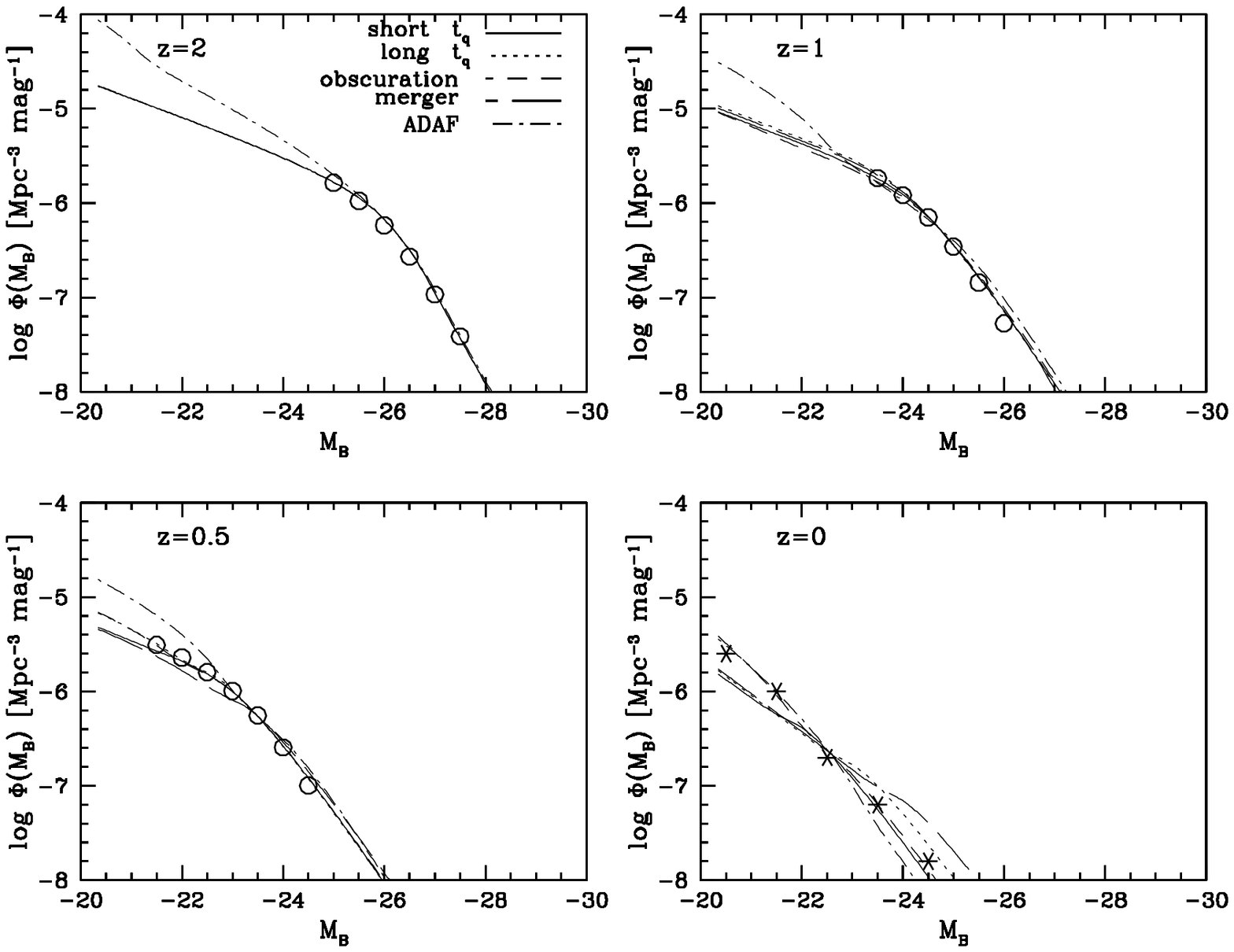}
\caption{Evolution of the $B$-band QLF for the five illustrative models 
discussed in \S\ref{sec:illus}.  In each panel, solid and dotted lines show 
results for two models dominated by thin-disk accretion, with short and long 
quasar lifetimes, respectively.  Short-dashed lines show a model with a large 
fraction of obscured quasars, long-dashed
lines a model in which mergers contribute substantially to the evolution 
of the black hole mass function, and dot-dashed lines a model with high 
probability of low $\dot{m}$ accretion, leading to significant black hole 
growth in an ADAF mode.  
Open circles in the $z=2$, 1, and 0.5 panels show the
\citet{boyle00} QLF fit over the observed range of luminosities at the
indicated redshift.  
Asterisks in the $z=0$ panel show the QLF estimate of \cite{wisotzki00}.
}
\label{fig:bbandevol}
\end{figure}

For our short quasar lifetime (\stq) model, we normalize the black hole
mass function at $z=2$ to $n_* M_* = 1.2\times 10^{-4}\dunits$.
Because of the high space density, the normalization of $p(\mdot)$
is relatively low, and the accretion weighted lifetime between
$z=2$ and $z=0$ is $\tacc=9.5\times 10^6$ yr, which implies
little growth of black hole masses over this redshift range
(see \S\ref{sec:lifetimes}).  For the \ltq\ model, we reduce the
black hole space density at $z=2$ by a factor of $\sim 6$,
to $n_*M_* = 2.012\times 10^{-5}\dunits$, 
thus continuing the growing phase model illustrated in the 
lower panels of Figure~\ref{fig:growth}.
The corresponding
value of $\tacc$ is $5.3\times 10^{7}$ yr, implying
about one $e$-fold of mass growth from $z=2$ to $z=0$
(compared to an order of magnitude growth from $z=5$ to $z=2$,
as shown in Figure~\ref{fig:growth}).
The factor of six difference in lifetimes is small compared to the
full range of quasar lifetimes discussed in the recent literature
(\citealt{martini03} and references therein), but the two models
straddle the boundary between significant low-$z$ accretion growth 
and minimal low-$z$ accretion growth.
Solid and dotted curves in Figure~\ref{fig:bbandevol} show the $B$-band
QLFs predicted by the short- and \ltq\ models, respectively.
By construction, they match the \cite{boyle00} data well at $z=2$, 1, 
and 0.5.
  
The $z=0$ data in Figure~\ref{fig:bbandevol} come from \cite{wisotzki00},
based on the Hamburg/ESO quasar survey.  They lie close to an extrapolation
of the \cite{boyle00} evolution model to $z=0$.
We have chosen $\gamma_p$ values so that each model passes through
these data at $M_B \approx -22$, but we have not
attempted to reproduce the shape, so the model predictions do not
nearly overlap as they do at higher redshift.  
The \stq\ model agrees well with the \cite{wisotzki00} data. 
The \ltq\ model
predicts a higher amplitude of the QLF at high luminosities,
mainly because greater black hole growth gives it a higher value 
of $M_*$ at $z=0$, and its luminosity function is steeper than the data.
The discrepancy 
could be reduced if we allowed $\mdotstar$ to
drop more rapidly between $z=0.5$ and 0, instead of
extrapolating the behavior that fits from $z=2$ to $z=0.5$.

For the obscured model, we essentially take the \stq\ model and 
multiply $p_*$ by five, assigning 20\% of the active systems at
each redshift the standard \cite{elvis94} SED and the remaining
80\% the obscured SED of Table~\ref{tbl:fnu}.
The obscured model's optical luminosity
function is shown by the dashed lines in Figure~\ref{fig:bbandevol}, and it 
also closely matches the observed evolution.  A steepening of the evolution 
of $p(\mdot)$ to lower accretion rates is required to balance the increased 
black hole growth from obscured accretion. Note that we could also have 
implemented the obscured scenario by increasing the $n(M)$ normalization
$n_*$ by a factor of five at $z=2$ and keeping $p_*$ the same, 
but then $\rhobh(z=0)$ would have been very high.  

For the merger model, we take the same initial parameters as the \stq\ 
model and use equation~(\ref{eqn:verysimplemerger}) to calculate
black hole growth, with a merging factor $f_m=1.8$ between each pair
of redshifts shown in Figure~\ref{fig:bbandevol} ($2\rightarrow 1$,
$1\rightarrow 0.5$, $0.5\rightarrow 0$).  This simple prescription for 
mergers is not fully self-consistent because 
the luminosity function necessarily implies some accretion growth as well.
At each redshift, therefore, we calculate the 
mean accretion rate adopted to produce the observed QLF and 
shift the black hole mass function by the corresponding amount.
With this additional accretion growth, $M_*$ increases by a factor of
two over each of the three redshift intervals,
and by a factor of eight over the full range $z=2-0$.
The merger model's optical luminosity 
function is shown by long dashed lines in Figure~\ref{fig:bbandevol}.  
Matching the observations requires steep evolution of $p(\mdot)$ to 
compensate for the large amount of black hole growth from mergers.  
The doubling of $M_*$ between $z=0.5$ and $z=0$ leaves the merger
model with a high amplitude tail of luminous systems at $z=0$.

The goal of our ADAF model, shown by the
dot-dashed lines in Figure~\ref{fig:bbandevol}, is 
to illustrate a case in which black holes
experience substantial growth through low efficiency 
accretion at low redshift.  This requires a high probability of having 
$\dot{m}<\dot{m}_{\rm crit}$, which is difficult to achieve while staying 
consistent with the observed QLF.  In particular, we 
are unable to find an acceptable fit to the optical QLF 
by simply adjusting the parameters $\mdotstar$ and $a$ of
our usual double power-law $p(\mdot)$.
After some experimentation, we settled on a model 
with the combination of a steeper low-$\mdot$ slope, $a=-1.5$, and 
a boost to the probability of accretion rates 
below $\dot{m}_{\rm crit}$ by a factor of sixteen.  
We slightly increased $M_*$ to $1.25\times 10^9 M_\odot$ to
improve the match to the QLF break given our changed $p(\mdot)$,
and we reduced $n_* M_*$ to $3\times 10^{-5}\dunits$ so that
there would be more overall growth of $n(M)$.
With these choices, $\rhobh$ grows by a factor of about six from $z=2$ 
to $z=0$, and roughly 
two-thirds of this growth comes from objects in an ADAF mode.  
The optical QLF is still significantly different from that of the other 
models, but mostly at luminosities below the observed range.
At $z=2$, 
the increased number density of low luminosity objects is mainly due to the 
steep slope of $p(\mdot)$, which produces many faint thin-disk systems,
but at lower redshifts the ADAF mode is 
directly responsible for this excess of faint systems.  
The fact that we had to adopt such an artificial $p(\mdot)$ to obtain
a model that is even approximately consistent with the optical QLF
already suggests that low-$z$ ADAF growth of black hole masses is not
important in the real universe, but it is interesting to explore
the predictions of such a model nonetheless.

\subsection{Black hole mass functions}
\label{sec:nofmz}

We expect the black hole mass function $n(M,z)$ to be a good discriminant
among our models because they involve different amounts of black hole
growth, and in some cases start from different $n(M)$ at $z=2$.
The best prospects for measurements of $n(M)$ are at $z=0$, using
the observed distribution of bulge luminosities or velocity dispersions
and the observed correlation of dynamical black hole masses with these
properties.  A number of authors have estimated the black hole mass 
function in this way (e.g., \citealt{salucci99,merritt01,yu02,aller02}),
and improving determinations of
the form and scatter of the $\Mbh-\sigma$ relation \citep{gebhardt00,merritt00}
and of the distribution of bulge dispersions \citep{sheth03}
should yield more accurate estimates in the near future.
Recent estimates of the average black hole mass density
at $z=0$ are $\rhobh=2-3 \times 10^5 (h/0.7)^2$ M$_{\odot}$ Mpc$^{-3}$ 
\citep{yu02,aller02}.  High redshift estimates of
$n(M)$ will be difficult, since achievable angular resolution is not
sufficient to measure dynamical masses of quiescent black holes.
The distribution of masses of {\it active}
systems at a given luminosity can be measured using reverberation
mapping or emission line widths, and
we discuss its diagnostic power in \S\ref{sec:activemass} below.
It may be possible to estimate the underlying $n(M)$
by establishing a correlation between $\Mbh$ and host 
galaxy properties using active systems, then proceeding as
at low redshift, but the observational uncertainties are likely
to remain considerable.  

\begin{figure}
\epsscale{1.0}
\plotone{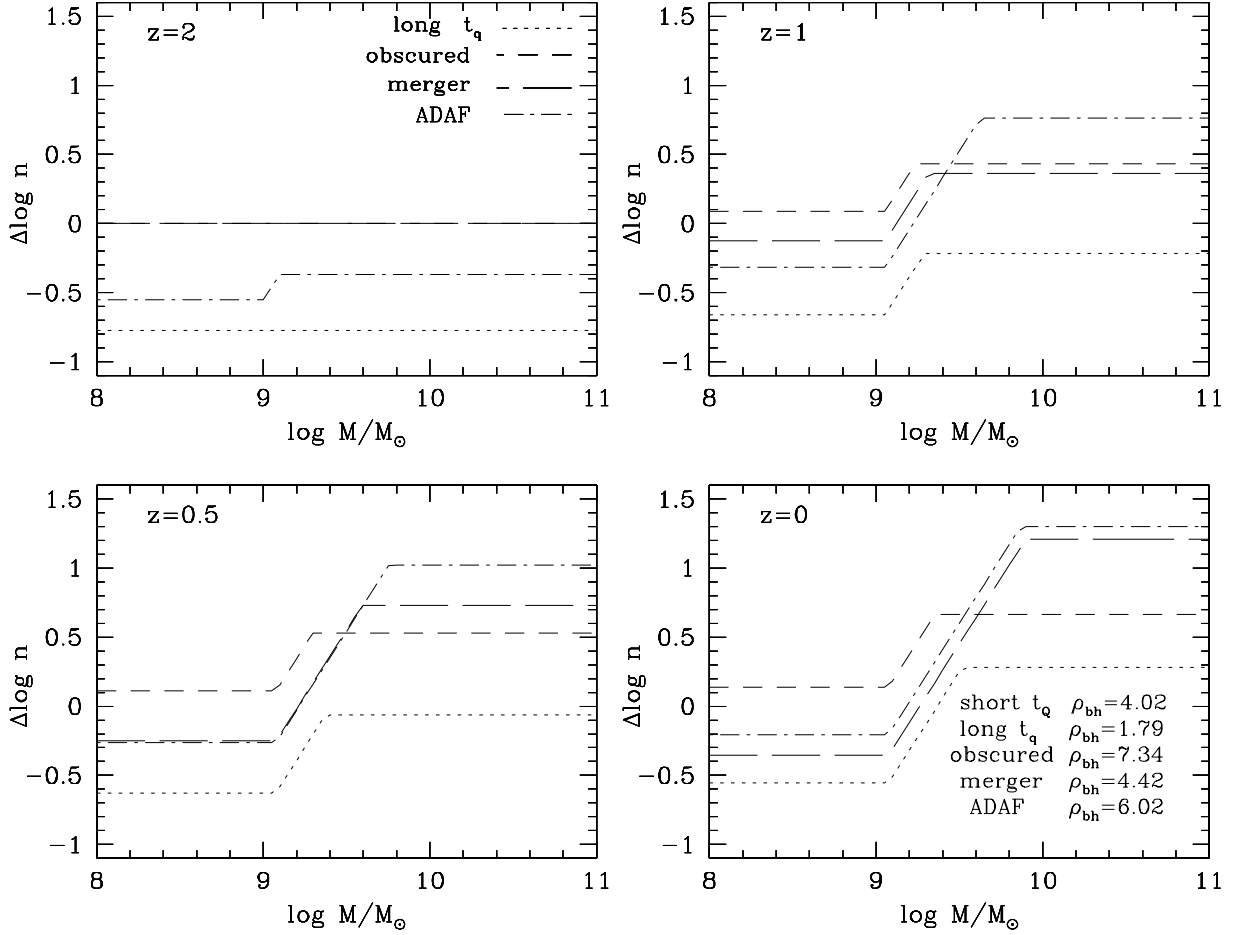}
\caption{ Evolution of the black hole mass function for the five models 
discussed in \S\ref{sec:illus}.  Lines in each panel show 
$\Delta \log n \equiv \log_{10} [n(M)/n_{s}(M)]$,
where $n_{s}(M)$ is the mass function of the \stq\ model at the 
indicated redshift and $n(M)$ is the mass function of the \ltq, obscured,
merger, or ADAF model (dotted, short-dashed, long-dashed, and dot-dashed, 
respectively).  The $z=0$ panel also lists the final value of the black hole 
mass density for each model in units of $10^5 M_{\odot}$ Mpc$^{-3}$.
All models have a double power-law mass function with slopes 
$\alpha=-1.5$ and $\beta=-3.4$ at all redshifts, and values of 
$M_*(z)$ are listed in Table~\ref{tbl:evol}.}
\label{fig:rationofmevol}
\end{figure}

We show the model black hole mass functions in Figure~\ref{fig:rationofmevol},
but to make differences more visible we divide through by the 
prediction of the \stq\ model and plot the logarithm of the
ratio, $\Delta \log n(M) = \log n(M) - [\log n(M)]_{{\rm short-}t_q}$.
The \stq\ model
has little growth of the black hole mass function, with a change from
$M_*=1 \times 10^9 M_{\odot}$ at $z=2$ to $M_*=1.17 \times 10^9 M_{\odot}$ at 
$z=0$.  The \ltq\ model at $z=2$ has $n(M)$ a factor of $\sim 6$ 
lower at all masses because it has the same value of 
$M_*$ and a lower value of $n_*$.  There is more growth in the \ltq\ 
model due to a higher $p(\mdot)$, and the values of $n(M)$ begin to catch up 
to the \stq\ case.  
Since the growth has the effect of shifting the break in the 
double power-law $n(M)$ to higher masses, the change in $n(M)$
is larger at the steep, high mass end and smaller at low masses,
producing the characteristic kinked shape of the lines in 
Figure~\ref{fig:rationofmevol}.  By $z=0$ the \ltq\ model has overtaken 
the \stq\ model at high masses, but it remains below at low masses.

The final black hole mass densities for the \stq\ and \ltq\ models are
$\rhobh=\int^{\infty}_{0} M n(M) dM = 4.02 \times 10^5 M_\odot {\rm Mpc}^{-3}$
and $1.80 \times 10^5 M_\odot {\rm Mpc}^{-3}$, respectively.
The mass density {\it added} between $z=2$ and $z=0.5$ is
nearly the same in the two models, $\Delta\rhobh \approx 0.7\times 10^5
M_\odot {\rm Mpc}^{-3}$.  This agreement is expected from the
\cite{soltan82} argument, which implies that 
$\Delta\rhobh = \int_{z_1}^{z_2} U(t)/(\epsilon c^2) dt$,
where $U(t)$ is the bolometric emissivity of the quasar population at time $t$.
Since mass growth in both models is dominated by thin-disk systems
with the \cite{elvis94} SED and $\epsd=0.1$, and both models reproduce
the observed optical QLF, they necessarily have similar emissivity
histories and mean efficiencies and therefore similar $\Delta\rhobh$.
The \ltq\ model adds significantly more mass between $z=0.5$ and $z=0$,
partly because it has a higher optical QLF at low redshift
(Fig.~\ref{fig:bbandevol}), and partly because its $\mdotstar$ falls
below $\mdotcrit$, increasing the amount of low efficiency accretion.
The \stq\ model has a final $\rhobh$ that is high in comparison with
recent estimates, though arguably in the range of their uncertainties.
The \ltq\ model's $\rhobh$ agrees well with these estimates, coming
in slightly on the low side.

The obscured model starts at $z=2$ with the same $n(M)$ as the
\stq\ model, but $n(M)$ grows quickly because of the 
large amount of optically invisible accretion.
The break mass $M_*$ and mass density $\rhobh$ grow by a factor of 2.25
between $z=2$ and $z=0$, with a final $\rhobh$
about a factor of two larger than the \stq\ case
and outside the range of recent estimates.
The amount of mass added, 
$\Delta\rhobh \approx 4\times 10^5 M_\odot {\rm Mpc}^{-3}$,
is about five times the amount for the \stq\ model, which is
as expected because they have similar optical QLFs while the obscured
model has four optically invisible systems for each unobscured system.
Since the observed optical QLF provides 
a fairly natural fit to the estimate $\rhobh$  on its own
(e.g., \citealt{yu02}), it is difficult to add a large amount of hidden 
accretion without overrunning these estimates.
Higher efficiency accretion, perhaps from spinning black holes, is one option 
\citep{elvis02}.  However, part of the solution is probably that 
our assumption of an 80\% obscured fraction at all redshifts 
and luminosities, taken from \cite{comastri95} and 
\cite{fabian99}, is too extreme.
Recent studies show that faint X-ray sources are generally at 
lower redshifts than synthesis models predict, in which case a smaller
fraction of obscured sources are needed to produce the hard X-ray background
\citep{barger02,ueda03}.

The merger model starts with the same black hole mass function as the \stq\
case, but it evolves primarily by merging two lower mass black holes 
that create one higher mass black hole.  
By $z=0$, high mass black holes are 
more numerous than in the \stq\ case by a factor of ten, and low mass black 
holes are less numerous by a factor of three.  
The {\it accretion} growth in this model is still dominated by 
thin-disk systems, and since mergers do not add mass to the
black hole population, the growth of $\rhobh$ tracks that of the \stq\ 
and \ltq\ models down to $z=0.5$.  Like the \ltq\ model, the merger
model adds more mass at $z<0.5$ because of its high QLF and low $\mdotstar$.

The ADAF model starts with  
$M_*=1.25 \times 10^9$ M$_{\odot}$ and a normalization of $n(M)$
a factor of four lower than in the \stq\ case.
The high amount of accretion at low $\mdot$ values, in both the
thin-disk and ADAF regimes, results in more
black hole growth than in any of the other pure accretion 
models.  By $z=1$, $n(M)$ is similar to that of the \stq\ case at low 
masses, and it is a factor of ten larger at high masses.  
(Recall that all masses grow by the same factor even in the ADAF
model, so this difference just reflects the slope of the mass
function in the two regimes.)
The final black hole mass density is 50\% larger than that of the \stq\ 
model, and it is difficult to make parameter changes 
that significantly reduce this value because of the emissivity
argument; the optical QLF of this model traces the observations at
observed luminosities, but it has a high amplitude tail at low
luminosity, and the mean efficiency is low because of the high
fraction of ADAF accretion.

\begin{figure}
\epsscale{1.0}
\plotone{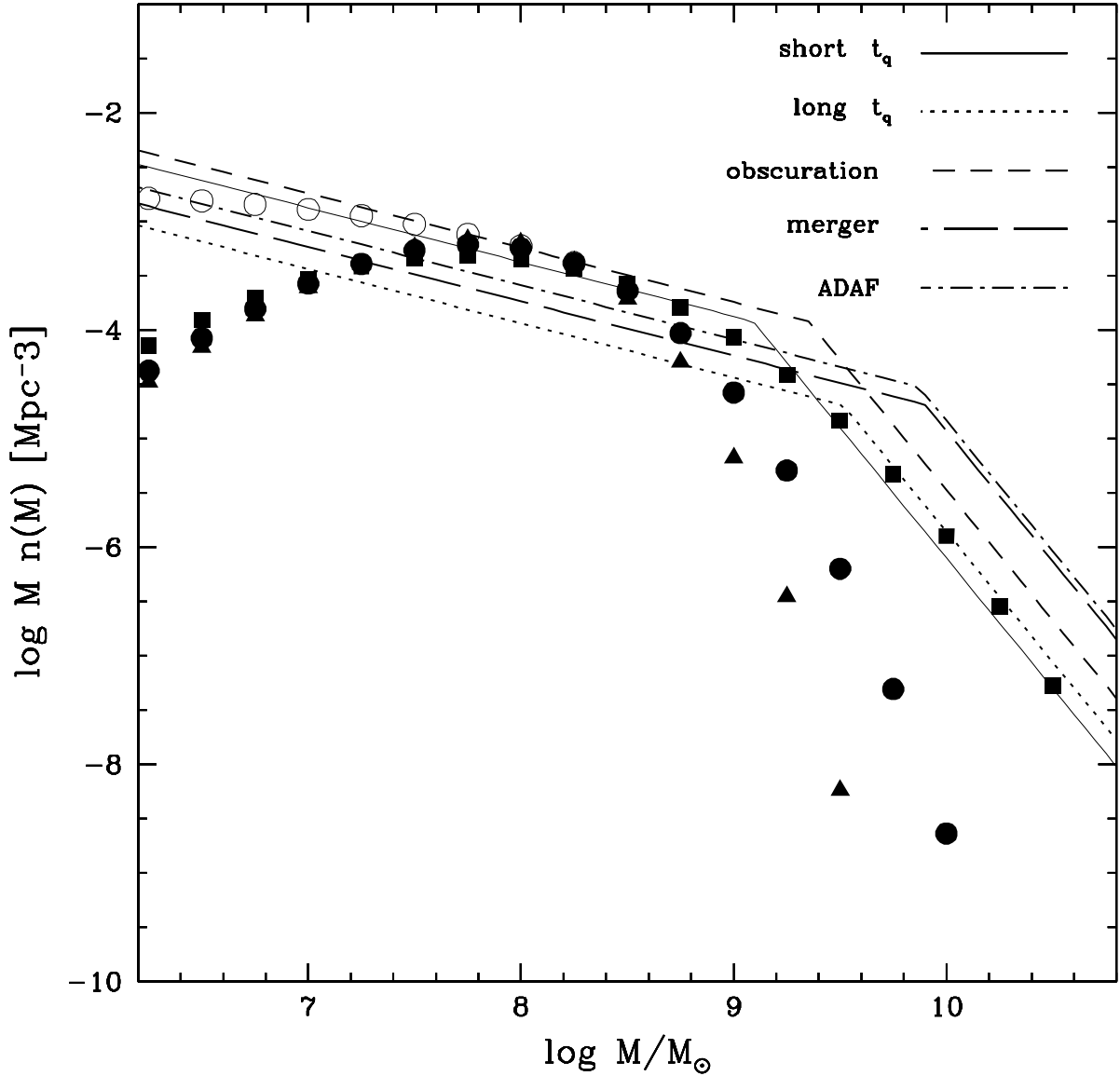}
\caption{ 
The black hole mass function at $z=0$.  
Lines represent the five model mass mass functions.
Solid points show the mass function derived by combining
the 
\citet{sheth03} velocity dispersion distribution of early-type galaxies 
with Tremaine et al.'s (\citeyear{tremaine02}) estimate of the $M-\sigma$
relation between black hole mass and bulge velocity dispersion.
Triangles, circles, and squares show results assuming no intrinsic
scatter in the $M-\sigma$ relation, 0.25-dex scatter, and 0.5-dex scatter,
respectively.  Open circles show the effect of adding
Aller \& Richstone's (\citeyear{aller02}) estimated contribution
from spiral galaxies to the 0.25-dex scatter (filled circle) estimate
for early-type galaxies.
}
\label{fig:comparenofm}
\end{figure}

Figure~\ref{fig:comparenofm} compares the $z=0$ black hole mass functions
of the five models to an estimate derived by combining the \cite{sheth03}
estimate of the distribution of early-type galaxy velocity dispersions
with the $M-\sigma$ relation found by \cite{tremaine02},
$\log M_{\rm bh}/M_\odot = 8.13 + 4.02 \log (\sigma/200\;\kms)$.
Filled triangles show the case where there is no intrinsic scatter
in the $M-\sigma$ relation, while filled circles and squares show
results assuming a log-normal $p(M|\sigma)$ distribution with intrinsic
scatter of 0.25 dex and 0.5 dex, respectively.
Above $10^9M_\odot$, the derived mass function is quite sensitive
to the assumed intrinsic scatter. \cite{tremaine02} argue that this
scatter is no larger than $\sim 0.25-0.3$ dex, but the data
are concentrated in the range $10^7-10^{8.5}M_\odot$, and the
scatter could increase or decrease with mass.
Above $10^{9.5}M_\odot$, the derived mass function also depends on 
extrapolating the mean $M-\sigma$ relation beyond the range of current 
observations.  Since \cite{sheth03} consider only early-type galaxies,
we show with open circles the effect of adding Aller \& Richstone's
(\citeyear{aller02}) estimate of the spiral galaxy contribution, which
becomes important below $\sim 10^{7.8} M_\odot$.  If we included their
S0 contribution as well (which might double count galaxies already
in the Sheth et al.\ sample), then the mass function would be higher
by 0.2 dex in this regime.

If the extrapolation of the $M-\sigma$ relation is correct and the scatter
is indeed $\la 0.3$ dex, then even the \stq\ model overpredicts
$n(M)$ above $10^9 M_\odot$, and our other models fare worse because of
their higher values of $M_*(z=0)$.  Bringing the models in line with this
estimate of $n(M)$ would require either reducing our initial $M_*(z=2)$
by 0.5-1 dex or changing our double power-law form of $n(M)$.
We selected $M_*(z=2)=10^9 M_\odot$ because the Eddington luminosity
$lM_*$ is then close to Boyle et al.'s (\citeyear{boyle00}) break
luminosity (more precisely, $lM_* = 1.7 \Lbrk$), 
making it straightforward to fit
the observed $\Phi(L)$.  However, given the interplay between $n(M)$
and $p(\mdot)$, there is at least some room to reduce $M_*$ and
continue to match the observed QLF without requiring super-Eddington
luminosities.  If we instead adopted an exponential
high-$M$ cutoff (or a steeper high-$M$ power-law slope), then we would
need a mass-dependent $p(\mdot)$ to reproduce the \cite{boyle00} data
at $z=2$, with more massive black holes having a higher probability
of being active.  We will explore these implications, and the tradeoff
with uncertainties in the data, in future work, where we run models
backwards from the $z=0$ mass function instead of forwards from the $z=2$ QLF.

\subsection{X-ray and FIR luminosity functions}
\label{sec:multilambda}

In three of our models (\stq, \ltq, and merger), the bolometric 
emission of the quasar population is dominated by systems with a
thin-disk SED.  We therefore expect them to make similar predictions
for the QLF in all wavebands (since they match in the $B$-band by
construction).  However, the obscured and ADAF models have large
populations of systems with different SED shapes, and they may be
distinguished by their X-ray or FIR luminosity functions.
Observationally, the X-ray luminosity function is not as well
characterized as the optical luminosity function, 
especially at high redshifts, but large surveys
following up sources from {\it ROSAT}, {\it ASCA}, {\it Chandra}, and
{\it XMM-Newton} are transforming the situation and yielding
much better constraints on the evolution of the X-ray QLF 
(e.g., \citealt{miyaji01,cowie03,fiore03,hasinger03,steffen03,ueda03}).
In the mid/far-IR, {\it SCUBA} detects the brightest sources
at 850\micron\ \citep{priddey03}, but the revolutionary instrument should
be {\it SIRTF}, with a much greater combination of wavelength range
and sensitivity than previously available.

\begin{figure}
\epsscale{1.0}
\plotone{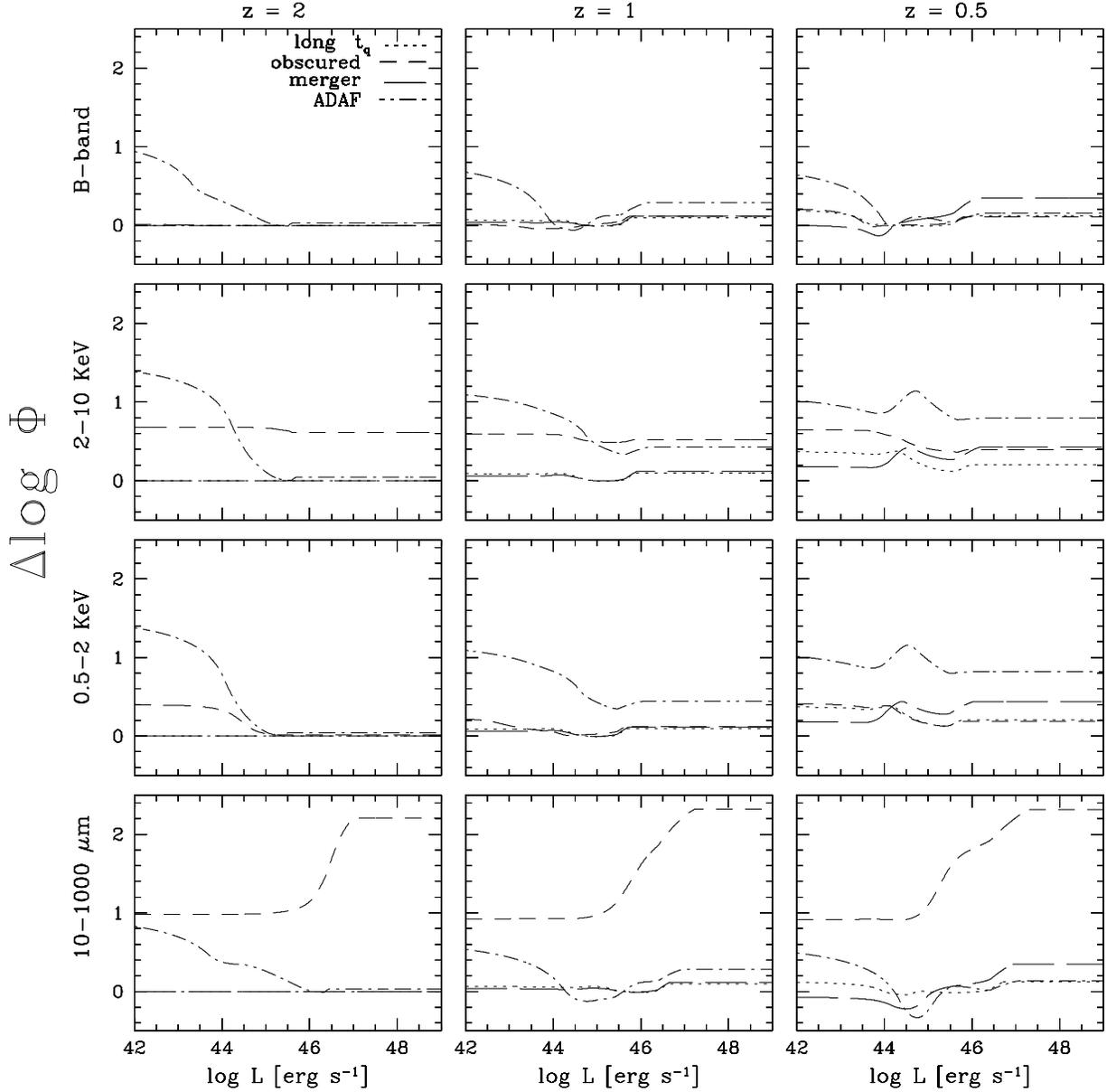}
\caption{Ratios of the luminosity functions of the \ltq, obscured,
merger, and ADAF models to those of the \stq\ model at $z=2, 1$, and $0.5$ 
(left, middle, right).  Lines in each panel show 
$\Delta \log \Phi \equiv \log_{10} [\Phi(L)/\Phi_{s}(L)]$
for $B$-band, 2-10 KeV, 0.5-2 KeV, and 10-1000 $\mu$m (top to bottom).  The 
X-ray bands are observed-frame with the redshift effects on $F_{\nu}$ shown in 
Table \ref{tbl:fnu} included.
}
\label{fig:ratioevol}
\end{figure}

We calculate X-ray and FIR luminosity functions of our five models
using the $F_\nu$ values in Table~\ref{tbl:fnu} for systems with
the various accretion modes.  For X-ray QLFs we use {\it observed-frame}
bandpasses of $0.5-2$ keV and $2-10$ keV at redshifts $z=2$, 1, and 0.5.
Results are shown in Figure~\ref{fig:ratioevol}.
To enhance the visibility of model differences, we again plot the log of the 
ratio of each model's predictions to those of the \stq\ model.  
The \stq\ model's predictions at $z=2$ are nearly identical to those shown
by the solid lines in Figure~\ref{fig:obsc.4panel}.  As expected,
results for the \ltq\ and merger models are very similar to \stq\ in all
bands because of the dominance of thin-disk SEDs, and they could probably
be brought closer still with slight adjustments of model parameters.
The low-$z$ X-ray luminosity functions of the \ltq\ and merger 
models are slightly enhanced
at low luminosities relative to $B$-band because their low $\mdotstar$
values (an indirect consequence of higher $M_*$) lead to more ADAF
accretion.

The obscured model, which initially has five times as many accreting
systems as the \stq\ model, shows nearly this full factor of five
enhancement in the $2-10$ keV band at $z=2$, where obscuration is
almost negligible.  This enhancement shrinks steadily towards lower $z$,
especially at higher luminosities, as the observed-frame $2-10$ keV band
becomes more affected by obscuration (see Table~\ref{tbl:fnu}).
The $0.5-2$ keV band shows a significant (0.4-dex) enhancement at
low luminosities at $z=2$, comprised of systems that have high bolometric
luminosity and are thus able to shine detectably at this wavelength
despite obscuration.  However, at lower redshift the $0.5-2$ keV band
is almost completely extinguished, and the obscured QLF is no different
from that of the \stq\ model.  The most dramatic feature of the 
obscured model is the booming FIR luminosity function, enhanced by
$1-2$ orders of magnitude at all redshifts because the numerous
obscured systems re-radiate all of their absorbed UV and soft X-ray
luminosity in the FIR.  This distinctive prediction of models with
a large obscured population should be easily testable with {\it SIRTF}.
The prediction holds regardless of whether obscured and unobscured
systems represent two separate populations or different orientations
of the same population, provided that the absorbed energy is indeed
re-radiated.

As previously noted, the ADAF model has some substantial differences
from the \stq\ model even in $B$-band.  At $z=2$, the steady rise
in low luminosity systems is a consequence of the steeper slope of
$p(\mdot)$, while the bump at $\log L \leq 10^{43.5} \erg\,\sec^{-1}$
reflects the boosted number of objects with $\mdot < \mdotcrit$
and thus consists of objects accreting in an ADAF mode.
At low redshifts, this ADAF bump moves to slightly higher luminosities.
The $B$-band differences reappear at other wavelengths, but there are
further differences in the X-ray bands that reflect the larger fraction
of the ADAF SED that emerges in these bands.  At high redshift,
the low luminosity boost to the QLF is larger in X-ray than in optical,
exceeding an order of magnitude.  At low redshift, the decreasing
value of $\mdotstar$ leads to a still higher probability of ADAF
accretion, and high mass black holes in ADAF mode boost the X-ray QLF
even at high luminosities.  Indeed, we can infer from 
Figure~\ref{fig:ratioevol} another prediction of our ADAF model:
at $z \la 1$, most X-ray selected AGN should be ADAF systems,
at every luminosity.  This does not appear to be the case in the
real universe, providing a further observational argument against
the importance of low efficiency accretion as a black hole growth
mechanism at low redshift.

\begin{figure}
\epsscale{1.0}
\plotone{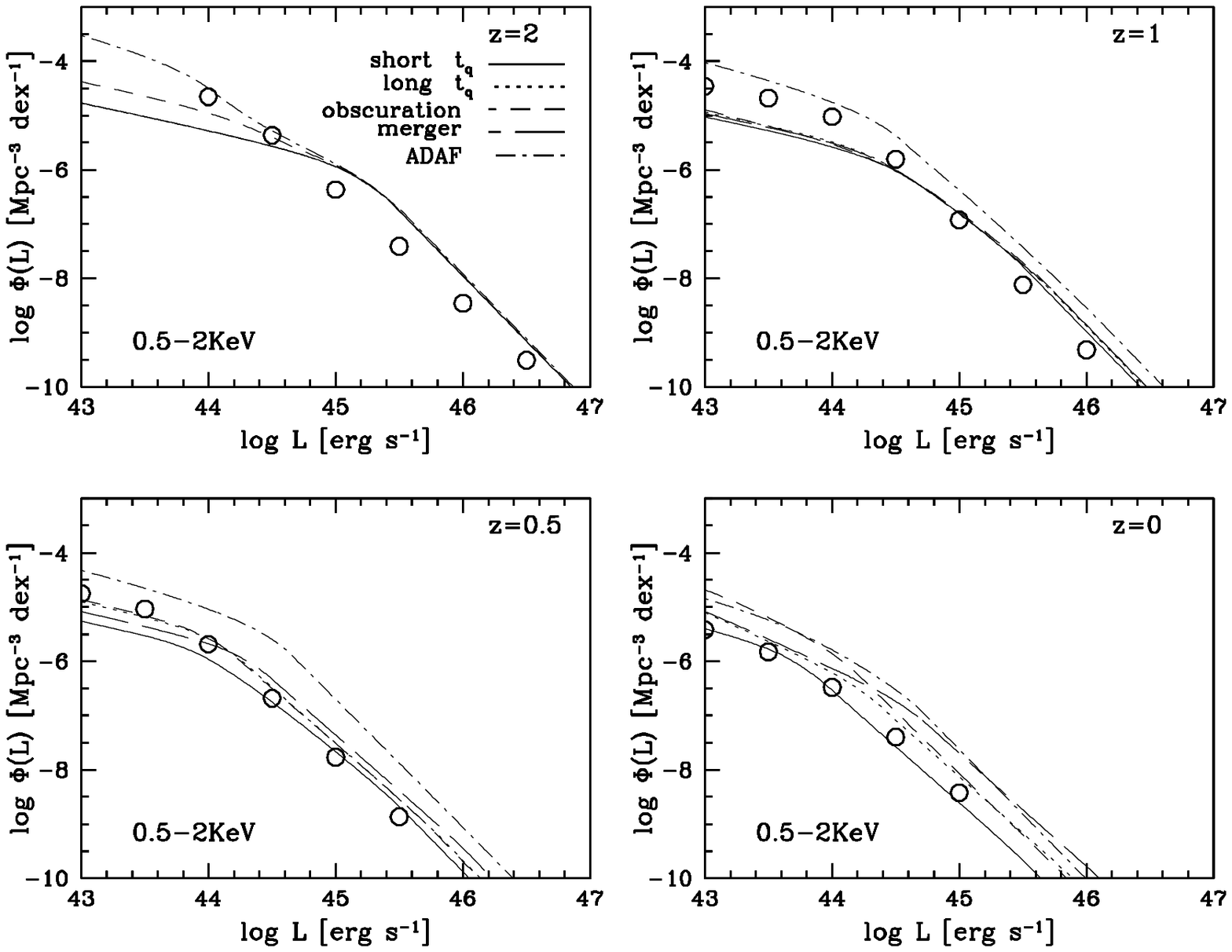}
\caption{
The observed frame soft X-ray luminosity function at $z=$2,1,0.5, and 0.
Lines show the predictions of our five models, as indicated.
Open circles show the evolutionary model fits to the ROSAT 0.5-2 keV
QLF from \citet{miyaji01}, for $\Omega_m=1$, $h=0.5$.
Points are plotted over the range of luminosities spanned by the
data at each redshift.
}
\label{fig:comparesxdata}
\end{figure}

\begin{figure}
\epsscale{1.0}
\plotone{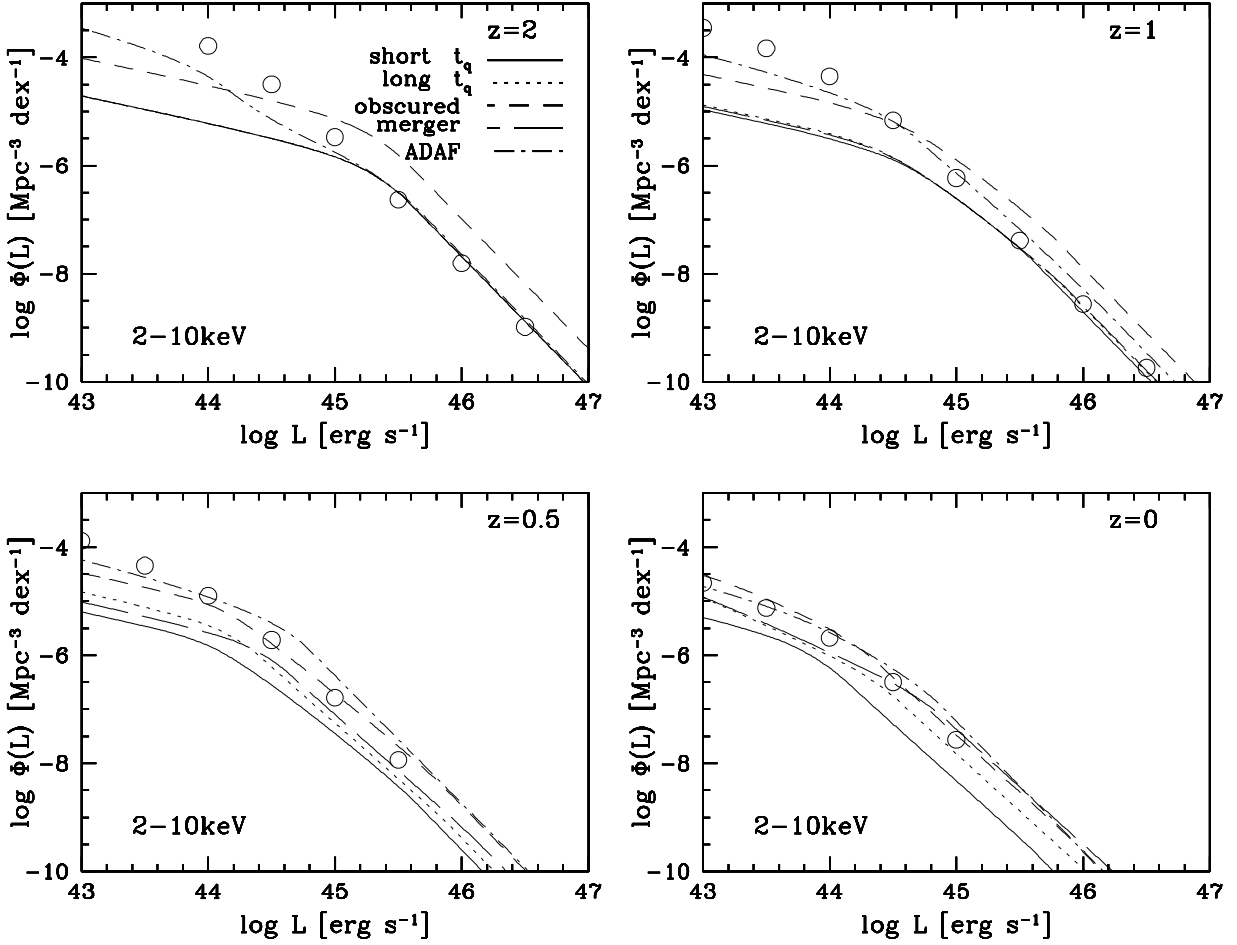}
\caption{
The intrinsic hard X-ray luminosity function at $z=$2, 1, 0.5, and 0.  
Lines show the predictions of our five models, as indicated.
Open circles show the evolutionary model fits to the 2-10 keV QLF from
\citet{ueda03}, based on {\it HEAO1}, {\it ASCA}, and {\it Chandra}
surveys.  \citet{ueda03} use spectral modeling to estimate the
rest-frame 2-10 keV luminosity corrected for obscuration.
We therefore use the 2-10 keV $F_\nu$ values from Table~\ref{tbl:fnu}
at each redshift, and we use the thin-disk values of $F_\nu$ for obscured
systems in the obscured model, since these represent the intrinsic
luminosities.
}
\label{fig:comparehxdata}
\end{figure}

Figures~\ref{fig:comparesxdata} and~\ref{fig:comparehxdata} compare
the model predictions to estimates of the soft and hard X-ray luminosity
functions from \citet{miyaji01} and \cite{ueda03}, respectively.
As with the optical QLF, we plot the authors' evolutionary model
fits over approximately the range covered by the observational data
at each redshift.  Miyaji et al.'s (\citeyear{miyaji01}) luminosity
function, is based on observed-frame, 0.5-2 keV luminosities
from the {\it ROSAT} All Sky Survey, with no correction for X-ray
obscuration.  \cite{ueda03}, on the other hand, use spectral shape
information to estimate the {\it intrinsic} (i.e., corrected for
obscuration), {\it rest-frame} 2-10 keV luminosity function, from a 
combination of {\it HEAO1}, {\it ASCA}, and {\it Chandra} surveys.
We compute the corresponding quantities from our models in each case.

At $z=2$, all models fit the high luminosity end of the 2-10 keV QLF,
except for the obscured model, which overpredicts by a factor of five.
However, all models overpredict the 0.5-2 keV QLF by a factor $\sim 3$.
At $z=1$ and $z=0.5$, models come into better agreement with the 
0.5-2 keV QLF (except for the ADAF model, which remains high), but
the thin-disk dominated models (\stq, \ltq, merger) fall below at 2-10 keV.
By $z=0$, the $M_*$ growth in the
\ltq\ and merger models has brought them back into better agreement
at 2-10 keV, but they are high at 0.5-2 keV, while the \stq\ model
is about right at 0.5-2 keV and well below at 2-10 keV.
The obscured model is in rough agreement with both X-ray QLFs at $z \leq 1$.

In brief, the situation is confusing, and there is no single obvious
change that would bring any of the models into agreement with both
X-ray luminosity functions and the $B$-band luminosity 
function (Fig.~\ref{fig:bbandevol}) at all redshifts.
A more sophisticated obscuration model, with soft X-ray and optical
obscuration becoming more important at low redshifts and low luminosities,
would certainly help, and it might explain why the faint end of the 
2-10 keV QLF rises above the $B$-band QLF at $z=2$.
However, the overprediction of the soft X-ray QLF at $z=2$ by models
that match the $B$-band and 2-10 keV QLF seems difficult to understand.
It is worth noting that Ueda et al.'s models, which incorporate
luminosity-dependent obscuration, also tend to overpredict the
soft X-ray QLF (see their figure 15), and that Miyaji et al.'s
redshift bins become large at high redshift (e.g., $z=1.6-2.3$ and $2.3-4.6$),
which may make the model interpolation to a given redshift less accurate.

\subsection{Masses and accretion rates of active black holes}
\label{sec:activemass}

While the underlying black hole mass function $n(M)$ may be difficult
to determine at $z>0$, the distribution of {\it active} black hole
masses is more accessible.  As discussed in \S\ref{sec:massdist},
the mass distribution of active systems depends on both $n(M)$ and $p(\mdot)$,
and its variation with luminosity can be a valuable discriminant of models.
Locally, the masses of active systems can be measured by combining
emission line widths with sizes of the emitting regions estimated
by reverberation mapping \citep{wandel99}.  This approach can in principle
be extended to high redshift, but fainter targets and longer variability
timescales make it difficult.  A more broadly applicable method is
to combine line widths (e.g., $H\beta$, or \ion{C}{4} at higher redshift)
with the average size-luminosity relation inferred from reverberation
mapping of local objects or from photoionization modeling
(e.g., \citealt{laor98,gebhardt00.2,mclure02,corbett03,vestergaard03}).
In the last few years, these methods have yielded 
black hole mass estimates over an increasing
range of redshift and luminosity (e.g., \citealt{woo02}).
Estimates of mass accretion rates are typically 
made by combining mass estimates with luminosities for an assumed efficiency,
so they are not independent of the mass estimates themselves.
However, quasar SEDs may also provide at least rough diagnostics
of accretion rates, in the broad categorization of thin-disk vs. ADAF
systems, for example, and perhaps in the finer distinction between
near-Eddington systems and significantly sub-Eddington systems 
\citep{kuraszkiewicz00,czerny03}.

\begin{figure}
\epsscale{1.0}
\plotone{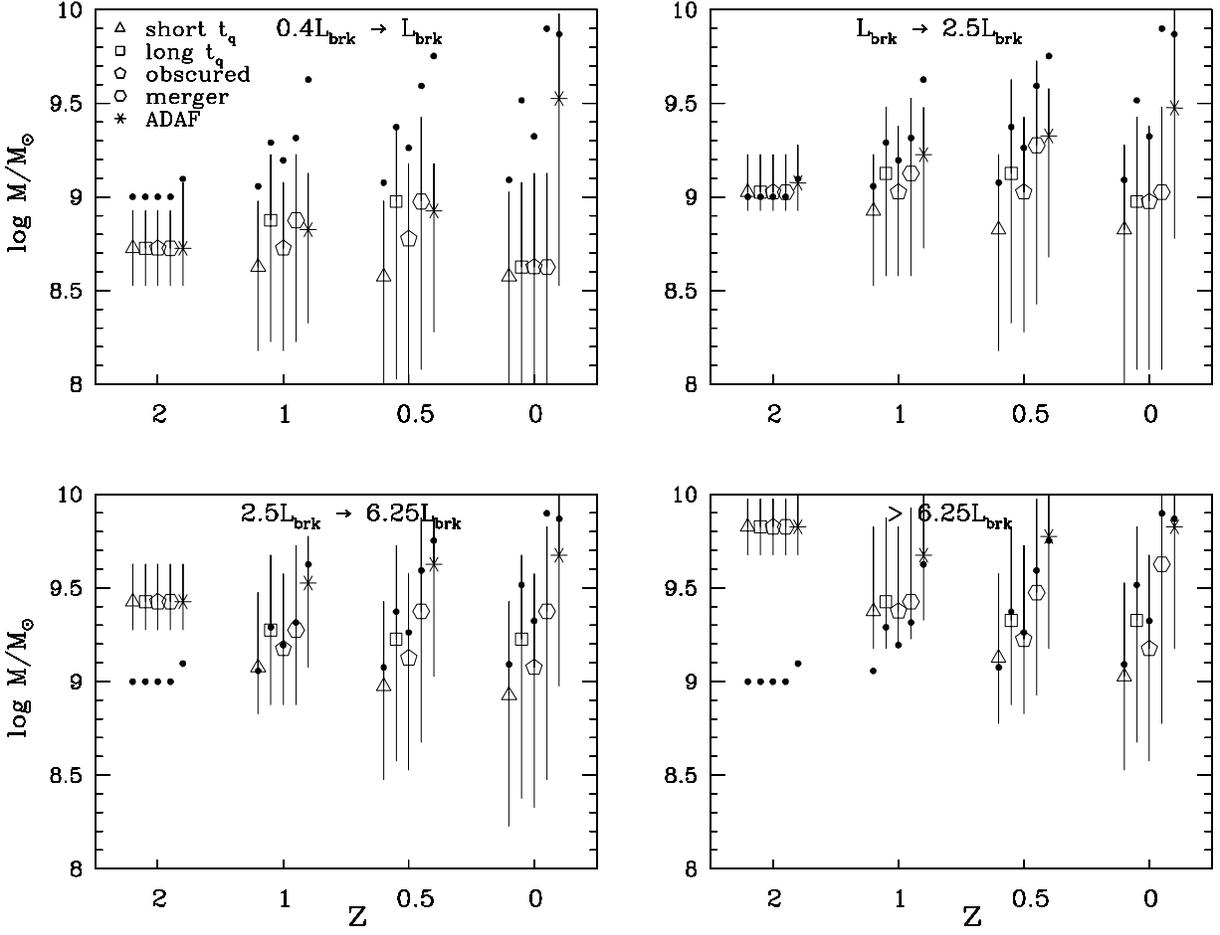}
\caption{Evolution of the mass distribution of active black holes as a 
function of luminosity for the five illustrative models discussed in 
\S\ref{sec:illus}.  Each panel corresponds to the range in $B$-band luminosity 
shown in the top center relative to the QLF break luminosity $\Lbrk(z)$.
The small black dots show the values of $M_*(z)$ for each model, and 
the open symbols
show the median mass of black holes contributing to the given luminosity
range for the \stq, \ltq, obscured, merger, and ADAF models 
(triangle, square, pentagon, circle, and
star).  The vertical bars show the 10\%-90\% range of the distribution.
The horizontal offset of the points from $z=2,1,0.5,$ and 0 is artificial and 
is done to distinguish the models from each other.
}
\label{fig:massbh}
\end{figure}

Figure~\ref{fig:massbh} illustrates the range of black hole masses 
that contribute to different ranges of $B$-band luminosity 
at $z=2$, 1, 0.5, and 0.  For each redshift and model, a symbol marks
the median mass $\Mmed$ of black holes that are active in this luminosity
range, and a vertical bar shows the 10\%--90\% range of masses
at this luminosity.  Small black dots show the model's value of $M_*$,
the mass of the break in the black hole mass function.
Different panels represent different luminosity ranges, which
we express in terms of $\Lbrk$, the break parameter in the observed
$B$-band QLF.  Bear in mind that the physical luminosity 
associated with $\Lbrk$ decreases towards low redshift, and that $\Lbrk$ drops 
increasingly below $lM_*$.  Furthermore, the $z=0$ results
here and in subsequent plots 
should be taken with a grain of salt because 
we do not require our models to match the shape of the QLF at $z=0$,
only the normalization at $\Lbrk$.

Figure~\ref{fig:massbh} is interesting both for the features that are
common to all of the models and for the features that distinguish them.
The key common feature is a change in the relation between mass and
luminosity from high redshift to low redshift.
At $z=2$, in all models,
the sequence of luminosity is also a sequence of black hole mass: 
the median active mass rises from 
$\Mmed \approx 5.6 \times 10^8 M_{\odot}$ in the lowest luminosity range to 
$\Mmed =6.8 \times 10^9 M_{\odot}$ 
in the 
highest luminosity range, roughly the same factor by which the 
luminosity itself rises.   Comparing the symbols to the black dots
shows that 
low luminosity quasars arise from sub-$M_*$ black holes and high
luminosity quasars from super-$M_*$ black holes.  
The 10\%--90\% range of masses in a given luminosity range is only
about 0.4-dex, similar to the width of the luminosity bin itself.
At low redshift, on the other hand, the trend of median mass with
luminosity is much weaker, and even $L>6.25\Lbrk$ systems
have median masses lower than $M_*$.  The distribution of masses
for a given luminosity range is much broader than at high 
redshift, typically close to an order of magnitude.
As we show more directly in Figure~\ref{fig:mdot} below, 
the luminosity function at low redshift represents
largely a sequence of accretion rate rather than black hole mass.

The differences between models largely trace the differences
in the growth of $M_*$.  At $z=2$, $\Mmed$ is similar at a given
luminosity for all models.  At lower redshifts, the \stq\ model,
which has the least $M_*$ growth, always has the lowest $\Mmed$
at a given luminosity, followed by the obscured model, which
has the next lowest growth of $M_*$.  The merger and ADAF models
have the most $M_*$ growth and the highest $\Mmed$ values at low
redshift.  The ADAF model is generally highest, even though the
merger model overtakes it in $M_*$ by $z=0$, because the high
probability of low $\mdot$ favors high mass black holes at a
given luminosity.  The 10\%--90\% spread at a given redshift
is generally similar among the models.  

The trends and model differences that we have shown here for 
$B$-band luminosity generally hold for luminosities defined in
other bands as well.  The most significant change is that the
median black hole mass at fixed (in $\erg\,\sec^{-1}$) FIR luminosity is
much smaller in the obscured model than in all other models because
a large fraction of the obscured SED emerges in the FIR band,
enabling low mass black holes to produce high luminosity.
The other significant change is that 
the median masses for the ADAF model are considerably lower in both X-ray 
bands because ADAF accretors emit a 
larger fraction of their bolometric energy in X-rays.

\begin{figure}
\epsscale{1.0}
\plotone{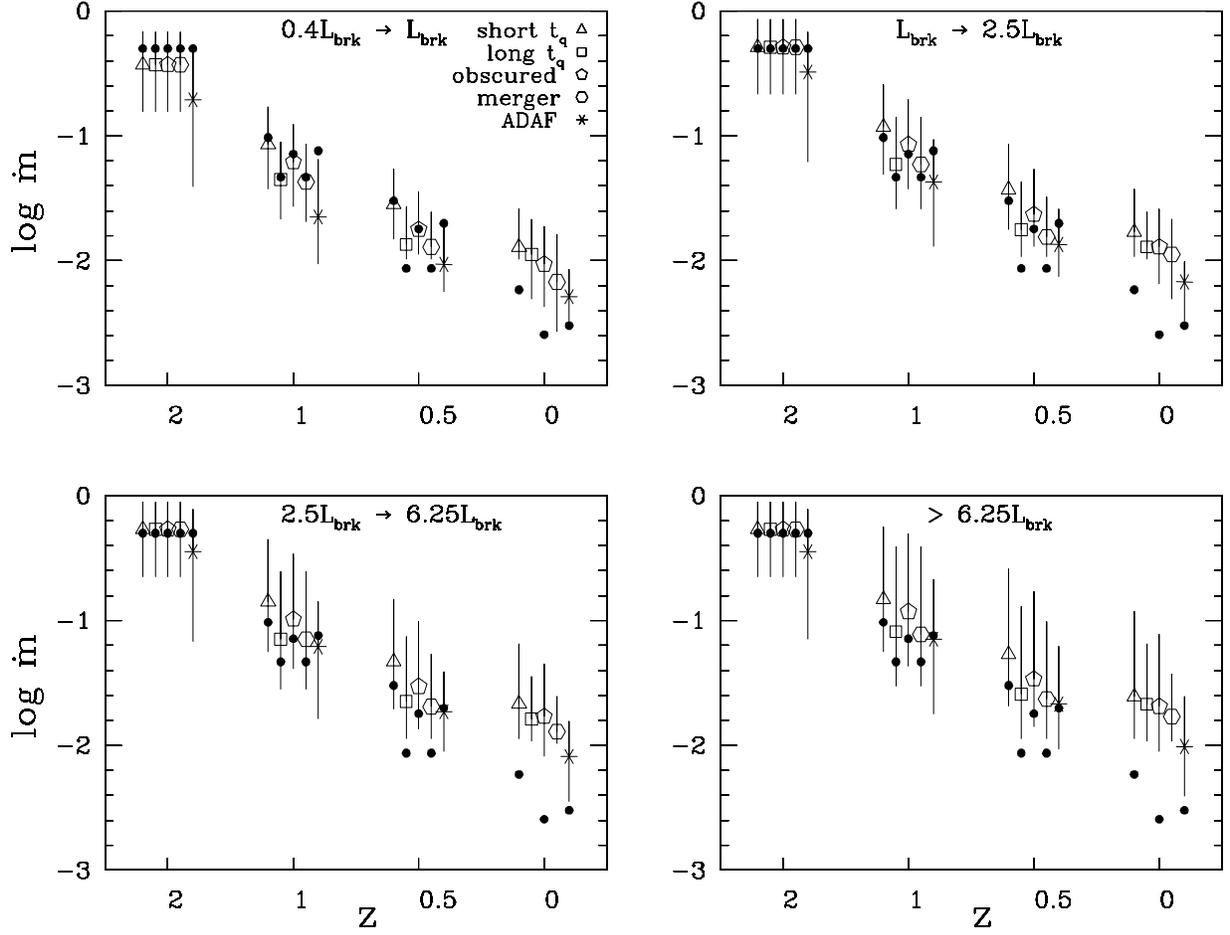}
\caption{Evolution of the accretion rate distribution of active black holes as 
a function of luminosity for the five illustrative models discussed in 
\S\ref{sec:illus}.  This figure is similar to Figure \ref{fig:massbh}, but 
presents results in terms of accretion rates rather
than black hole masses. Each panel corresponds to the range in $B$-band 
luminosity shown in the top center. The small black dots show the value of 
$\dot{m}_*(z)$ for each model, open symbols show the median accretion rate 
contributing to the given luminosity range, 
and the vertical bars show the 10\%-90\% range of $\mdot$.
The offset of the points from $z=2,1,0.5,$ and 0 is 
artificial and is done to distinguish the models from each other.  The values 
of $\dot{m}_*$ at $z=0$ for the \ltq\ and merger model fall well below 
$10^{-3}$ and are off the bottom of the plot.
}
\label{fig:mdot}
\end{figure}

While the distributions of black hole masses and of accretion rates associated 
with a given luminosity provide essentially the same information, it is
helpful to look directly at both distributions.
Figure~\ref{fig:mdot} is similar in spirit to Figure~\ref{fig:massbh}, but it 
shows the distribution of accretion rates at a given luminosity rather than
the distribution of black hole masses.  Symbols
mark the median value of $\dot{m}$, vertical bars show the 10\%--90\% 
range of the distribution, and small black dots show each model's
$\mdotstar$ at each redshift.
A model-to-model comparison shows essentially the reverse behavior
from Figure~\ref{fig:massbh}, with higher median black hole masses 
corresponding to lower median accretion rates.  
The key results are that the median $\mdot$ is close to $\mdotstar$
at all luminosities at $z=2$, in all of the models, while at low
redshift the median $\mdot$ is an increasing function of luminosity.
Furthermore, even the low luminosity systems tend to have $\mdot>\mdotstar$
at low redshift.  

The ADAF model is the outlier in this plot because its $p(\mdot)$
distribution is strongly skewed to favor low accretion rates.
It consistently has the lowest median accretion rate at a given
luminosity and redshift, and the spread in accretion rates is large.
At $z=0.5$, the median $\mdot$ in the $0.4\Lbrk - \Lbrk$ luminosity
bin is equal to the critical value $\mdotcrit$ at which ADAF accretion
sets in, indicating that about half of optically systems selected in this
luminosity range are predicted to be ADAF accretors.
A similar conclusion holds even for the highest luminosity bin at $z=0$.
The situation is more extreme for X-ray selection, where the 
ADAF model predicts a median $\mdot$ in the ADAF range in {\it all} 
luminosity ranges at $z \leq 1$.
As already noted in \S\ref{sec:multilambda}, this prediction that
a large fraction of luminous X-ray quasars are ADAF systems appears
to be an empirical failure of this model.

\begin{figure}
\epsscale{1.0}
\plotone{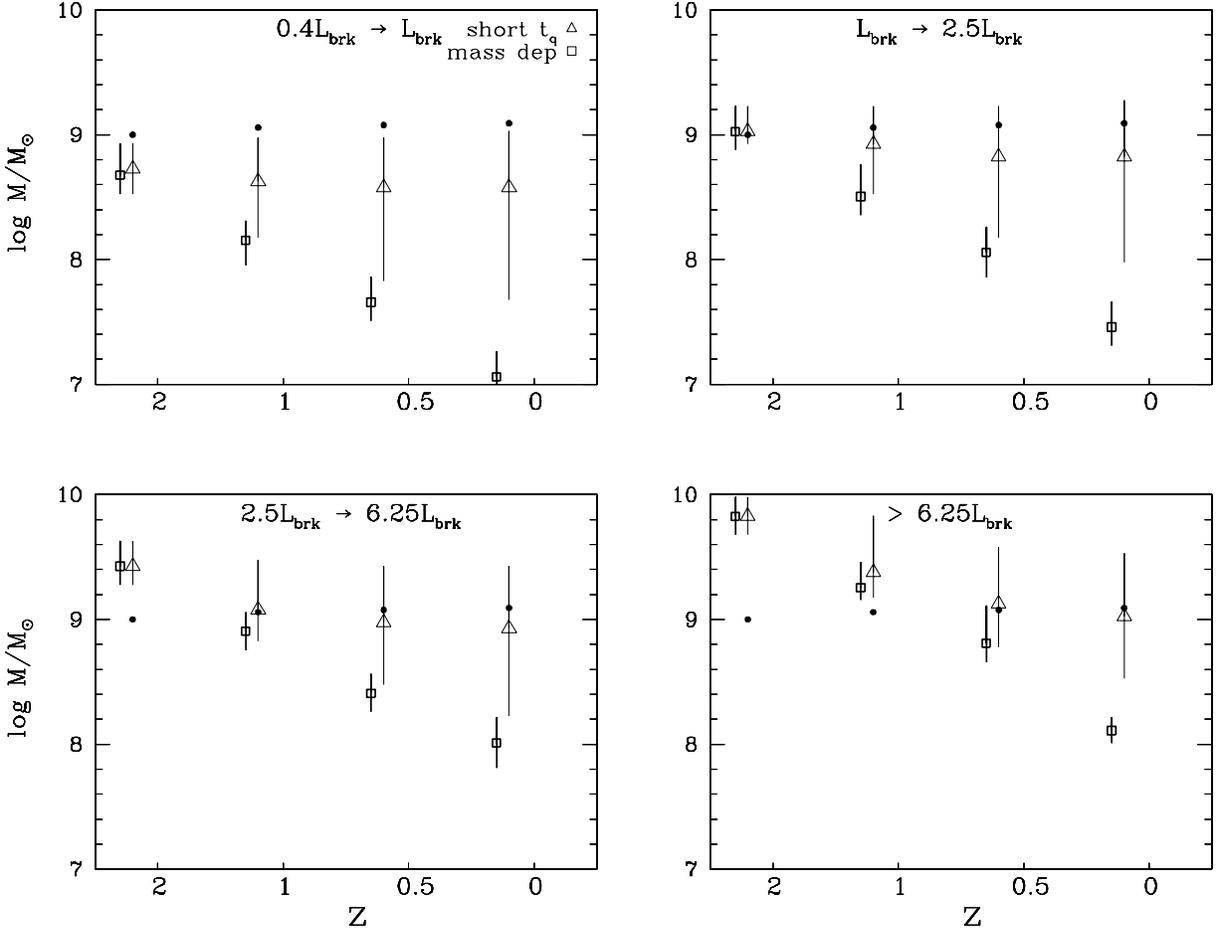}
\caption{Evolution of the mass distribution of active black holes as a 
function of luminosity for the \stq\ model (triangles) and the mass-dependent 
$p(\mdot|M)$ model of \S\ref{sec:massdepdeclining} (squares).
The format is similar to that of Figure~\ref{fig:massbh}.
Small black dots show
values of $M_*(z)$, which are nearly identical for both models.  Open points
show the median mass of active black holes in the indicated
luminosity range, and vertical bars show the 10\%-90\% mass range.
}
\label{fig:shortvsmassdep}
\end{figure}

All five of our models assume $p(\mdot)$ independent of mass,
and the transition in behavior from high redshift to low redshift
is a consequence of the declining evolution of $\mdotstar$ that is required
to match the \cite{boyle00} luminosity evolution in any such model.
However, as shown in Figure~\ref{fig:massdepevol}, it is also possible
to reproduce the \cite{boyle00} results with a model in
which $\mdotstar$ is constant but $p(\mdot|M,z)$ is mass-dependent.
Since this model is dominated by thin-disk accretors, its luminosity
function is similar to that of our \stq\ model in every band.
However, Figure~\ref{fig:shortvsmassdep} shows that its predicted
distribution of active black hole masses is strikingly different
at low redshift.  Because the mass-dependent model reproduces the
downward evolution of $\Lbrk$ by preferentially reducing activity in 
massive systems, it predicts much lower median black hole masses
at any luminosity at low redshift.  Furthermore, the luminosity
function remains primarily a sequence of black hole mass even
at low redshift, with the median mass rising by a factor of ten
between our lowest and highest luminosity bins at $z=0$, compared
to only a factor of three for the \stq\ model.
Because of the tight link between black hole mass and luminosity
in this model, the spread in masses at a given luminosity remains
small even at low redshift.  The evolution of the distribution
of active black hole masses thus provides an excellent tool for
deciding whether the observed decline of $\Lbrk$ reflects decreasing
characteristic accretion rates or a preferential drop in activity
among more massive black holes.

Recent studies using line widths to estimate black hole masses for
large data samples (e.g., \citealt{mclure03,vestergaard03}) provide
the kind of data needed to test the predictions in 
Figure~\ref{fig:massbh}--\ref{fig:shortvsmassdep}.
Vestergaard's (\citeyear{vestergaard03}) Figure 5a bins estimated
black hole masses by luminosity and redshift, allowing a
qualitative comparison.  At $z\sim 2$, the typical estimated masses
for $L \sim \Lbrk$ are $\sim 10^9 M_\odot$, in agreement with our
model initial conditions, and there is a clear trend of estimated
black hole mass with luminosity, though perhaps less strong
than the nearly linear relation predicted by our models.
At $z\sim 0$, there is a broader spread in estimated masses at
a given luminosity, as predicted by our standard models in which
$\mdotstar$ declines at low $z$.  However, the typical mass at
$L \sim \Lbrk$ is $\sim 10^{7.5}-10^{8} M_\odot$, which is
in between the predictions of the mass-independent $p(\mdot)$ and
mass-dependent $p(\mdot)$ models shown in Figure~\ref{fig:shortvsmassdep}.
Careful assessment and modeling of the statistical
errors is needed to draw reliable conclusions from a more quantitative
comparison, since the random errors in the mass estimates are large
enough (a factor $\sim 3$) to distort the underlying mass distributions
significantly.  

\subsection{Space densities of quasar hosts}
\label{sec:hostdensity}

Some of our models have a low space density of black holes and a
high duty cycle --- i.e., low $n(M)$ and high $p(\mdot)$ --- while
others have more numerous black holes and lower duty cycles.
Unfortunately, neither the luminosity function nor the distribution
of active black hole masses distinguishes these cases, since both
depend only on the product $n(M) p(\mdot|M)$
(see eqs.~\ref{eqn:lumfun} and~\ref{eqn:relmassdist}).
However, if the locally observed correlations between black hole mass
and bulge velocity dispersion or luminosity
continue to higher redshifts, they offer a tool 
for diagnosing, at least approximately, the underlying space 
density of black holes that shine at a given luminosity.
If the space density is low and the duty cycle high, then
quasars should reside in rare host galaxies with 
luminous bulges.  If the space density is high, then host
galaxies should include later type and less luminous systems.

To translate this idea into a precisely defined observable, we
find the median mass of black holes that produce quasars in 
a given luminosity range (the symbols in Figure~\ref{fig:massbh}),
then compute the space density of black holes with this mass or greater.
The corresponding observational program would require measuring
the median host galaxy luminosity $\Lhostmed(L_q)$ of quasars in
the same luminosity range, then measuring the galaxy luminosity
function at the same redshift and computing $\Phi(L>\Lhostmed)$,
the space density of galaxies brighter than $\Lhostmed(L_q)$.
The predicted and observable quantities are directly comparable
if the scatter between $\Mbh$ and $\Lhost$ is negligible.
Note that this condition does not imply negligible scatter between $L_q$
and $\Lhost$, since $\mdot$ variations still produce variations
in quasar luminosity.  Since black hole mass appears to be most
directly correlated with bulge properties, one would ideally use
host bulge luminosity and the bulge luminosity function rather
than total luminosities.  Accurate bulge-disk decomposition at
high redshift may be impractical, however, especially in the presence
of an active nucleus, so the next best thing is to use a red
passband that is sensitive to old stellar populations.
While the space density of hosts is not as informative as an
actual measurement of the black hole mass function $n(M,z)$, it 
is less demanding observationally, since it does not require
{\it calibration} of the $\Mbh-\Lhost$ relation at high
redshift, only the existence of a relation with relatively small scatter.

\begin{figure}
\epsscale{0.7}
\plotone{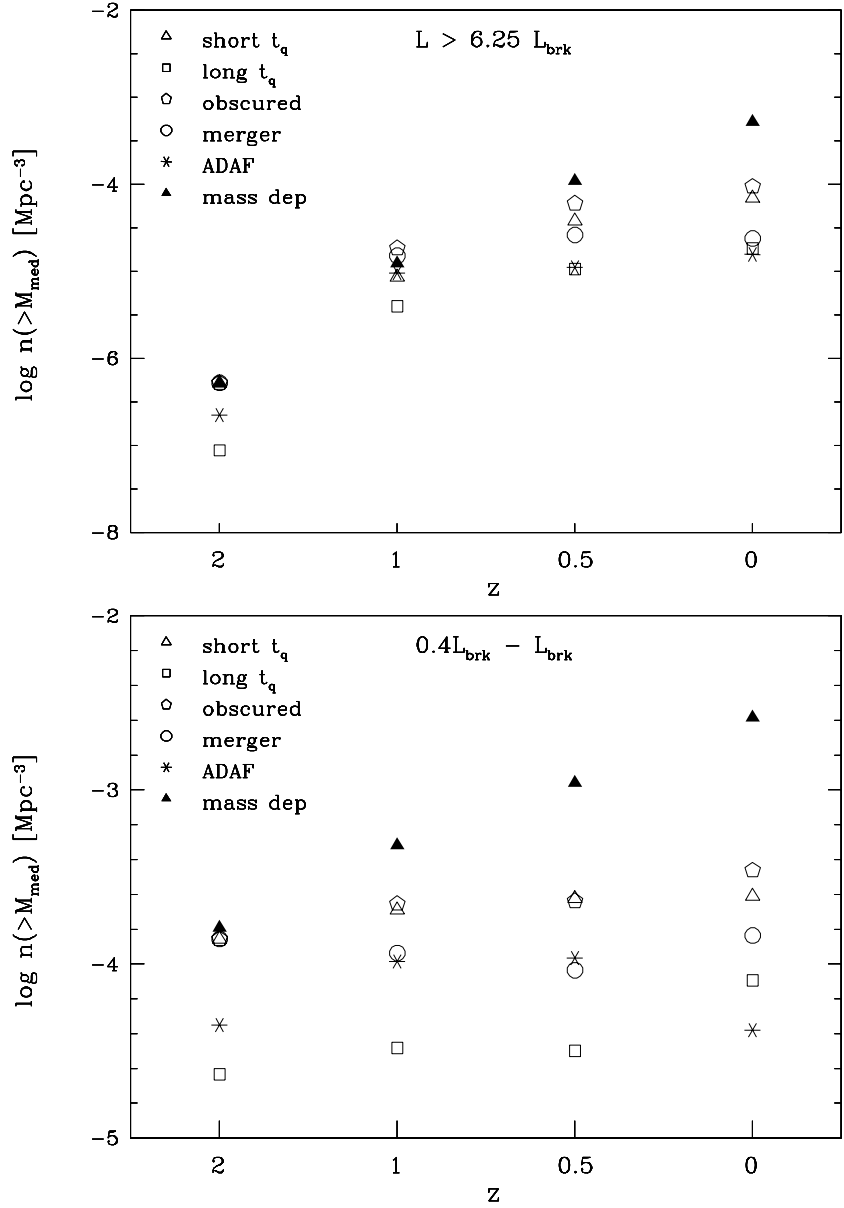}
\caption{
The comoving space density of black holes with mass above the median
mass of active systems with $B$-band luminosity
$0.4 \Lbrk < L < \Lbrk$ (top) or $L>6.25\Lbrk$ (bottom).
Open symbols show results for the five models discussed in
\S\ref{sec:illus}, as indicated, and filled triangles show
results for the mass-dependent $p(\mdot)$ model illustrated
in Figure~\ref{fig:massdepevol}.  These predictions can be tested
by studies of quasar hosts or quasar clustering.
Note the different $y$-axis ranges of the top and bottom plots.
}
\label{fig:spacedens}
\end{figure}

Figure~\ref{fig:spacedens} shows the model results for the luminosity
ranges $0.4 \Lbrk < L < \Lbrk$ and $L> 6.25\Lbrk$ at redshifts
$z=2$, 1, 0.5, and 0.  As expected, the \ltq\ model starts with
a host space density that is a factor of ten below that of the \stq\
model, and although it catches up at lower redshift because of 
the greater amount of black hole growth, a considerable gap remains.
The obscured model starts with an $n(M)$ close to that of the
\stq\ model, and its predictions for host space densities remain
close to it at all redshifts, with the greater growth of black
hole masses largely compensated by the faster decline in $\mdotstar$.
Note, however, that this model's prediction would have been quite
different if we had implemented it by boosting the black hole space
density instead of the duty cycle.

The merger model is identical to the \stq\ model at $z=2$, 
but its distinctive evolution of $n(M)$, with
low mass black holes transforming into high mass black holes,
leads to different behavior of the predicted space densities
with redshift and with luminosity.  In particular, the low-$z$ depletion
of the low end of $n(M)$ leads to a relatively low space density, 
especially at low luminosity.  
The ADAF model starts with a relatively low $n(M)$ (intermediate between
\ltq\ and \stq) and thus a relatively low host space density.
However, its mass function overtakes that of \stq\ at high masses by $z=1$
(see Fig.~\ref{fig:rationofmevol}), and the predicted host space densities
are similar for $L>6.25\Lbrk$.  At low luminosities and higher redshifts,
the ADAF model has a lower space density despite its 
large amount of black hole growth because its $p(\mdot)$ distribution favors
rarer, higher mass black holes at a given luminosity.

The model that stands out most distinctively in Figure~\ref{fig:spacedens},
at least in terms of its redshift dependence,
is the mass-dependent $p(\mdot)$ model, shown by the filled triangles.
Because this model matches the QLF by shifting activity preferentially 
towards low mass black holes at low redshift, the predicted host space
density climbs rapidly.  In this model, the hosts of moderate luminosity 
quasars at low redshift should include relatively late-type galaxies
and low luminosity ellipticals.  In effect, the properties of hosts
offer an indirect way to detect the sharp decline in typical active 
black hole mass shown in Figure~\ref{fig:shortvsmassdep}.

If the masses of black holes are correlated with the masses of the
dark matter halos in which they reside, a natural expectation given
the observed correlation with bulge velocity dispersion, then 
quasar clustering provides another observational tool for inferring
their space density.  A low black hole space density implies that 
the host halos are rare, massive systems that tend to be strongly clustered
\citep{kaiser84,mo96}, while a high space density implies more common,
less strongly clustered hosts.  This is the idea behind the proposals of
\cite{haiman01}, \cite{martini01}, and \cite{kauffmann02} to 
constrain quasar lifetimes (hence duty cycles, hence black hole space 
densities) using the quasar correlation function or the quasar-galaxy
cross-correlation function (see also \citealt{hnr98}).
In Figure~\ref{fig:spacedens}, models with low points predict strong
quasar clustering and models with high points predict weaker clustering.
To a first approximation, one could calculate the expected bias (relative to
mass clustering) for quasars in the two luminosity ranges by
reading off the space density from the figure, finding the 
mass threshold for halos that have this space density given an
assumed cosmological model, and calculating the bias of halos
above this mass threshold using the methods of 
Mo \& White (\citeyear{mo96}; 
see \citeauthor{martini01} [\citeyear{martini01}] for
a more detailed description of this approach).
One could also carry out a more thorough calculation, weighting
the contribution to the bias by the fraction of black holes of a 
given mass contributing to the luminosity range, but we suspect
that the results would not be very different.

Recent results from the 2dF Quasar Redshift Survey favor a relatively
low comoving correlation length, $s_0 \approx 5.8 h^{-1}\;$Mpc in redshift
space, with no clear evidence for dependence on redshift or luminosity
\citep{croom03}.  Using the \cite{martini01} model, which assumes 
simple on-off quasar activity with a monotonic relation between quasar
luminosity and host halo mass, the implied lifetime is short, 
$t \sim 10^6$ years.  The current data set is not quite large enough
to allow a precise clustering measurement for a volume-limited 
subset of quasars at $z\sim 2-3$, which is what one would ideally like
to use for the lifetime analysis.  However, if future results continue
to show a low correlation length at high redshift, and no significant
dependence on luminosity, then they may indicate that there is 
substantial scatter in the relation between quasar luminosity and
host halo mass.  This scatter could in turn indicate that the
correlation between black hole mass and halo mass at high redshift is
much weaker than the measured correlation between black hole mass
and bulge mass at low redshift, or else that the luminous quasars have a
wide range of $L/\Ledd$ even at high-$z$.  
Clustering analyses of the full 2dF quasar
survey and of the SDSS quasar survey should yield interesting insights
on these questions over the next few years.

\section{Summary}
\label{sec:summary}

In the framework developed here, the central actor in black hole
and quasar evolution is the accretion rate distribution
$p(\mdot|M,z)$, the probability that a black hole of mass $M$
accretes at a rate $\mdot$ (in Eddington units) at redshift $z$.
Given a model for the accretion efficiency as a function of $\mdot$,
which can be inferred from observations and theoretical considerations
that are largely independent of QLF evolution {\it per se},
the combination of $p(\mdot|M)$ and the black hole mass function
$n(M)$ determines the bolometric luminosity function via 
equation~(\ref{eqn:lumfun}).  Furthermore, in the absence of mergers,
$p(\mdot|M,z)$ determines the evolution of $n(M,z)$ given
a ``boundary value'' of $n(M)$ at some redshift.  Mergers can 
complicate the picture by changing $n(M)$ independently of $p(\mdot|M,z)$.
We have generally made the plausible but not incontrovertible
assumption that black holes accreting in the range $0.01 < \mdot < 1$
have ``thin-disk'' efficiencies $\epsd\approx 1$ and that
efficiencies decrease at higher (super-Eddington) and lower (ADAF)
accretion rates (eq.~\ref{eqn:effofmdot}).  
We have derived many of our results under the
mathematically simplifying assumption that $p(\mdot)$ is independent
of mass.  While this assumption is unlikely to hold to high accuracy,
it may be a reasonable approximation at redshifts near the peak
of quasar activity, since black holes that grow faster than their
peers tend to reduce their accretion rates in Eddington units 
and {\it vice versa}.
Most of our specific examples assume that $n(M)$ and $p(\mdot)$
are double power-laws with breaks at $M_*$ and $\mdotstar$,
respectively, with $p(\mdot)$ truncated at $\mdotmin=10^{-4}$
and $\mdotmax=10$.

\subsection{Basic Results}
\label{sec:summary1}

Our framework yields a number of mathematical results that give
insight into the relations among the black hole mass function,
the accretion rate distribution, and the QLF.
When $p(\mdot)$ is independent of mass, the convolution 
integral~(\ref{eqn:lumfun}) for $\Phi(L)$ can be understood as follows:
for each range $\mdot \rightarrow \mdot+d\mdot$, the mass function $n(M)$
is mapped to a luminosity $L=\epsd \mdot l M$, multiplied by
$p(\mdot)d\mdot$, and added to a running total.
If $n(M)$ is a double power-law and the range of $\mdot$ is bounded,
then the asymptotic slopes of $\Phi(L)$ must equal the low and high
mass slopes of $n(M)$, since high luminosity
objects must come from
black holes with $M>M_*$ and low luminosity objects
must come from black holes with $M<M_*$.
In the intermediate luminosity regime, where black holes above and
below $M_*$ can both contribute, the QLF turns over in a way that
depends on the slopes of $n(M)$ and the shape of $p(\mdot)$.
For our usual double power-law $p(\mdot)$, the QLF break occurs 
at a luminosity $\Lbrk \sim \mdotstar l M_*$.
If $\mdotstar$ is close to one, then the break luminosity
corresponds roughly to the Eddington luminosity of $M_*$ objects,
and the slope above the break corresponds to the high mass slope
of $n(M)$.  However, if
$\mdotstar$ is low, then the turnover may be associated largely with the 
change in slope of $p(\mdot)$, and the asymptotic regime where the high-$L$
slope of $\Phi(L)$ matches the high-$M$ slope of $n(M)$ may only be reached
beyond the observed range of luminosities.  
For most of the models that we have presented here,
which are designed to match the \cite{boyle00} optical luminosity function,
the first case applies at high redshift and the second at low redshift.

A key feature of our model of accretion physics is that objects do not
radiate at super-Eddington luminosities --- instead, we assume that the 
efficiency is $\epsd = \mdot^{-1}$ for $\mdot > 1$, so that 
super-Eddington accretors radiate at $L=\Ledd$.
Although we have generally adopted $\mdotstar = 0.5-1$ for fitting
data at $z \geq 2$, our results would not be very different if we
took $\mdotstar > 1$, or if we changed the maximum accretion rate
or the high-$\mdot$ slope, because the change in efficiency would
cut off the luminosity distribution at $\Ledd$ anyway.
Our results would also not be very different if we assumed that
black holes fed at a super-Eddington rate by their host galaxy 
regulate their accretion 
by outflows or convection \citep{blandford99,quataert00}
so that they gain mass at $\mdot\approx 1$ and radiate at $L \approx \Ledd$.
If such flows drove gas out into the galaxy halo,
they would effectively truncate $p(\mdot)$ at $\mdot=1$, while if they
returned gas to a reservoir from which it would eventually be accreted,
they would transform the $\mdot>1$ tail of $p(\mdot)$ into a
spike at $\mdot=1$.  With our standard efficiency assumptions, the 
latter scenario would increase the average radiative efficiency
of high-$\mdot$ accretion, by moving it from the super-Eddington
regime to the thin-disk regime, thus yielding more luminosity
for a given amount of black hole growth.
Changing the accretion physics to allow $\epsd \approx 1$
with $\mdot > 1$, and thus to allow substantially super-Eddington
luminosities when the accretion rate is high \citep{begelman02},
would have a more drastic effect on our results.  In this case, only a
truncation of $p(\mdot)$ would cut off the luminosity distribution
at a given black hole mass, and the predicted luminosity function
would therefore be sensitive to the values of $\mdotstar$
and $\mdotmax$ and to the shape of $p(\mdot)$ at $\mdot > 1$.

The fraction of black holes of mass $M$ that are active at a luminosity $L$
is proportional to $n(M) p\left(\mdot=\frac{L}{\epsd \mdot l M}\right)$,
the underlying black hole space density times the probability of 
having the accretion rate required to shine at $L$.
The same product appears in the integrand for the luminosity function
itself (eq.~\ref{eqn:lumfun}), but by constraining the integrand
rather than the integral, measurements of active black hole masses
can discriminate among models that
produce similar $\Phi(L)$ with different $p(\mdot)$ and $n(M)$.
In terms of our double power-law
models, the active mass distribution is particularly useful for
distinguishing cases with high $M_*$ and low $\mdotstar$ from
models with low $M_*$ and high $\mdotstar$.

A general mass-dependent $p(\mdot)$ can be written in the form
$p(\mdot|M) = p_0 (\mdot) D(M | \mdot)$, with $p_0(\mdot) \equiv p(\mdot|M_0)$
for some fiducial mass $M_0$ and $D(M_0|\mdot) \equiv 1$.
With such a mass dependence,
we can still understand the convolution integral for $\Phi(L)$ by
considering the contribution from each range 
$\mdot \rightarrow \mdot + d\mdot$, but where the sum before involved
only horizontally and vertically
shifted versions of the mass function $n(M)$, now the mass
function can be tilted or distorted by $D(M | \mdot)$ before
being added to the running total.
Mass-dependence of $p(\mdot)$ thus breaks the tight link between
$n(M)$ and $\Phi(L)$ and adds freedom to models of the QLF.
Specifically, for a model with a given $n(M)$ and a
mass-independent $p(\mdot)$, there is a family of models with different
$n(M)$ and mass-dependent $p(\mdot)$ that yield identical
predictions for $\Phi(L)$ and the distribution of {\it active} black
hole masses.  However, this degeneracy applies only at a single
redshift; the evolution of models with mass-dependent $p(\mdot)$ 
is different because the shape of $n(M)$ changes with time.
Furthermore, while the active black hole mass distributions are the same,
the underlying black hole mass functions are different,
and a measurement of the full $n(M)$ at $z=0$
may be sufficient to diagnose the mass-dependence of $p(\mdot)$
at higher redshifts.
If two models have similar underlying $n(M)$, then the masses of
active black holes or space densities of their host galaxies are
powerful diagnostics for mass dependence of $p(\mdot)$, as illustrated
in Figures~\ref{fig:shortvsmassdep} and~\ref{fig:spacedens}.

For evolutionary calculations, a crucial simplification 
(eq.~\ref{eqn:acc})
is that the
accretion driven growth of $n(M)$ depends only on the mean accretion
rate $\langle \mdot(M,z) \rangle \equiv 
\int_0^\infty \mdot p(\mdot|M,z) d\mdot$,
not on the full form of the distribution function.
Between times $t_1$ and $t_2$, a black hole with accretion rate $\mdot(t)$
grows in mass by a factor $\exp (\tacc/t_s)$, where 
$\tacc = \int_{t_1}^{t_2} \mdot dt$ is the accretion weighted 
lifetime (eq.~\ref{eqn:tacc}) and $t_s$ is the Salpeter time
(eq.~\ref{eqn:tsdef}).
In any particular mass bin, some of the black holes grow faster than average
and some grow slower, but $n(M)$ evolves as if all of them grew at the
average rate for that bin.  If $\avmdot$ is independent of mass,
then the growth factor is the same for all mass bins, and the
mass function simply shifts in a self-similar fashion, preserving
its shape (eq.~\ref{eqn:accgrowth}).  One might expect this self-similar
behavior to emerge, approximately, as a ``fixed point'' solution
during the epoch of rapid black hole growth.  If $\avmdot$ depends
on mass, then the shape of $n(M)$ changes with time, becoming steeper
if low mass black holes grow more rapidly and shallower if high mass
black holes grow more rapidly.

Our results also imply a number of critical values for the logarithmic
slopes of $p(\mdot)$ and $n(M)$, where the character of the solutions
changes.  We summarize these critical slopes in Appendix~\ref{sec:appx2}.
The most important result of this sort is that growth by accretion 
drives $n(M)$ at fixed mass up if the mass function is steeper than $M^{-1}$
and down if it is shallower, and that the corresponding critical slope
for merger driven growth is $-2$ rather than $-1$.  Generically, one
expects mergers to increase $n(M)$ at high masses and decrease it at
low masses.

\subsection{Implications of the Observed Luminosity Evolution}
\label{sec:summary2}

We have kept the comparison to observations in this paper rather
loose, saving detailed tests of models against
multi-wavelength luminosity functions and estimates of black hole mass 
distributions for future work.  
However, we are able to draw some interesting general
conclusions from our efforts to reproduce the observed evolution
of the optical luminosity function.

At $z\leq 2$, we have focused on the \cite{boyle00} evolution results,
and particularly on their finding that the QLF has a break that shifts
to lower luminosities at lower redshifts.  In any model where 
$p(\mdot)$ is independent of mass, reproducing this result requires
a shift towards lower typical accretion rates over time 
(Fig.~\ref{fig:dblpowevol}).
Decreasing
the duty cycle uniformly by reducing the amplitude of $p(\mdot)$ can
lower the normalization of $\Phi(L)$, but on its own it cannot shift
the break to lower luminosities, since black hole masses, and thus
the location of the break in $n(M)$, can only increase with time.
Lowering the break luminosity requires decreasing
the probability of high accretion rates {\it relative} to 
low accretion rates.  In the context of our
double power-law models, this change is achieved by reducing $\mdotstar$,
from slightly below unity at $z\sim 2$ to far below unity at $z \sim 0-0.5$.
Physically, such a change could arise because of decreasing gas
supplies and increasing dynamical times in galaxies at lower redshifts
\citep{kauffmann00}.

A consequence of this reduction in $\mdotstar$ is
a change in the nature of the QLF between high and low redshift.
At high redshift, the sequence of quasar luminosities is primarily
a sequence of black hole masses.  Because $\mdotstar$ is close to unity
and $p(\mdot)$ is shallow below the break, most high luminosity quasars
are produced by the more numerous low mass black holes accreting 
at $\mdot \approx 1$ rather than extremely rare 
high mass black holes with low accretion rates.  
The typical mass at a given luminosity is $M \approx L/l$, and
a factor of ten increase in luminosity roughly corresponds 
to a factor of ten increase in black hole mass.
However, once $\mdotstar$ falls well below unity, there is a large
range $\mdotstar < \mdot < 1$ over which $p(\mdot)$
is steep ($\propto \mdot^{-3}$ for most of our models), thus increasing
the probability that a given luminosity is generated by
a high mass black hole accreting at a low $\mdot$ close to $\mdotstar$.
The typical active black hole mass still increases with luminosity,
but at low redshift a factor of ten increase in $L$ corresponds to
only a factor $\sim 3$ increase in median black hole mass, and the
range in black hole mass at fixed luminosity is much larger than
at high redshift.  At low redshift, the sequence of quasar luminosities
remains partly a sequence of black hole mass, but it is also in large part
a sequence of $\mdot$ (Figs.~\ref{fig:massbh} and~\ref{fig:mdot}).

This change in character arises only because
we reduce the radiative efficiency in the super-Eddington regime,
forcing systems with $\mdot > 1$ to radiate at the Eddington luminosity.
Without this transition in the accretion physics, there would be
no preferred scale to change the relative importance of mass and $\mdot$
between high and low redshift, since we assume the same double power
law {\it form} of $p(\mdot)$ at all redshifts and it is only the
value of $\mdotstar$ relative to unity that changes.
If the efficiency does not decrease in the super-Eddington regime,
then the shape of the luminosity function depends in detail on
the form and cutoff of $p(\mdot)$ at $\mdot > 1$.
However, the limited data on masses of high redshift black holes
generally shows systems with $L/\Ledd \approx 1$ but not
substantially larger \citep{mclure03,vestergaard03}.  
This result suggests that
there is indeed a break in the radiative efficiency at $\mdot \approx 1$,
or else that black hole accretion self-regulates to enforce
$\mdot \la 1$, since otherwise it would require $p(\mdot)$ to
cut off coincidentally at $\mdot > 1$ and be relatively high
at $\mdot$ just below one.

If $p(\mdot)$ is mass-dependent, then there is a fundamentally different
alternative for explaining the shift of the QLF break to lower luminosities:
instead of a decline in characteristic accretion rates, the mass-dependence
can itself evolve so that activity is preferentially suppressed in high
mass black holes at low redshift (Fig.~\ref{fig:massdepevol}).  
Physically, such behavior could
arise because high mass black holes reside in early type galaxies
with large bulges, which tend to exhaust their gas supplies earlier.
In this scenario,
quasars at a given position on the
luminosity function (relative to $\Lbrk$) are always associated with
the same distribution of accretion rates, but the associated black hole
mass declines with redshift as the high mass black holes turn off.
In comparison with the mass-independent $p(\mdot)$ models, the median
black hole mass at fixed luminosity is lower, and the range of
black hole masses is smaller, with near-Eddington accretors making
a large contribution to the high end of the luminosity function at
all redshifts (see Figure~\ref{fig:shortvsmassdep}).
The low black hole mass implies a high space density
of hosts, so this model predicts that a larger fraction of low redshift
quasars reside in late type or low luminosity galaxies.

Based on anecdotal evidence, it is hard to say whether a decline
in characteristic accretion rates or a decline in the activity of
high mass black holes is more important in producing the observed
decline of $\Lbrk$ at $z<2$.  The first scenario's prediction of
a wider range of $L/\Ledd$ at lower redshifts seems in qualitative
agreement with studies of black hole masses \citep{woo02,vestergaard03}.
The second scenario seems in qualitative agreement with
the relative quiescence of the black holes in most massive ellipticals
(such as M87), though \cite{kauffmann03} find that the most luminous
AGN in the local universe do reside in massive, early type systems.
The two scenarios make quantitatively different predictions, 
and careful comparison to studies of active black hole masses
and host galaxy properties should be able to show whether one
mechanism dominates or both are comparably important.
The results of \cite{vestergaard03} suggest that active black hole
masses at low redshift are intermediate between the values predicted
by the two models shown in Figure~\ref{fig:shortvsmassdep}.

At high redshifts, constraints on the form of the QLF are weaker;
in particular, the SDSS measurements of \cite{fan01} probe only
the high luminosity end of $\Phi(L)$.  For matching the observed
evolution over the range $z\sim 5$ to $z\sim 2$, we find one 
acceptable solution in which $p(\mdot)$ is roughly constant in
Eddington units and the growth of the black holes themselves
drives the growth in amplitude of the QLF (Fig.~\ref{fig:growth}).
With our adopted 
parameters, the black holes grow by a factor $\sim 10$ between
$z=5$ and $z=2$, and the space density at $z=2$ is 
$n_* M_* = 2.012 \times 10^{-5} \dunits$ at $M_*=10^9 M_\odot$.
Significantly 
lower normalizations of $n(M)$ at $z=2$ are not allowed because
they would require a negative black hole mass density at $z=5$.
Higher normalizations are allowed, in which case the quasar duty cycle
is shorter, the rate of black hole growth is smaller, and the
growth of the QLF is driven by a steady increase in $p(\mdot)$.
While a solution with a high black hole space density
can give an acceptable match to the QLF, 
it seems physically unattractive because it requires that most of 
the black hole mass density was already in place at $z=5$, before the
main epoch of quasar activity.  In this latter scenario, the luminous
phases of quasars represent the addition of a small amount of mass
to already formed black holes.  If we assume instead that the observed
optical QLF {\it does} trace the growth of black holes, 
and that the former model is therefore more realistic,
then our analysis predicts a black hole space density 
$Mn(M) \sim 2-3\times 10^{-5}\dunits$ at $M=10^9 M_\odot$ at $z=2$,
and continuation to $z=0$ implies a similar space density
at $M \sim 2\times 10^9 M_\odot$.  However, we have not investigated
the sensitivity of this prediction to our specific choices of
parameters, such as the double power-law form and adopted slopes of
$p(\mdot)$ and $n(M)$.  Black hole mergers and obscured accretion 
could also alter the prediction significantly, especially at low redshift.

\subsection{Distinguishing Scenarios}
\label{sec:summary3}

Many of the qualitative results mentioned above could have been anticipated
without detailed calculations.  Our framework, however, allows one to compute
quantitative predictions of concrete models that illustrate distinct ideas
about the nature of black hole and quasar evolution.  We did this in 
\S\ref{sec:illus} for the low redshift $(z \leq 2)$ regime, adopting as our
baseline a model with quasar emission and black hole growth dominated by 
unobscured thin-disk accretion and a normalization of $n(M,z=2)$ implying
a short quasar lifetime, and, consequently, little growth of black hole masses
from $z=2$ to $z=0$.  We compared this model to four variants: one with a
lower $n(M,z=2)$ and correspondingly longer quasar lifetime, one with a
4:1 ratio of obscured to unobscured systems, one with a 
large amount of merger driven growth of black hole masses, and one with a
boosted probability of low-$\mdot$ accretion leading to substantial ADAF
growth of black holes.  For each scenario, we are able to find parameters
that acceptably reproduce the \cite{boyle00} optical QLF at $z=2$, 1, and 0.5.

Relative to the \stq\ model, the \ltq\ model starts at $z=2$ with a factor
$\sim 6$ lower $n(M)$ at every mass.  Because of the larger 
amount of accretion per black hole, however, the \ltq\ $n(M)$ 
overtakes the \stq\ $n(M)$ at high
masses by $z=0$, while remaining below it at low masses.  The two models
have similar QLFs at every wavelength, since they match in the optical by
construction and are dominated by systems with the \cite{elvis94} SED.
However, the growth of $M_*$ in the \ltq\ model requires a more rapid
decline of $\mdotstar$ to compensate, so at lower redshifts it predicts
lower median $\mdot$ and higher median black hole mass at a given luminosity.
The two models can thus be distinguished observationally by the $z=0$
black hole mass function, by the mass distributions of active black holes
at $z\sim 0.5-1$, and by the space densities of host systems, which are
lower in the \ltq\ model at every luminosity and redshift.

In the obscured model, the large amount of obscured accretion produces more
black hole growth than in the \stq\ model, leading to a higher $n(M)$ and
$\rhobh$ at low redshift.  By $z=0$, $n(M)$ is higher by a factor $\sim 6$
at high masses.  However, the median black hole mass and host space density
at a given optical luminosity are only slightly higher.  The clearest
distinguishing feature of this model is the relative amplitude of luminosity
functions in different wavelength bands, a consequence of the different
SED shapes of obscured and unobscured accretors.  At $z=2$, the 2-10 keV
luminosity function is elevated by nearly a factor of five, the ratio of
all systems to unobscured systems.  This boost decreases towards low redshift
because of the increasing importance of obscuration in the (observed-frame)
2-10 keV band.  The 0.5-2 keV band is heavily obscured at $z<2$, so the
soft X-ray luminosity function of the obscured model is similar to that of
the \stq\ and \ltq\ models, except at $z=2$ where it has a higher amplitude
at low luminosity.  The strongest departure of all is in the FIR, where the 
re-radiated emission of obscured accretors boosts $\Phi(L)$ by factors
of ten (low luminosity) to one hundred (high luminosity),
relative to the \stq\ and \ltq\ models.
This crucial prediction should soon be testable by {\it SIRTF} and by
other sub-mm and mm-wavelength observations.

In the merger model, low redshift mergers strongly distort the initial black
hole mass function, depleting it at low masses and boosting it at high masses,
with a factor of 16 increase in the high mass end at $z=0$ relative to 
the \stq\ model. The quasar population of this model is still dominated by 
systems with a thin-disk SED, and since it matches the observed optical 
luminosity function by construction, its predictions at other wavelengths 
are close to those of the \stq\ and \ltq\ models.
However, the high space density of high mass black holes leads to a 
high median mass of active black holes at fixed luminosity, similar to that
of \ltq\ at low luminosities and higher still at high luminosities.
Merger driven distortions of the mass function also lead to distinctive 
redshift and luminosity dependence of the host space density.  

Finding parameters that yield significant ADAF growth and an acceptable
match to the optical luminosity function proves quite difficult, requiring
an artificial boost to the probability of accretion rates below $\mdotcrit$.
With our adopted parameters, the ADAF model predicts a large amount of
black hole growth between $z=2$ and $z=0$, and thus a high $n(M)$ and
$\rhobh$ at $z=0$.  With the combination of high $M_*$ and high
probability of low $\mdot$, the ADAF model predicts the largest median
black hole masses and lowest median accretion rates at fixed luminosity,
with the median accretion rate approaching $\mdotcrit$ even for 
high luminosity AGN at $z \leq 0.5$.  The high X-ray fraction of the
ADAF SED boosts the soft and hard X-ray luminosity functions relative
to the optical, especially at low redshift, and the model predicts that
a majority of X-ray selected systems at $z\sim 0.5$ should be ADAF
accretors, even at high luminosities.  This prediction appears observationally 
untenable, and the model requires rather implausible parameter choices
in the first place, so our results suggest that ADAFs are unlikely to
make an important contribution to black hole growth in the real
universe, even at $z<2$ (\citeauthor{hnr98} [\citeyear{hnr98}]
reach a similar conclusion).
Low radiative efficiency at low $\mdot$ may
nonetheless help explain the remarkable quiescence of most black holes 
in the local universe \citep{narayan98}.

\subsection{Prospects}

Our results illustrate how a variety of observational constraints can
be brought to bear on the key questions of quasar and black hole evolution.
In particular, we have extended the ideas of \cite{soltan82} and \cite{small92}
to show that incorporating the link between luminous accretion and black
hole growth allows one to construct concrete physical models that are
simultaneously constrained by multi-wavelength luminosity function
measurements and estimates of black hole masses and accretion rates.
For our models in \S\ref{sec:illus}, we chose a plausible but not unique
set of initial conditions at $z=2$ and evolved them forward in time
under varying assumptions, always matching the observed optical QLF.
The models then make distinguishable predictions for other observables.

With our current parameter choices, all of our models face some 
difficulty when confronted with recent estimates of X-ray luminosity
functions and the local black hole mass function, as illustrated
in Figures~\ref{fig:comparenofm}, \ref{fig:comparesxdata}, and
\ref{fig:comparehxdata}.
Models that fit the \cite{ueda03} hard X-ray QLF generally do not
fit the \cite{miyaji01} soft X-ray QLF at the same redshift,
and vice versa.  A model incorporating luminosity and redshift
dependence of the obscured quasar fraction might fare better,
though some of the problem may still lie with the observational
estimates themselves.  Combining Tremaine et al.'s (\citeyear{tremaine02})
estimate of the $M-\sigma$ relation with Sheth et al.'s (\citeyear{sheth03})
estimate of the distribution of galaxy velocity dispersions yields a 
mass function that lies well below our model predictions for 
$M > 10^9 M_\odot$, unless the intrinsic scatter of the $M-\sigma$
relation is $\sim 0.5$ dex, compared to Tremaine et al.'s estimate
of $\leq 0.3$ dex.  Repairing this discrepancy would require either
reducing the break mass at $z=2$ substantially below $M_*=10^9 M_\odot$
or dropping our assumed double power-law form of $n(M)$ and adopting a
mass-dependent $p(\mdot)$ to reproduce the \cite{boyle00} QLF.
For the present, we do not want to draw strong conclusions from
these discrepancies, since we have not thoroughly assessed the
observational uncertainties, and we have not investigated the
extent to which a failing model can be ``fixed up'' by adjusting its
parameters (e.g., the initial shape of the black hole mass function), 
while retaining its essential features (e.g., a large fraction of
obscured systems).

The $z=0$ black hole mass function can be reasonably well estimated
from current data, at least at masses $M \leq 10^9 M_\odot$ where the
form and scatter of the $M-\sigma$ relation are well constrained,
and it is a fundamental boundary constraint on
any evolution model.  For a comprehensive attempt to match observations,
therefore, it probably makes sense to impose this constraint {\it a priori}
on all models, and integrate the evolutionary equations {\it backward} 
in time.  In our framework, this approach is just as easy as integrating
forward, even if it is less intuitive.  In effect, one takes the 
known black hole mass function today, infers the accretion rate distribution
by matching the luminosity function, steps backward by removing
the implied amount of mass from each bin of the mass function, 
and repeats.  
We will apply this approach to available observations in future work.
Given the inevitable uncertainties in
the observational data, radiative efficiencies, and bolometric corrections, 
there are likely to be
some degeneracies in the solutions, but we can hope that models
that differ in fundamental rather than incidental features will
remain observationally distinguishable.  Based on the results found here,
we suspect that the primary source of uncertainty 
will be the mass dependence of $p(\mdot|M)$, which requires accurate
measurements of both the QLF and the masses of active black holes
to pin down empirically.  
Mergers look like the other most difficult problem, though in this case
there are good theoretical ideas about what the merger rates of dark
halos and galaxies should be (e.g., \citealt{taylor01}), and these
can be incorporated into model calculations.

The traditional picture of quasars as a population of supermassive black
holes growing by accretion seems more secure than ever.
Many open questions remain about the roles of black hole mass, accretion
rate, radiative efficiency, and SED shape in determining quasar luminosities,
about the properties of accretion flows at low and high accretion rates,
about the importance of black hole mergers and obscured accretion as
drivers of black hole growth, and about the relations among populations
observed at different wavelengths.  
Our work highlights a number of areas where observational advances will be
crucial to answering these questions.
These include improved determination of the local black hole mass function,
better understanding of the dependence of radiative efficiency and
SED shape on accretion rate,
measurements of the luminosity function at different wavelengths 
over the widest achievable range in luminosity and redshift,
estimates of masses and accretion rates of active black holes
as a function of redshift and luminosity,
and indirect estimates of black hole space densities from
host galaxy and quasar clustering studies.
Fortunately, the observational situation is advancing rapidly,
and many of these areas have seen substantial progress in the
last few months alone, as discussed in \S\ref{sec:illus}.
It is worth emphasizing the value of luminosity function determinations
and black hole mass estimates that traverse the break in the QLF
and extend as far below as possible.
Accurate characterization of this regime is crucial for separating
the roles of $n(M)$ and $p(\mdot)$ in shaping the luminosity function,
which in turn is necessary for understanding the contribution of 
sub-Eddington accretion rates to black hole mass evolution.
These lower luminosities are also where optical and X-ray evolution
appear to be radically different, and better measurements of the
joint X-ray, optical, and IR luminosity functions
are needed to pin down the origin of these differences.
The emerging data on black hole and quasar evolution are complex,
complementary, and rich.  We hope that the physical modeling
approach described in this paper will prove useful in exploiting
their power.

\acknowledgments

We thank Jordi Miralda-Escud\'e for crucial suggestions during the early
stages of this work.  We also thank Jordi and Martin Haehnelt, Paul Martini,
and Smita Mathur for providing detailed comments on the manuscript that
led to significant improvements.
We have also benefited from discussions with many other colleagues, including
Ramesh Narayan, Patrick Osmer, Bradley Peterson, Richard Pogge,
Hans-Walter Rix, and Marianne Vestergaard.
The discussion of quasar hosts in \S\ref{sec:hostdensity} was stimulated
by conversations with Hans-Walter Rix.
This work was supported by NSF grants AST-0098515 and AST-0098584.

\clearpage
\appendix

\section{Luminosity Function for a Double Power-Law $p(\mdot)$}
\label{sec:appx1}

For the double power-law $p(\mdot)$ and double power-law $n(M)$,
the convolution integral~(\ref{eqn:lumfun}) for the luminosity
function must be broken into three different regimes to 
account for the different efficiencies of the accretion modes.  
The total QLF is the sum of the QLFs produced by each accretion mode, 
$\Phi(L)_{{\rm Total}}=\Phi(L)_{{\rm SE}}+\Phi(L)_{{\rm TD}}+
\Phi(L)_{{\rm ADAF}}$. 
The calculation is analogous to that in \S\ref{sec:powerlawpofmdot},
though more tedious, and we omit the details.  The solution depends
on whether $\mdotstar$ lies in the thin-disk, super-Eddington, or ADAF
regimes.  For the first case, which is the one usually relevant to our
models, the results for the three accretion modes are
\be
\Phi(L)_{\rm SE}= \left\{ \begin{array}{ll}
 { \nstar \pstar \over l (b+1) \mdotstar^b} \left[ 10^{b+1} - 1 \right] 
 \left({ L \over l \Mstar}\right)^{\alpha} & \mbox{ $L<l \Mstar$}  \\ [3mm]
 { \nstar \pstar \over l (b+1) \mdotstar^b} \left[ 10^{b+1} - 1 \right] 
 \left({ L \over l \Mstar}\right)^{\beta} & \mbox{ $L>l \Mstar$~,}
 \end{array}
 \right.
\label{eqn:sol:dblpow:se}
\ee
\be
\Phi(L)_{\rm TD}= \left\{ \begin{array}{ll}
 {\nstar \pstar \over l} 
 \left[{\mdotstar^{a-\alpha}-0.01^{a-\alpha} \over (a-\alpha) \mdotstar^a } + 
 {1-\mdotstar^{b-\alpha} \over (b-\alpha) \mdotstar^b }\right]
 \left( {L \over l \Mstar}\right)^\alpha & \mbox{ $ L<0.01 l \Mstar$} \\ [3mm]
 
 {\nstar \pstar \over l} 

 \left[{ (\beta-\alpha) \mdotstar^{-a} \over (a-\alpha)(a-\beta)} 
 \left({L \over l \Mstar}\right)^a + 
 {(b-a) \mdotstar^{-\alpha} \over (a-\alpha)(b-\alpha)}
 \left({L \over l \Mstar}\right)^{\alpha} \right. & \\ [3mm]

 \left. \;\;\;\;\;\; + { 1 \over (b-\alpha) \mdotstar^b} 
 \left({L \over l \Mstar}\right)^{\alpha} - 
 {0.01^{a-\beta} \over (a - \beta) \mdotstar^a} 
 \left({L \over l \Mstar}\right)^{\beta}\right] & 
 \mbox{ $0.01 l \Mstar<L<\mdotstar l \Mstar$} \\ [3mm]

 {\nstar \pstar \over l} 
 \left[{ (\beta-\alpha) \mdotstar^{-b} \over (b-\alpha)(b-\beta)} 
 \left({L \over  l \Mstar}\right)^b + 
 {(b-a) \mdotstar^{-\beta} \over (a-\beta)(b-\beta)}
 \left({L \over l \Mstar}\right)^{\beta} \right. & \\ [3mm]

 \left. \;\;\;\;\;\; + { 1 \over (b-\alpha) \mdotstar^b} 
 \left({L \over l \Mstar}\right)^{\alpha} -  
 { 0.01^{a-\beta}\over (a - \beta) \mdotstar^a} 
 \left({L \over l \Mstar}\right)^{\beta}\right] & 
 \mbox{ $\mdotstar l \Mstar<L< l \Mstar$} \\ [3mm]

 {\nstar \pstar \over l} 
 \left[{\mdotstar^{a-\beta}-0.01^{a-\beta} \over (a-\alpha) \mdotstar^a } + 
 {1-\mdotstar^{b-\beta} \over (b-\alpha) \mdotstar^b }\right]
 \left( {L \over l \Mstar}\right)^\beta & \mbox{ $ L > l \Mstar$~,} 
 
 \end{array}
 \right.
\label{eqn:sol:dblpow:td}
\ee
and
\be
\Phi(L)_{\rm ADAF}= \left\{ \begin{array}{ll}
 {\nstar \pstar 0.01^{\alpha+1} \over l (a-2\alpha-1) \mdotstar^a}
 \left[ 0.01^{a-2\alpha-1} - 10^{-4(a-2\alpha-1)}\right] 
 \left( {L \over l \Mstar}\right)^\alpha &
\mbox{ $ L < 10^{-4} l \Mstar$} \\ [3mm]

 {\nstar \pstar \over l \mdotstar^a} 
 \left[ {2(\beta-\alpha) \over (a-2\beta-1)(a-2\alpha-1)} 
 \left({L \over l \Mstar}\right)^{a-1 \over 2} \right. & \\ [1mm]

 \left. \;\;\;\;\;\;\;\; - {10^{-4(a-2\beta-1)} \over a-2\beta-1}
 \left({L \over l \Mstar}\right)^\beta + 
 {0.01^{a-2\alpha-1} \over a-2\alpha-1}
 \left({L \over l \Mstar}\right)^\alpha \right] & 
 \mbox{ $10^{-4} l \Mstar <L< 0.01 l \Mstar$} \\ [3mm]

 {\nstar \pstar 0.01^{\beta+1} \over l (a-2\beta-1) \mdotstar^a}
 \left[ 0.01^{a-2\beta-1} - 10^{-4(a-2\beta-1)}\right] 
 \left( {L \over l \Mstar}\right)^\beta &
\mbox{ $ L > 0.01 l \Mstar$ ~,}
 \end{array}
 \right.
\label{eqn:sol:dblpow:adaf}
\ee
where the values $\mdotcrit=0.01$, $\mdotmax=10$, and 
$\mdotmin=10^{-4}$ are explicitly included in the solutions. 
The three regimes differ because of the different dependence of
$\epsd$ on $\mdot$ and because $\mdotstar$ lies
in the thin-disk regime, breaking that integral into more parts.
Note that the slopes at the high and low luminosity end of each mode
are equal to the mass function slopes ($\alpha$ and $\beta$),
for the reasons discussed in \S\ref{sec:powerlawpofmdot}.

\section{Critical Slopes}
\label{sec:appx2}

We gain some insight into results for general $p(\mdot)$ and $n(M)$
by considering cases in which $p(\mdot)=\mdot^a$ is a pure power-law
in some range $\mdotmin - \mdotmax$ and $n(M)=M^\alpha$ is a pure
power-law in some range $\Mmin - \Mmax$.  Analysis of such cases
reveals a number of critical slopes where the character
of the solutions changes.  For the critical $p(\mdot)$ slope $a=-2$, each 
logarithmic range of $\mdot$ contributes equally to black hole growth
and, if $\epsd$ is constant, to emissivity of the quasar population.
When $a \ll -2$, growth and emissivity are dominated by low accretion rates
(near $\mdotmin$),
and when $a \gg -2$ high accretion rates (near $\mdotmax$) dominate.
Our double power-law models have a low-$\mdot$ slope $a>-2$ and a
high-$\mdot$ slope $b<-2$, so that the largest contributions to
growth and emissivity are from accretion rates near $\mdotstar$,
which we usually take (at least at high redshift)
to lie in the thin-disk range $0.01 < \mdotstar < 1$.
These parameter choices make our results relatively insensitive
to our assumptions about efficiencies and the form of $p(\mdot)$ in
the ADAF and super-Eddington regimes.  However, low-$\mdot$ or
high-$\mdot$ accretion may be more important in the real universe,
at least at some redshifts, in which case the form of $p(\mdot)$
and behavior of the accretion physics in these regimes would
have a larger impact on observable properties of the quasar population.

The distribution of active black hole masses at a fixed luminosity
depends on both the $p(\mdot)$ and $n(M)$ slopes.  
When $\alpha < a+1$ and $\epsd={\rm constant}$, 
the black hole mass function is steep enough that low mass black holes 
with high accretion rates predominate (eq.~\ref{eqn:relmassdist}) ---
i.e., the most common black holes at a given luminosity $L$ are either those
with $M=\Mmin$ or those with the maximum accretion rate
and $M=L/(\epsd l \mdotmax)$.  Conversely, when $\alpha > a+1$, high
mass black holes with low accretion rates predominate.
For more general forms of $n(M)$ and $p(\mdot)$, the roles of
``minimum'' and ``maximum'' values are in practice played by
values where the slope of the distribution changes or where
there is a break in efficiency.
For example, if $\alpha < a+1$ and the power-law behavior of $p(\mdot)$ 
extends to $\mdot\approx 1$, 
then decreasing efficiency in the super-Eddington regime comes into
play, and the luminosity function is dominated by black holes
radiating near the Eddington luminosity.
This is the typical behavior for our double power-law models 
at high redshift, where $\mdotstar$ is close to unity and the
low-$\mdot$ slope of $p(\mdot)$ and high mass slope of $n(M)$
easily satisfy $\beta < a+1$.  At low redshift we have low values
of $\mdotstar$, so the high-$\mdot$ slope of $p(\mdot)$ becomes
important.  Even here we usually have $\beta < b+1$, so that
systems with $\mdot \approx \mdotstar$ dominate the
high luminosity end of $\Phi(L)$, but because the slopes that
we adopt ($b=-3$, $\beta=-3.4$) are not so far from the
critical relation, systems at a given luminosity span
a wide range of black hole masses and accretion rates.

Two more critical slopes arise when we ask whether 
the space density of black holes of mass $M$ increases or decreases with time.
If accretion drives the evolution of $n(M)$, then the critical slope
is $-1$: for a mass function steeper than $n(M) \propto M^{-1}$,
accretion increases $n(M)$, while for a shallower slope the
number of black holes ``lost'' to higher masses exceeds the 
number gained from lower masses, driving $n(M)$ down with time.
For merger driven growth, the corresponding critical slope is $-2$,
at least if objects merge with others of equal mass as in the
simple models considered here.  The difference in slopes arises
because accretion adds mass to the black hole population while
mergers do not, and the critical slope for mergers is steeper still
if black hole ejection or gravitational radiation reduce the mass
of the surviving merger product below the combined mass of its progenitors.
For black hole mass functions whose asymptotic
slopes match the asymptotic slopes of the \cite{boyle00} luminosity
function, mergers drive $n(M)$ up in the high mass regime and
down in the low mass regime.  Accretion increases $n(M)$ in both
regimes, but the increase is slow for low masses and rapid for
high masses.

\end{document}